\documentclass[11pt]{report}

\usepackage[T1]{fontenc}
\usepackage[utf8]{inputenc}
\usepackage{microtype}
\usepackage{rotating} 
\usepackage{newpxtext}
\usepackage{amsmath,amssymb}
\usepackage{authblk}
\usepackage[scaled=0.94]{helvet}      
\renewcommand{\sfdefault}{phv}
\usepackage{booktabs}
\usepackage{multirow}
\usepackage{longtable}

\usepackage{geometry}
\usepackage{xcolor}
\definecolor{ESAblue}{HTML}{002147}
\definecolor{ESAgold}{HTML}{D4AA00}
\definecolor{lightgray}{gray}{0.92}

\usepackage{fancyhdr}
\usepackage{titlesec}
\titleformat{\section}[hang]
  {\large\sffamily\bfseries\color{ESAblue}}
  {\thesection}{0.75em}{}
\titleformat{\subsection}[hang]
  {\normalsize\sffamily\bfseries\color{ESAblue}}
  {\thesubsection}{0.75em}{}

\usepackage[most]{tcolorbox}
\newtcolorbox{executivebox}{
  colback=lightgray,
  colframe=ESAblue,
  left=12pt,right=12pt,top=10pt,bottom=10pt,
  boxrule=0pt,
  borderline west={3pt}{0pt}{ESAblue},
  sharp corners,
  breakable
}

\newtcolorbox{keyquestions}{
  colback=lightgray!20,
  colframe=ESAblue,
  left=10pt, right=10pt, top=8pt, bottom=8pt,
  boxrule=0.5pt,
  sharp corners,
  breakable,
  borderline west={3pt}{0pt}{ESAblue},
  fonttitle=\sffamily\bfseries\color{ESAblue},
  title=Key Questions
}

\usepackage{graphicx,booktabs,tabularx,subcaption}
\usepackage{caption}
\usepackage{natbib}
\usepackage{aas_macros}
\usepackage[colorlinks=true,
            linkcolor=ESAblue,
            citecolor=ESAblue,
            urlcolor=ESAgold]{hyperref}

\usepackage{xparse}
\NewDocumentCommand{\mychapter}{m m O{}}{%
  \clearpage
  \refstepcounter{chapter}
  \addcontentsline{toc}{chapter}{\thechapter\ #1}
  \vspace*{2em}
  {\Large\sffamily\bfseries\color{ESAblue}\thechapter\quad #1}\par
  \vspace{1em}
  \IfValueT{#3}{\label{#3}}
  \input{#2}
}

\begin{document}
\begin{titlepage}
\centering
\vspace*{2cm}

\includegraphics[width=1.0\textwidth]{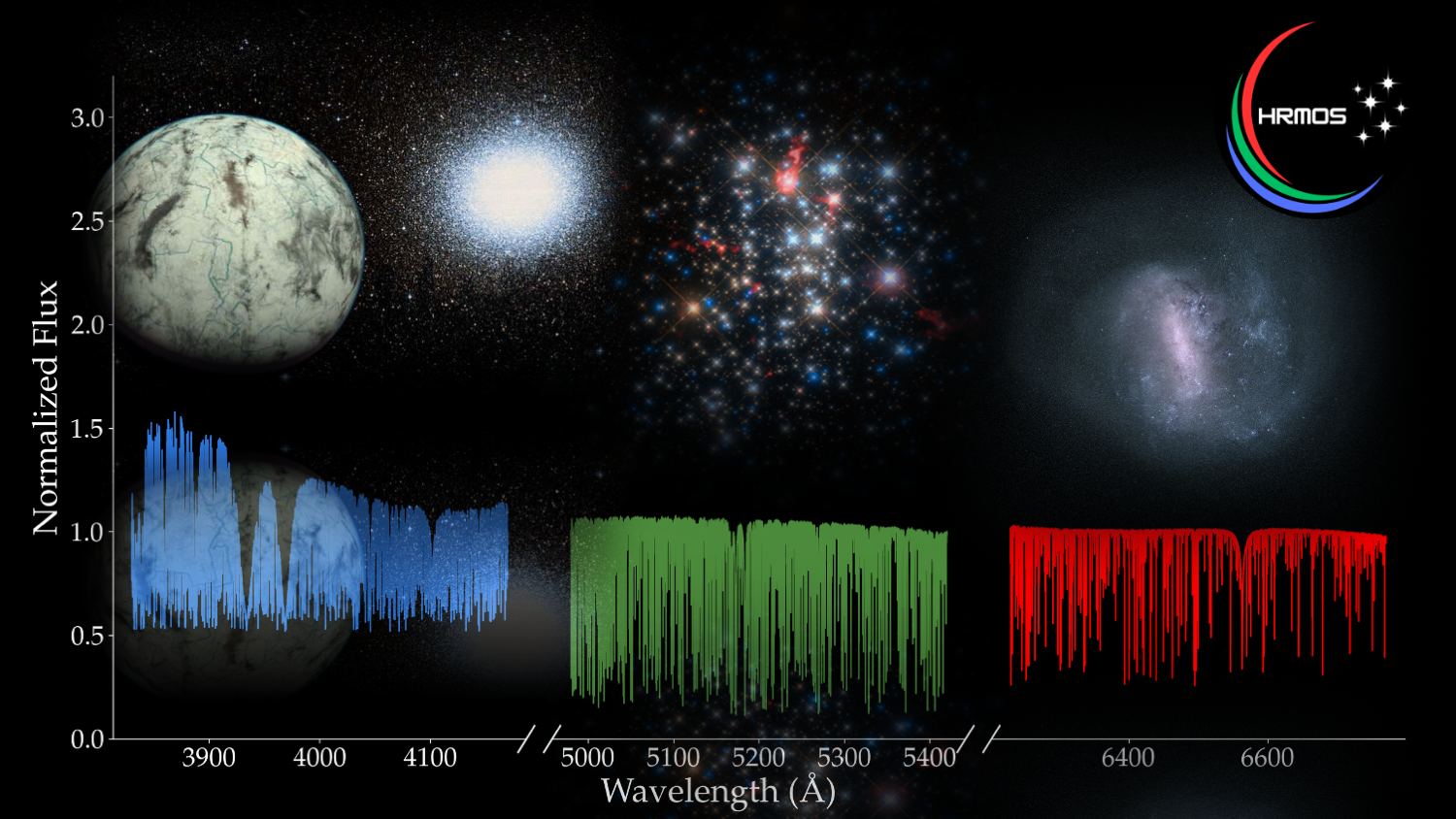}\\[0.5cm]

\color{ESAblue}\rule{\linewidth}{2pt}\color{black}\\[0.6cm]

{\fontsize{70}{84}\sffamily\bfseries HRMOS}\\[0.35cm]
{\large\sffamily A High-Resolution Multi-Object Spectrograph for the VLT}\\[0.9cm]
{\Large\sffamily White Paper}\\[0.35cm]
{\large\sffamily Science Motivation}\\[1.5cm]

\color{ESAblue}\rule{0.6\linewidth}{1.5pt}\color{black}\\[1.6cm]

{\large\sffamily \today}
\vfill

\clearpage
\fancyhead[R]{\textsf{Authors}}
\thispagestyle{fancy}
\phantomsection

\section*{Authors}

\begingroup
\small
\raggedright

Laura Magrini$^{1}$, 
Thomas Bensby$^{2}$, 
Sofia Randich$^{1}$, 
Andrea Bianco$^{3}$, 
Oscar Gonzalez$^{4}$, 
Emma Fern\'andez-Alvar$^{5,6}$, 
Sergio G. Sousa$^{7}$, 
Letizia Caito$^{8}$, 
Marco Riva$^{3}$, 
Vardan Adibekyan$^{7,9}$
% ...

\endgroup

\end{titlepage}

\fancyhead[R]{\textsf{Executive summary}}
\begin{executivebox}
\textbf{\large Executive Summary}

This White Paper presents the updated scientific rationale and instrument concept 
for {\sc HRMOS} (High-Resolution Multi-Object Spectrograph), a next generation instrument that will be proposed for the European Southern Observatory (ESO) Very Large Telescope (VLT) in the context of the VLT 2030 call and roadmap. 

Over the past decades, large spectroscopic surveys have transformed our understanding of Galactic and extragalactic astrophysics, while ultra-high-resolution spectrographs have delivered unprecedented precision for individual targets, opening new frontiers in precision radial velocity for exoplanet detection and detailed chemical profiling. New multi-object spectrographs (MOS) and surveys are starting or are proposed; however, no existing or planned instrument at ESO and, more in general in the international landscape, combines these complementary capabilities: MOS capability and very high resolution. This is not only a technical gap, but a scientific need.

{\sc HRMOS} is conceived to bridge this gap; by combining very high spectral resolution, multi-object observing capability, and stability, it will enable statistically significant studies, in a variety of Galactic and extra-galactic environments, requiring both spectral fidelity and observational efficiency.

The baseline design delivers a resolving power of R$=80\,000$, radial velocity precision of 10\,m\,s$^{-1}$ (with a goal of 5\,m\,s$^{-1}$), simultaneous observations of 50–60 targets, and broad wavelength coverage across three optical spectral windows down to 385\,nm. These capabilities will allow highly precise measurements of elemental abundances, isotopic ratios, line profiles, and radial velocities for large samples of stars, including in crowded stellar fields,  members of star clusters, the Galactic bulge, and nearby dwarf galaxies that will not practically be feasible with current or upcoming very high-resolution facilities.

{\sc HRMOS} will solve fundamental questions in astrophysics that we expect to be still open when it will be operating. Specifically, it will enable the precise and accurate determination of the ages of the oldest globular clusters through nucleocosmochronology, and thereby place independent and robust constraints on the age of the Galaxy, the timeline of cosmic evolution, and H$_0$; it will allow surveys of hot Jupiters in complete samples of star cluster members, testing how planets form and survive in different environments; it will probe hierarchical galaxy assembly in the Milky Way satellite galaxies, identifying relics of ancient accretion events; it will trace the astrophysical sites responsible for the production of the heaviest elements, unveiling the contribution of neutron star mergers; it will advance studies of star clusters and stellar physics; it will characterize the interstellar and circumgalactic gas through absorption spectroscopy, revealing the chemical composition, physical conditions, and gas flows that regulate the exchange of matter between stars, galaxies, and their surroundings.

{\sc HRMOS} occupies a unique region of the parameter space, complementing both current and forthcoming multi-object survey facilities and single-object high-resolution spectrographs. In particular, it will provide an essential bridge between large spectroscopic surveys and the capabilities of the next generation of extremely large telescopes, maximizing scientific return through strong synergies with facilities such as 4MOST, {\em Gaia}, TESS, PLATO, the proposed Haydn space mission and the future ELT instrumentation.

Noticeably, to ensure the success of this science framework, the {\sc HRMOS} initiative proactively integrates two key components directly into its deployment: namely, {\it i} laboratory astrophysics and a collaboration with atomic theory and experimental groups to deliver highly accurate atomic data before first light; {\it ii} advanced atmospheric modelling: Expanding grids of full 3D Radiation-Hydrodynamics (RHD) stellar atmospheres and multi-element non-LTE codes.

Finally, {\sc HRMOS} represents the natural technological evolution and heritage of the VLT/FLAMES instrument, upgrading its capabilities to meet the challenges of the 2030s. The proposed instrument builds upon proven technologies, while introducing innovative solutions for fibre positioning, atmospheric dispersion correction, and fibre scrambling. Together, these developments enable the radial velocity stability and spectroscopic precision required by the ambitious science goals, while maintaining operational efficiency and compatibility with the VLT infrastructure. Overall, we believe that it represents a strategic investment in the future of European astronomy.
\end{executivebox}

\clearpage
\fancyhead[R]{\textsf{Table-of-contents}}
\tableofcontents
\clearpage

\mychapter{Introduction}{intro.tex}[chapt:intro]
\mychapter{Exoplanets across Galactic environments}{exoplanets.tex}[chapt:exoplanets]
\mychapter{Origin of the elements and nucleosynthesis}{nucleosynthesis.tex}[chapt:origin]
\mychapter{Hierarchical formation and evolution of nearby galaxies}{Satellites.tex}[chapt:satellites]
\mychapter{Nucleocosmochronology: Constraining Cosmology from the Ages of the Oldest Milky Way Stars}{nucleocosmo.tex}[chapt:nucleocosmo]
\mychapter{High-resolution spectroscopy of star clusters}{clusters.tex}[chapt:clusters]
\mychapter{Galactic and extragalactic gas reservoirs}{galactic_and_extragalactic_gas.tex}[chapt:gas]
\mychapter{Atomic data for high-precision spectroscopy}{atomic_data.tex}[chapt:atomic]
\mychapter{Advanced stellar atmosphere and radiative transfer models for high-precision spectroscopy}{model_atmo.tex}[chapt:atmo]
\mychapter{Synergies and complementarities with present and forthcoming instrumentation}{sinergy_v2.tex}[chapt:synergy]
\mychapter{Target selection: Crowded stellar fields in the Milky Way and nearby dwarf galaxies}{targets.tex}[chapt:targets]
\mychapter{Appendix}{appendix.tex}[chapt:appendix]

\clearpage
\phantomsection
\addcontentsline{toc}{chapter}{List of Acronyms}
\fancyhead[R]{\textsf{List of acronyms}}

\section*{List of acronyms} \label{sec:acronyms}
1D - One dimensional\\
2D - Two dimensional\\
3D - Three dimensional\\
4MOST - 4-metre Multi-Object Spectroscopic Telescope \\
AAT - Anglo-Australian Telescope \\
ADC -  Atmospheric dispersion corrector\\
AGB - Asymptotic giant branch \\
AGN- Active Galactic Nucleus\\
ANDES - AmazoNes high Dispersion Echelle Spectrograph\\
APOGEE - Apache Point Observatory Galactic Evolution Experiment \\
ASD - Atomic Spectra Database \\
BEBOP -Binaries Escorted By Orbiting Planets\\
CA -Core Accretion\\
CBP - Circumbinary planet \\
CCD -  Charge-coupled device\\
CCF - Cross-Correlation Function\\
ccSN - Core-Collapse SuperNovae\\
CDM - Cold Dark Matter\\
CDPS - the Cluster Difference Imaging Photometric Survey\\
CEMP - Carbon-enhanced metal-poor \\
CEMP-rs - Carbon-enhanced metal-poor with r- and s- elements\\
CHEOPS - CHaracterising ExOPlanet Satellite\\
CMB - Cosmic Microwave Background\\
CMOS - Complementary Metal–Oxide–Semiconductor\\
CTL - Candidate Target List\\
CRIRES+ - CRyogenic high-resolution InfraRed Echelle Spectrograph \\ 
CVZ - Continuous Viewing Zone \\
DAZ - D = White Dwarf, A = hydrogen-dominated atmosphere, Z = metals present.\\
DBZ - D = White Dwarf, B = helium-dominated atmosphere, Z = metals present.\\
DESI -  Dark Energy Spectroscopic Instrument\\
DI - Disc Instability\\
DM - Dark Matter\\
DSPH - Dwarf spheroidal \\
DZ -  D = White Dwarf, Z = metals present \\
ELT -  Extremely Large Telescope\\
EMP -  Extremely metal-poor \\
EP - Excitation potential \\
ESA - European Space Agency\\
ESO - European Southern Observatory\\
ESPRESSO - Echelle SPectrograph for Rocky Exoplanet and Stable Spectroscopic Observations\\
EUV- Extreme UltraViolet\\
EW - Equivalent Width \\
FLAMES - Fibre Large Array Multi Element Spectrograph\\
FDM - Fuzzy Dark Matter\\
FFIs - Full Frame Images\\ 
FOV - Field-Of-View \\
FTS - Fourier Transform Spectroscopy \\
FUV - Far UltraViolet \\
FNX - Fornax \\
FWHM - Full Width Half Maximum\\
GALAH - GALactic Archaeology with HERMES\\
GAPS - Global Architecture of Planetary Systems\\ 
GC - Globular Cluster \\
GEMS - Galaxy Evolution via Montecarlo Sampling\\
HARPS - The High Accuracy Radial velocity Planet Searcher\\
HAYDN - High-precision AsteroseismologY of DeNse stellar fields\\
HB- Horizontal Branch\\
HCL- Hollow cathode lamps \\ 
HERMES -  The high efficiency and resolution Mercator echelle spectrograph\\
HR - High Resolution \\
HRMOS - High-Resolution Multi-Object Spectrograph \\
HST - Hubble Space Telescope \\
K2 - Kepler Extended Mission \\
KMOS - K-band Multi Object Spectrograph\\
IC - Index Catalogue\\
iCCF - Interpolated Cross-Correlation Function\\
LAMOST - The Large Sky Area Multi-Object Fibre Spectroscopic Telescope \\
LBT - The Large Binocular Telescope\\
LFC - Laser Frequency Comb \\
LIM - Low- and intermediate-mass \\
LMC -  Large Magellanic Cloud \\
LR - Low Resolution \\
LSST - Large Synoptic Survey Telescope \\
LTE- Local Thermodynamic Equilibrium\\
MARCS - Model Atmospheres with a Radiative and Convective Scheme\\
mas - milliarcsecond\\
MCDHF - Multi-Configuration Dirac– Hartree–Fock \\
MDF - Metallicity Distribution Function\\
MHD - Magneto-HydroDynamics \\
MICADO - Multi-AO Imaging Camera for Deep Observations\\
MOONS - Multi-Object Optical and Near-infrared Spectrograph \\
MOS - Multi-Object Spectrograph \\
MOSAIC - the Multi-Object Spectrograph for the ELT \\
MP - Multiple Population\\
MS - Main Sequence \\
MSE - Maunakea Spectroscopic Explorer \\
MSTO - Main Sequence Turn Off\\
MW - Milky Way \\
MWM- Milky Way Mapper\\
NIR - Near InfraRed \\
NIST - National Institute of Standards and Technology\\
NGC- New General Catalogue\\
NLTE- Non Local Thermodynamic Equilibrium\\
NSM - Neutron star merger \\
OC - Open cluster \\
PATHOS - A PSF-based Approach to TESS High quality data Of Stellar clusters\\
PEPSI - Potsdam Echelle Polarimetric and Spectroscopic Instrument \\
PIE - Proton-Ingestion Event\\
PISN - Pair-instability SuperNova \\
PLATO - PLAnetary Transits and Oscillations of stars \\
PMS - Pre-main sequence \\
PSF - Point Spread Function\\
R - Resolving power \\
RAVE - The Radial Velocity Experiment \\
RC - Red Clump \\
RCI-  Relativistic Configuration Interaction \\
RCW - Rodgers, Campbell, Whiteoak Catalogue \\
RGB - Red Giant Branch\\
RHD - Radiation-HydroDynamics \\ 
RV - Radial velocity \\
SB1 - Spectroscopic Binary Type 1\\
SB2 - Spectroscopic Binary Type 2\\
SDSS -  Sloan Digital Sky Survey \\
SEGUE - Sloan Extension for Galactic Understanding and Exploration \\
SGB -  Sub Giant Branch \\
SGR -  Sagittarius \\
SMC -  Small Magellanic Cloud \\
SMHM - Stellar Mass-Halo Mass \\
SN - SuperNova \\
SNIa - SuperNova Type Ia\\ 
SNR - Signal-to-noise ratio \\
STC - Science and Technical Committee \\
TATOOINE -The Attempt to Observe Outer-planets In Non-single-stellar Environments \\
TESS - Transiting Exoplanet Survey Satellite \\
TLR - Top Level Requirements \\
TOIs - TESS Objects of Interest \\
TNG - Telescopio Nazionale Galileo \\
THYME - TESS Hunt for Young and Maturing Exoplanets\\
UES - The Utrecht Echelle Spectrograph \\
UCLES - University College London Echelle Spectrograph \\
UFD - Ultra-faint dwarf \\
UVES - The UltraViolet-Visual Echelle Spectrograph\\
VLT - Very Large Telescope \\
VPH - Volume Phase Holographic  \\
WEAVE -  WHT Enhanced Area Velocity Explorer\\
WD -White Dwarf\\
WDM - Warm Dark Matter\\
WFIRST - The Wide Field Infrared Survey Telescope\\
WHT - William Herschel Telescope \\
WST - The Wide-Field Spectroscopic telescope \\
ZAMS - Zero-age main sequence \\

\clearpage
\fancyhead[R]{\textsf{References}}
\phantomsection
\addcontentsline{toc}{chapter}{References}
\bibliography{clean_utf8}

\begin{thebibliography}{786}
\expandafter\ifx\csname natexlab\endcsname\relax\def\natexlab#1{#1}\fi

\bibitem[{{Abazajian} {et~al.}(2009){Abazajian}, {Adelman-McCarthy},
  {Ag{\"u}eros}, {Allam}, {Allende Prieto}, {An}, {Anderson}, {Anderson},
  {Annis}, {Bahcall}, \& et~al.}]{abazajian2009}
{Abazajian}, K.~N., {Adelman-McCarthy}, J.~K., {Ag{\"u}eros}, M.~A., {et~al.}
  2009, \apjs, 182, 543

\bibitem[{{Abbott} {et~al.}(2017){Abbott}, {Abbott}, {Abbott}, {Acernese},
  {Ackley}, {Adams}, {Adams}, {Addesso}, {Adhikari}, {Adya}, {Affeldt},
  {Afrough}, {Agarwal}, {Agathos}, {Agatsuma}, {Aggarwal}, {Aguiar}, {Aiello},
  {Ain}, {Ajith}, {Allen}, {Allen}, {Allocca}, {Altin}, {Amato}, {Ananyeva},
  {Anderson}, {Anderson}, {Angelova}, {Antier}, {Appert}, {Arai}, {Araya},
  {Areeda}, {Arnaud}, {Arun}, {Ascenzi}, {Ashton}, {Ast}, {Aston}, {Astone},
  {Atallah}, {Aufmuth}, {Aulbert}, {AultONeal}, {Austin}, {Avila-Alvarez},
  {Babak}, {Bacon}, {Bader}, {Bae}, {Bailes}, {Baker}, {Baldaccini},
  {Ballardin}, {Ballmer}, {Banagiri}, {Barayoga}, {Barclay}, {Barish},
  {Barker}, {Barkett}, {Barone}, {Barr}, {Barsotti}, {Barsuglia}, {Barta},
  {Barthelmy}, {Bartlett}, {Bartos}, {Bassiri}, {Basti}, {Batch}, {Bawaj},
  {Bayley}, {Bazzan}, {B{\'e}csy}, {Beer}, {Bejger}, {Belahcene}, {Bell},
  {Berger}, {Bergmann}, {Bernuzzi}, {Bero}, {Berry}, {Bersanetti}, {Bertolini},
  {Betzwieser}, {Bhagwat}, {Bhandare}, {Bilenko}, {Billingsley}, {Billman},
  {Birch}, {Birney}, {Birnholtz}, {Biscans}, {Biscoveanu}, {Bisht}, {Bitossi},
  {Biwer}, {Bizouard}, {Blackburn}, {Blackman}, {Blair}, {Blair}, {Blair},
  {Bloemen}, {Bock}, {Bode}, {Boer}, {Bogaert}, {Bohe}, {Bondu}, {Bonilla},
  {Bonnand}, {Boom}, {Bork}, {Boschi}, {Bose}, {Bossie}, {Bouffanais}, {Bozzi},
  {Bradaschia}, {Brady}, {Branchesi}, {Brau}, {Briant}, {Brillet}, {Brinkmann},
  {Brisson}, {Brockill}, {Broida}, {Brooks}, {Brown}, {Brown}, {Brunett},
  {Buchanan}, {Buikema}, {Bulik}, {Bulten}, {Buonanno}, {Buskulic}, {Buy},
  {Byer}, {Cabero}, {Cadonati}, {Cagnoli}, {Cahillane}, {Calder{\'o}n
  Bustillo}, {Callister}, {Calloni}, {Camp}, {Canepa}, {Canizares}, {Cannon},
  {Cao}, {Cao}, {Capano}, {Capocasa}, {Carbognani}, {Caride}, {Carney},
  {Carullo}, {Casanueva Diaz}, {Casentini}, {Caudill}, {Cavagli{\`a}},
  {Cavalier}, {Cavalieri}, {Cella}, {Cepeda}, {Cerd{\'a}-Dur{\'a}n},
  {Cerretani}, {Cesarini}, {Chamberlin}, {Chan}, {Chao}, {Charlton}, {Chase},
  {Chassande-Mottin}, {Chatterjee}, {Chatziioannou}, {Cheeseboro}, {Chen},
  {Chen}, {Chen}, {Cheng}, {Chia}, {Chincarini}, {Chiummo}, {Chmiel}, {Cho},
  {Cho}, {Chow}, {Christensen}, {Chu}, {Chua}, \& {Chua}}]{Abbott2017}
{Abbott}, B.~P., {Abbott}, R., {Abbott}, T.~D., {et~al.} 2017, \prl, 119,
  161101

\bibitem[{{Adamo} {et~al.}(2024){Adamo}, {Bradley}, {Vanzella}, {Claeyssens},
  {Welch}, {Diego}, {Mahler}, {Oguri}, {Sharon}, {Abdurro'uf}, {Hsiao}, {Xu},
  {Messa}, {Lassen}, {Zackrisson}, {Brammer}, {Coe}, {Kokorev}, {Ricotti},
  {Zitrin}, {Fujimoto}, {Inoue}, {Resseguier}, {Rigby}, {Jim{\'e}nez-Teja},
  {Windhorst}, {Hashimoto}, \& {Tamura}}]{Adamo2024}
{Adamo}, A., {Bradley}, L.~D., {Vanzella}, E., {et~al.} 2024, \nat, 632, 513

\bibitem[{{Adams}(2010)}]{adams2010}
{Adams}, F.~C. 2010, \araa, 48, 47

\bibitem[{{Adams} {et~al.}(2004){Adams}, {Hollenbach}, {Laughlin}, \&
  {Gorti}}]{adams2004}
{Adams}, F.~C., {Hollenbach}, D., {Laughlin}, G., \& {Gorti}, U. 2004, \apj,
  611, 360

\bibitem[{{Adibekyan}(2019)}]{Adibekyan2019Geosc...9..105A}
{Adibekyan}, V. 2019, Geosciences, 9, 105

\bibitem[{{Aguado} {et~al.}(2022){Aguado}, {Molaro}, {Caffau}, {Gonz{\'a}lez
  Hern{\'a}ndez}, {Zapatero Osorio}, {Bonifacio}, {Allende Prieto}, {Rebolo},
  {Damasso}, {Su{\'a}rez Mascare{\~n}o}, {Howell}, {Furlan}, {Cristiani},
  {Cupani}, {Di Marcantonio}, {D'Odorico}, {Lovis}, {Martins}, {Milakovi},
  {Murphy}, {Nunes}, {Pepe}, {Santos}, {Schmidt}, \& {Sozzetti}}]{Aguado2022}
{Aguado}, D.~S., {Molaro}, P., {Caffau}, E., {et~al.} 2022, \aap, 668, A86

\bibitem[{{Aguado} {et~al.}(2023){Aguado}, {Salvadori}, {Sk{\'u}lad{\'o}ttir},
  {Caffau}, {Bonifacio}, {Vanni}, {Gelli}, {Koutsouridou}, \&
  {Amarsi}}]{aguado2023}
{Aguado}, D.~S., {Salvadori}, S., {Sk{\'u}lad{\'o}ttir}, {\'A}., {et~al.} 2023,
  \mnras, 520, 866

\bibitem[{{Aguilera-G{\'o}mez} {et~al.}(2020){Aguilera-G{\'o}mez},
  {Chanam{\'e}}, \& {Pinsonneault}}]{aguilera20LSb}
{Aguilera-G{\'o}mez}, C., {Chanam{\'e}}, J., \& {Pinsonneault}, M.~H. 2020,
  \apjl, 897, L20

\bibitem[{{Aguilera-G{\'o}mez} {et~al.}(2016){Aguilera-G{\'o}mez},
  {Chanam{\'e}}, {Pinsonneault}, \& {Carlberg}}]{aguilera16LSb}
{Aguilera-G{\'o}mez}, C., {Chanam{\'e}}, J., {Pinsonneault}, M.~H., \&
  {Carlberg}, J.~K. 2016, \apj, 829, 127

\bibitem[{{Aguilera-G{\'o}mez} {et~al.}(2023){Aguilera-G{\'o}mez}, {Jones}, \&
  {Chanam{\'e}}}]{aguilera23LSb}
{Aguilera-G{\'o}mez}, C., {Jones}, M.~I., \& {Chanam{\'e}}, J. 2023, \aap, 670,
  A73

\bibitem[{{Aicken Davies} \& {Worley}(2025)}]{AickenDaviesWorley2025}
{Aicken Davies}, Q. \& {Worley}, C.~C. 2025, arXiv e-prints, arXiv:2510.22487

\bibitem[{{Alencastro Puls} {et~al.}(2025){Alencastro Puls}, {Kuske}, {Hansen},
  {Lombardo}, {Visentin}, {Arcones}, {Fernandes de Melo}, {Reichert},
  {Bonifacio}, {Caffau}, \& {Fritzsche}}]{AlencastroPuls2025}
{Alencastro Puls}, A., {Kuske}, J., {Hansen}, C.~J., {et~al.} 2025, \aap, 693,
  A294

\bibitem[{{Alexeeva} {et~al.}(2023){Alexeeva}, {Wang}, {Zhao}, {Wang}, {Wu},
  {Wang}, {Yan}, \& {Shi}}]{alexeeva2023}
{Alexeeva}, S., {Wang}, Y., {Zhao}, G., {et~al.} 2023, \apj, 957, 10

\bibitem[{{Allende Prieto} {et~al.}(2002){Allende Prieto}, {Asplund},
  {Garc{\'\i}a L{\'o}pez}, \& {Lambert}}]{allendeprieto2002b}
{Allende Prieto}, C., {Asplund}, M., {Garc{\'\i}a L{\'o}pez}, R.~J., \&
  {Lambert}, D.~L. 2002, \apj, 567, 544

\bibitem[{{Almeida} {et~al.}(2023){Almeida}, {Anderson},
  {Argudo-Fern{\'a}ndez}, {Badenes}, {Barger}, {Barrera-Ballesteros}, {Bender},
  {Benitez}, {Besser}, {Bird}, {Bizyaev}, {Blanton}, {Bochanski}, {Bovy},
  {Brandt}, {Brownstein}, {Buchner}, {Bulbul}, {Burchett}, {Cano D{\'\i}az},
  {Carlberg}, {Casey}, {Chandra}, {Cherinka}, {Chiappini}, {Coker}, {Comparat},
  {Conroy}, {Contardo}, {Cortes}, {Covey}, {Crane}, {Cunha}, {Dabbieri},
  {Davidson}, {Davis}, {de Andrade Queiroz}, {De Lee}, {M{\'e}ndez Delgado},
  {Demasi}, {Di Mille}, {Donor}, {Dow}, {Dwelly}, {Eracleous}, {Eriksen},
  {Fan}, {Farr}, {Frederick}, {Fries}, {Frinchaboy}, {G{\"a}nsicke}, {Ge},
  {Gonz{\'a}lez {\'A}vila}, {Grabowski}, {Grier}, {Guiglion}, {Gupta}, {Hall},
  {Hawkins}, {Hayes}, {Hermes}, {Hern{\'a}ndez-Garc{\'\i}a}, {Hogg},
  {Holtzman}, {Ibarra-Medel}, {Ji}, {Jofre}, {Johnson}, {Jones}, {Kinemuchi},
  {Kluge}, {Koekemoer}, {Kollmeier}, {Kounkel}, {Krishnarao}, {Krumpe},
  {Lacerna}, {Lago}, {Laporte}, {Liu}, {Liu}, {Liu}, {Lopes}, {Macktoobian},
  {Majewski}, {Malanushenko}, {Maoz}, {Masseron}, {Masters}, {Matijevic},
  {McBride}, {Medan}, {Merloni}, {Morrison}, {Myers}, {M{\'e}sz{\'a}ros},
  {Negrete}, {Nidever}, {Nitschelm}, {Oravetz}, {Oravetz}, {Pan}, {Peng},
  {Pinsonneault}, {Pogge}, {Qiu}, {Ramirez}, {Rix}, {Fern{\'a}ndez Rosso},
  {Runnoe}, {Salvato}, {Sanchez}, {Santana}, {Saydjari}, {Sayres},
  {Schlaufman}, {Schneider}, {Schwope}, {Serna}, {Shen}, {Sobeck}, {Song},
  {Souto}, {Spoo}, {Stassun}, {Steinmetz}, {Straumit}, {Stringfellow},
  {S{\'a}nchez-Gallego}, {Taghizadeh-Popp}, {Tayar}, {Thakar}, {Tissera},
  {Tkachenko}, {Hernandez Toledo}, {Trakhtenbrot}, {Fern{\'a}ndez-Trincado},
  {Troup}, {Trump}, {Tuttle}, {Ulloa}, {Vazquez-Mata}, {Vera Alfaro},
  {Villanova}, {Wachter}, {Weijmans}, {Wheeler}, {Wilson}, {Wojno}, {Wolf},
  {Xue}, {Ybarra}, {Zari}, \& {Zasowski}}]{mwm23}
{Almeida}, A., {Anderson}, S.~F., {Argudo-Fern{\'a}ndez}, M., {et~al.} 2023,
  \apjs, 267, 44

\bibitem[{{Amarsi} {et~al.}(2018{\natexlab{a}}){Amarsi}, {Barklem}, {Asplund},
  {Collet}, \& {Zatsarinny}}]{amarsi2018b}
{Amarsi}, A.~M., {Barklem}, P.~S., {Asplund}, M., {Collet}, R., \&
  {Zatsarinny}, O. 2018{\natexlab{a}}, \aap, 616, A89

\bibitem[{{Amarsi} {et~al.}(2019{\natexlab{a}}){Amarsi}, {Barklem}, {Collet},
  {Grevesse}, \& {Asplund}}]{amarsi2019a}
{Amarsi}, A.~M., {Barklem}, P.~S., {Collet}, R., {Grevesse}, N., \& {Asplund},
  M. 2019{\natexlab{a}}, \aap, 624, A111

\bibitem[{{Amarsi} {et~al.}(2025){Amarsi}, {Li}, {Grevesse}, \&
  {Jurewicz}}]{amarsi2025}
{Amarsi}, A.~M., {Li}, W., {Grevesse}, N., \& {Jurewicz}, A.~J.~G. 2025, \aap,
  703, A35

\bibitem[{{Amarsi} {et~al.}(2022){Amarsi}, {Liljegren}, \&
  {Nissen}}]{amarsi2022}
{Amarsi}, A.~M., {Liljegren}, S., \& {Nissen}, P.~E. 2022, \aap, 668, A68

\bibitem[{{Amarsi} {et~al.}(2020){Amarsi}, {Lind}, {Osorio}, {Nordlander},
  {Bergemann}, {Reggiani}, {Wang}, {Buder}, {Asplund}, {Barklem}, {Wehrhahn},
  {Sk{\'u}lad{\'o}ttir}, {Kobayashi}, {Karakas}, {Gao}, {Bland-Hawthorn}, {de
  Silva}, {Kos}, {Lewis}, {Martell}, {Sharma}, {Simpson}, {Zucker},
  {{\v{C}}otar}, {Horner}, \& {Galah Collaboration}}]{amarsi2020b}
{Amarsi}, A.~M., {Lind}, K., {Osorio}, Y., {et~al.} 2020, \aap, 642, A62

\bibitem[{{Amarsi} {et~al.}(2019{\natexlab{b}}){Amarsi}, {Nissen}, \&
  {Sk{\'u}lad{\'o}ttir}}]{amarsi2019b}
{Amarsi}, A.~M., {Nissen}, P.~E., \& {Sk{\'u}lad{\'o}ttir}, {\'A}.
  2019{\natexlab{b}}, \aap, 630, A104

\bibitem[{{Amarsi} {et~al.}(2018{\natexlab{b}}){Amarsi}, {Nordlander},
  {Barklem}, {Asplund}, {Collet}, \& {Lind}}]{amarsi2018}
{Amarsi}, A.~M., {Nordlander}, T., {Barklem}, P.~S., {et~al.}
  2018{\natexlab{b}}, \aap, 615, A139

\bibitem[{{Amarsi} {et~al.}(2024){Amarsi}, {Ogneva}, {Buldgen}, {Grevesse},
  {Zhou}, \& {Barklem}}]{amarsi2024}
{Amarsi}, A.~M., {Ogneva}, D., {Buldgen}, G., {et~al.} 2024, \aap, 690, A128

\bibitem[{{Andrae} {et~al.}(2023){Andrae}, {Rix}, \& {Chandra}}]{Andrae2023}
{Andrae}, R., {Rix}, H.-W., \& {Chandra}, V. 2023, \apjs, 267, 8

\bibitem[{{Andrews} {et~al.}(2018){Andrews}, {Huang}, {P{\'e}rez}, {Isella},
  {Dullemond}, {Kurtovic}, {Guzm{\'a}n}, {Carpenter}, {Wilner}, {Zhang}, {Zhu},
  {Birnstiel}, {Bai}, {Benisty}, {Hughes}, {{\"O}berg}, \&
  {Ricci}}]{andrews2018}
{Andrews}, S.~M., {Huang}, J., {P{\'e}rez}, L.~M., {et~al.} 2018, \apjl, 869,
  L41

\bibitem[{{Arav} {et~al.}(2018){Arav}, {Liu}, {Xu}, {Stidham}, {Benn}, \&
  {Chamberlain}}]{Arav2018ApJ...857...60A}
{Arav}, N., {Liu}, G., {Xu}, X., {et~al.} 2018, \apj, 857, 60

\bibitem[{{Arcones} \& {Thielemann}(2023)}]{ArconesThielemann2023}
{Arcones}, A. \& {Thielemann}, F.-K. 2023, \aapr, 31, 1

\bibitem[{{Arellano-C{\'o}rdova} \&
  {Rodr{\'\i}guez}(2020)}]{ArellanoCordova2020MNRAS.497..672A}
{Arellano-C{\'o}rdova}, K.~Z. \& {Rodr{\'\i}guez}, M. 2020, \mnras, 497, 672

\bibitem[{{Armitage} {et~al.}(2013){Armitage}, {Simon}, \&
  {Martin}}]{armitage2013}
{Armitage}, P.~J., {Simon}, J.~B., \& {Martin}, R.~G. 2013, \apjl, 778, L14

\bibitem[{{Armstrong} {et~al.}(2014){Armstrong}, {Osborn}, {Brown}, {Faedi},
  {G{\'o}mez Maqueo Chew}, {Martin}, {Pollacco}, \& {Udry}}]{Armstrong2014}
{Armstrong}, D.~J., {Osborn}, H.~P., {Brown}, D.~J.~A., {et~al.} 2014, \mnras,
  444, 1873

\bibitem[{{Arnould} \& {Goriely}(2003)}]{ArnouldGoriely2003}
{Arnould}, M. \& {Goriely}, S. 2003, \physrep, 384, 1

\bibitem[{{Arnould} {et~al.}(2007){Arnould}, {Goriely}, \&
  {Takahashi}}]{arnould2007}
{Arnould}, M., {Goriely}, S., \& {Takahashi}, K. 2007, \physrep, 450, 97

\bibitem[{{Aros} {et~al.}(2021){Aros}, {Sippel}, {Mastrobuono-Battisti},
  {Bianchini}, {Askar}, \& {van de Ven}}]{Aros2021}
{Aros}, F.~I., {Sippel}, A.~C., {Mastrobuono-Battisti}, A., {et~al.} 2021,
  \mnras, 508, 4385

\bibitem[{{Arroyo-Polonio} {et~al.}(2023){Arroyo-Polonio}, {Battaglia},
  {Thomas}, {Irwin}, {McConnachie}, \& {Tolstoy}}]{arroyo-polonio2023}
{Arroyo-Polonio}, J.~M., {Battaglia}, G., {Thomas}, G.~F., {et~al.} 2023, \aap,
  677, A95

\bibitem[{{Asplund} {et~al.}(2021){Asplund}, {Amarsi}, \&
  {Grevesse}}]{asplund2021}
{Asplund}, M., {Amarsi}, A.~M., \& {Grevesse}, N. 2021, arXiv e-prints,
  arXiv:2105.01661

\bibitem[{{Azhari} {et~al.}(2025){Azhari}, {Matsuno}, {Aoki}, {Ishigaki}, \&
  {Tolstoy}}]{Azhari2025}
{Azhari}, A., {Matsuno}, T., {Aoki}, W., {Ishigaki}, M.~N., \& {Tolstoy}, E.
  2025, \aap, 699, A276

\bibitem[{{Bacon} {et~al.}(2024){Bacon}, {Maineiri}, {Randich}, {Cimatti},
  {Kneib}, {Brinchmann}, {Ellis}, {Tolstoi}, {Smiljanic}, {Hill}, {Anderson},
  {Sanchez Saez}, {Opitom}, {Bryson}, {Dierickx}, {Garilli}, {Gonzalez}, {de
  Jong}, {Lee}, {Mieske}, {Otarola}, {Schipani}, {Travouillon}, {Vernet},
  {Bryant}, {Casali}, {Colless}, {Couch}, {Driver}, {Fontana}, {Lehnert},
  {Magrini}, {Montet}, {Pasquini}, {Roth}, {Sanchez-Janssen}, {Steinmetz},
  {Tresse}, {Yeche}, \& {Ziegler}}]{bacon24}
{Bacon}, R., {Maineiri}, V., {Randich}, S., {et~al.} 2024, in Society of
  Photo-Optical Instrumentation Engineers (SPIE) Conference Series, Vol. 13094,
  Ground-based and Airborne Telescopes X, ed. H.~K. {Marshall},
  J.~{Spyromilio}, \& T.~{Usuda}, 130941O

\bibitem[{{Bacon} {et~al.}(2023){Bacon}, {Roth}, {Amico}, {Hernandez}, \& {WST
  Consortium}}]{bacon2023}
{Bacon}, R., {Roth}, M.~M., {Amico}, P., {Hernandez}, E., \& {WST Consortium},
  t. 2023, arXiv e-prints, arXiv:2308.16064

\bibitem[{Badnell(2016)}]{2016ascl.soft12014B}
Badnell, N.~R. 2016, AUTOSTRUCTURE: General program for calculation of atomic
  and ionic properties, Astrophysics Source Code Library

\bibitem[{{Bai}(2016)}]{bai2016}
{Bai}, X.-N. 2016, \apj, 821, 80

\bibitem[{{Balsalobre-Ruza} {et~al.}(2025){Balsalobre-Ruza}, {Lillo-Box},
  {Silva}, {Grouffal}, {Aceituno}, {Castro-Gonz{\'a}lez}, {Cifuentes},
  {Standing}, {Faria}, {Figueira}, {Santerne}, {Marfil}, {Abreu}, {Aguichine},
  {Gonz{\'a}lez-Ram{\'\i}rez}, {Morales}, {Santos}, {Hu{\'e}lamo}, {Delgado
  Mena}, {Barrado}, {Adibekyan}, {Barros}, {Berihuete}, {Morales-Calder{\'o}n},
  {Nagel}, {Solano}, {Sousa}, {Ag{\"u}{\'\i} Fern{\'a}ndez}, {Azzaro},
  {Bergond}, {Cikota}, {Fern{\'a}ndez-Mart{\'\i}n}, {Flores}, {G{\'o}ngora},
  {Guijarro}, {Hermelo}, {Pinter}, \& {Vico Linares}}]{Balsalobre-Ruza2025}
{Balsalobre-Ruza}, O., {Lillo-Box}, J., {Silva}, A.~M., {et~al.} 2025, \aap,
  694, A15

\bibitem[{{Banerjee} {et~al.}(2018){Banerjee}, {Qian}, \&
  {Heger}}]{Banerjee2018}
{Banerjee}, P., {Qian}, Y.-Z., \& {Heger}, A. 2018, \apj, 865, 120

\bibitem[{Bar-Shalom {et~al.}(2001)Bar-Shalom, Klapisch, \&
  Oreg}]{BarShalom2001HULLAC}
Bar-Shalom, A., Klapisch, M., \& Oreg, J. 2001, Journal of Quantitative
  Spectroscopy and Radiative Transfer, 71, 169

\bibitem[{{Baranne} {et~al.}(1996){Baranne}, {Queloz}, {Mayor}, {Adrianzyk},
  {Knispel}, {Kohler}, {Lacroix}, {Meunier}, {Rimbaud}, \& {Vin}}]{Baranne1996}
{Baranne}, A., {Queloz}, D., {Mayor}, M., {et~al.} 1996, \aaps, 119, 373

\bibitem[{{Baratella} {et~al.}(2020){Baratella}, {D'Orazi}, {Carraro},
  {Desidera}, {Randich}, {Magrini}, {Adibekyan}, {Smiljanic}, {Spina},
  {Tsantaki}, {Tautvai{\v{s}}ien{\.{e}}}, {Sousa}, {Jofr{\'e}},
  {Jim{\'e}nez-Esteban}, {Delgado-Mena}, {Martell}, {Van der Swaelmen},
  {Roccatagliata}, {Gilmore}, {Alfaro}, {Bayo}, {Bensby}, {Bragaglia},
  {Franciosini}, {Gonneau}, {Heiter}, {Hourihane}, {Jeffries}, {Koposov},
  {Morbidelli}, {Prisinzano}, {Sacco}, {Sbordone}, {Worley}, {Zaggia}, \&
  {Lewis}}]{Baratella20}
{Baratella}, M., {D'Orazi}, V., {Carraro}, G., {et~al.} 2020, \aap, 634, A34

\bibitem[{{Baratella} {et~al.}(2021){Baratella}, {D'Orazi}, {Sheminova},
  {Spina}, {Carraro}, {Gratton}, {Magrini}, {Randich}, {Lugaro}, {Pignatari},
  {Romano}, {Biazzo}, {Bragaglia}, {Casali}, {Desidera}, {Frasca}, {de Silva},
  {Melo}, {Van der Swaelmen}, {Tautvai{\v{s}}ien{\.{e}}},
  {Jim{\'e}nez-Esteban}, {Gilmore}, {Bensby}, {Smiljanic}, {Bayo},
  {Franciosini}, {Gonneau}, {Hourihane}, {Jofr{\'e}}, {Monaco}, {Morbidelli},
  {Sacco}, {Sbordone}, {Worley}, \& {Zaggia}}]{Baratella2021}
{Baratella}, M., {D'Orazi}, V., {Sheminova}, V., {et~al.} 2021, \aap, 653, A67

\bibitem[{{Barklem}(2016)}]{barklem2016}
{Barklem}, P.~S. 2016, \aapr, 24, 9

\bibitem[{{Barklem} {et~al.}(2011){Barklem}, {Belyaev}, {Guitou}, {Feautrier},
  {Gad{\'e}a}, \& {Spielfiedel}}]{barklem2011}
{Barklem}, P.~S., {Belyaev}, A.~K., {Guitou}, M., {et~al.} 2011, \aap, 530, A94

\bibitem[{{Barnes} \& {Collier Cameron}(2001)}]{barnes2001}
{Barnes}, J.~R. \& {Collier Cameron}, A. 2001, \mnras, 326, 950

\bibitem[{{Barnes} {et~al.}(2000){Barnes}, {Collier Cameron}, {James}, \&
  {Donati}}]{barnes2000}
{Barnes}, J.~R., {Collier Cameron}, A., {James}, D.~J., \& {Donati}, J.~F.
  2000, \mnras, 314, 162

\bibitem[{{Barnes} {et~al.}(2017){Barnes}, {Jeffers}, {Anglada-Escud{\'e}},
  {Haswell}, {Jones}, {Tuomi}, {Feng}, {Jenkins}, \& {Petit}}]{barnes2017}
{Barnes}, J.~R., {Jeffers}, S.~V., {Anglada-Escud{\'e}}, G., {et~al.} 2017,
  \mnras, 466, 1733

\bibitem[{{Barnes} {et~al.}(2011){Barnes}, {Jeffers}, \& {Jones}}]{barnes2011}
{Barnes}, J.~R., {Jeffers}, S.~V., \& {Jones}, H.~R.~A. 2011, \mnras, 412, 1599

\bibitem[{{Barrag{\'a}n} {et~al.}(2019){Barrag{\'a}n}, {Aigrain}, {Kubyshkina},
  {Gandolfi}, {Livingston}, {Fridlund}, {Fossati}, {Korth}, {Parviainen},
  {Malavolta}, {Palle}, {Deeg}, {Nowak}, {Rajpaul}, {Zicher}, {Antoniciello},
  {Narita}, {Albrecht}, {Bedin}, {Cabrera}, {Cochran}, {de Leon},
  {Eigm{\"u}ller}, {Fukui}, {Granata}, {Grziwa}, {Guenther}, {Hatzes},
  {Kusakabe}, {Latham}, {Libralato}, {Luque},
  {Monta{\~n}{\'e}s-Rodr{\'\i}guez}, {Murgas}, {Nardiello}, {Pagano}, {Piotto},
  {Persson}, {Redfield}, \& {Tamura}}]{Barragan2019}
{Barrag{\'a}n}, O., {Aigrain}, S., {Kubyshkina}, D., {et~al.} 2019, \mnras,
  490, 698

\bibitem[{{Barrag{\'a}n} {et~al.}(2024){Barrag{\'a}n}, {Yu}, {Freckelton},
  {Meech}, {Cretignier}, {Mortier}, {Aigrain}, {Klein}, {O'Sullivan}, {Gillen},
  {Nielsen}, {Mallorqu{\'\i}n}, \& {Zicher}}]{Barrag2024MNRAS.531.4275B}
{Barrag{\'a}n}, O., {Yu}, H., {Freckelton}, A.~V., {et~al.} 2024, \mnras, 531,
  4275

\bibitem[{{Bastian} \& {Lardo}(2018)}]{BastianLardo2018}
{Bastian}, N. \& {Lardo}, C. 2018, \araa, 56, 83

\bibitem[{{Baumgardt} {et~al.}(2023){Baumgardt}, {H{\'e}nault-Brunet},
  {Dickson}, \& {Sollima}}]{Baumgardt2023}
{Baumgardt}, H., {H{\'e}nault-Brunet}, V., {Dickson}, N., \& {Sollima}, A.
  2023, \mnras, 521, 3991

\bibitem[{{Baumgardt} \& {Hilker}(2018)}]{Baumgardt2018}
{Baumgardt}, H. \& {Hilker}, M. 2018, \mnras, 478, 1520

\bibitem[{{Baumgardt} {et~al.}(2019){Baumgardt}, {Hilker}, {Sollima}, \&
  {Bellini}}]{2019MNRAS.482.5138B}
{Baumgardt}, H., {Hilker}, M., {Sollima}, A., \& {Bellini}, A. 2019, \mnras,
  482, 5138

\bibitem[{{Baycroft} {et~al.}(2025){Baycroft}, {Santerne}, {Triaud}, {Heidari},
  {Sebastian}, {Davis}, {Correia}, {Sairam}, {Freckelton}, {Adamson}, {Boisse},
  {Coleman}, {Dransfield}, {Faria}, {Grouffal}, {Hara}, {H{\'e}brard},
  {Kunovac}, {Martin}, {Maxted}, {Nelson}, {Scott}, {Scutt}, \&
  {Standing}}]{Baycroft2025}
{Baycroft}, T.~A., {Santerne}, A., {Triaud}, A. H.~M.~J., {et~al.} 2025,
  \mnras, 541, 2801

\bibitem[{{Bedin} {et~al.}(2004){Bedin}, {Piotto}, {Anderson}, {Cassisi},
  {King}, {Momany}, \& {Carraro}}]{Bedin2004}
{Bedin}, L.~R., {Piotto}, G., {Anderson}, J., {et~al.} 2004, \apjl, 605, L125

\bibitem[{Belmonte {et~al.}(2025)Belmonte, Mar, DjuroviÄ‡, MenÃ©ndez, \&
  Gavanski}]{Belmonte2025}
Belmonte, M.~T., Mar, S., DjuroviÄ‡, S., MenÃ©ndez, J.~A., \& Gavanski, L.
  2025, Spectrochimica Acta Part B: Atomic Spectroscopy, 107190

\bibitem[{Belmonte {et~al.}(2017)Belmonte, Pickering, Ruffoni, Hartog, Lawler,
  Guzman, \& Heiter}]{Belmonte2017}
Belmonte, M.~T., Pickering, J.~C., Ruffoni, M.~P., {et~al.} 2017, arXiv

\bibitem[{{Belokurov} {et~al.}(2018){Belokurov}, {Erkal}, {Evans}, {Koposov},
  \& {Deason}}]{belokurov2018}
{Belokurov}, V., {Erkal}, D., {Evans}, N.~W., {Koposov}, S.~E., \& {Deason},
  A.~J. 2018, \mnras, 478, 611

\bibitem[{{Benisty} {et~al.}(2022){Benisty}, {Dominik}, {Follette}, {Garufi},
  {Ginski}, {Hashimoto}, {Keppler}, {Kley}, \& {Monnier}}]{Benisty2022a}
{Benisty}, M., {Dominik}, C., {Follette}, K., {et~al.} 2022, arXiv e-prints,
  arXiv:2203.09991

\bibitem[{{Benito} {et~al.}(2020){Benito}, {Criado}, {H{\"u}tsi}, {Raidal}, \&
  {Veerm{\"a}e}}]{benito2020}
{Benito}, M., {Criado}, J.~C., {H{\"u}tsi}, G., {Raidal}, M., \& {Veerm{\"a}e},
  H. 2020, \prd, 101, 103023

\bibitem[{{Benito} {et~al.}(2025){Benito}, {H{\"u}tsi}, {M{\"u}{\"u}rsepp},
  {S{\'a}nchez Almeida}, {Urrutia}, {Vaskonen}, \& {Veerm{\"a}e}}]{benito2025}
{Benito}, M., {H{\"u}tsi}, G., {M{\"u}{\"u}rsepp}, K., {et~al.} 2025, Physics
  of the Dark Universe, 49, 102010

\bibitem[{{Bensby} {et~al.}(2019){Bensby}, {Bergemann}, {Rybizki}, {Lemasle},
  {Howes}, {Kovalev}, {Agertz}, {Asplund}, {Barklem}, {Battistini},
  {Casagrande}, {Chiappini}, {Church}, {Feltzing}, {Ford}, {Gerhard},
  {Kushniruk}, {Kordopatis}, {Lind}, {Minchev}, {McMillan}, {Rix}, {Ryde}, \&
  {Traven}}]{bensby2019}
{Bensby}, T., {Bergemann}, M., {Rybizki}, J., {et~al.} 2019, The Messenger,
  175, 35

\bibitem[{{Berdyugina} {et~al.}(2008){Berdyugina}, {Berdyugin}, {Fluri}, \&
  {Piirola}}]{berdyugina2008}
{Berdyugina}, S.~V., {Berdyugin}, A.~V., {Fluri}, D.~M., \& {Piirola}, V. 2008,
  \apjl, 673, L83

\bibitem[{{Bergemann} {et~al.}(2019){Bergemann}, {Gallagher}, {Eitner},
  {Bautista}, {Collet}, {Yakovleva}, {Mayriedl}, {Plez}, {Carlsson},
  {Leenaarts}, {Belyaev}, \& {Hansen}}]{bergemann2019}
{Bergemann}, M., {Gallagher}, A.~J., {Eitner}, P., {et~al.} 2019, \aap, 631,
  A80

\bibitem[{Bergstrom {et~al.}(1988)Bergstrom, Faris, Hallstadius, Lundberg,
  Persson, \& Wahlstr{\"o}m}]{Bergstrom1988}
Bergstrom, H., Faris, G., Hallstadius, H., {et~al.} 1988, Zeitschrift f{\"u}r
  Physik D Atoms, Molecules and Clusters, 8, 17

\bibitem[{{Bernab{\`o}} {et~al.}(2022){Bernab{\`o}}, {Turrini}, {Testi},
  {Marzari}, \& {Polychroni}}]{Bernabo2022}
{Bernab{\`o}}, L.~M., {Turrini}, D., {Testi}, L., {Marzari}, F., \&
  {Polychroni}, D. 2022, \apjl, 927, L22

\bibitem[{{Bertelli Motta} {et~al.}(2018){Bertelli Motta}, {Pasquali},
  {Richer}, {Michaud}, {Salaris}, {Bragaglia}, {Magrini}, {Randich}, {Grebel},
  {Adibekyan}, {Blanco-Cuaresma}, {Drazdauskas}, {Fu}, {Martell},
  {Tautvai{\v{s}}ien{\.{e}}}, {Gilmore}, {Alfaro}, {Bensby}, {Flaccomio},
  {Koposov}, {Korn}, {Lanzafame}, {Smiljanic}, {Bayo}, {Carraro}, {Casey},
  {Costado}, {Damiani}, {Franciosini}, {Heiter}, {Hourihane}, {Jofr{\'e}},
  {Lardo}, {Lewis}, {Monaco}, {Morbidelli}, {Sacco}, {Sousa}, {Worley}, \&
  {Zaggia}}]{bertellimotta17}
{Bertelli Motta}, C., {Pasquali}, A., {Richer}, J., {et~al.} 2018, \mnras, 478,
  425

\bibitem[{{Bertran de Lis} {et~al.}(2022){Bertran de Lis}, {Allende Prieto},
  {Ludwig}, \& {Koesterke}}]{bertrandelis2022}
{Bertran de Lis}, S., {Allende Prieto}, C., {Ludwig}, H.~G., \& {Koesterke}, L.
  2022, \aap, 661, A76

\bibitem[{{Besla} {et~al.}(2010){Besla}, {Kallivayalil}, {Hernquist}, {van der
  Marel}, {Cox}, \& {Kere{\v{s}}}}]{besla2010}
{Besla}, G., {Kallivayalil}, N., {Hernquist}, L., {et~al.} 2010, \apjl, 721,
  L97

\bibitem[{{Bianchini} {et~al.}(2018){Bianchini}, {van der Marel}, {del Pino},
  {Watkins}, {Bellini}, {Fardal}, {Libralato}, \& {Sills}}]{Bianchini2018}
{Bianchini}, P., {van der Marel}, R.~P., {del Pino}, A., {et~al.} 2018, \mnras,
  481, 2125

\bibitem[{{Biazzo} {et~al.}(2022){Biazzo}, {D'Orazi}, {Desidera}, {Turrini},
  {Benatti}, {Gratton}, {Magrini}, {Sozzetti}, {Baratella}, {Bonomo}, {Borsa},
  {Claudi}, {Covino}, {Damasso}, {Di Mauro}, {Lanza}, {Maggio}, {Malavolta},
  {Maldonado}, {Marzari}, {Micela}, {Poretti}, {Vitello}, {Affer}, {Bignamini},
  {Carleo}, {Cosentino}, {Fiorenzano}, {Giacobbe}, {Harutyunyan}, {Leto},
  {Mancini}, {Molinari}, {Molinaro}, {Nardiello}, {Nascimbeni}, {Pagano},
  {Pedani}, {Piotto}, {Rainer}, \& {Scandariato}}]{Biazzoetal2022}
{Biazzo}, K., {D'Orazi}, V., {Desidera}, S., {et~al.} 2022, \aap, 664, A161

\bibitem[{{Bijavara Seshashayana} {et~al.}(2025){Bijavara Seshashayana},
  {J{\"o}nsson}, {D'Orazi}, {Bragaglia}, {Jian}, {Andreuzzi}, \& {Dal
  Ponte}}]{Bijavara2025}
{Bijavara Seshashayana}, S., {J{\"o}nsson}, H., {D'Orazi}, V., {et~al.} 2025,
  arXiv e-prints, arXiv:2511.05115

\bibitem[{{Bischetti} {et~al.}(2022){Bischetti}, {Feruglio}, {D'Odorico},
  {Arav}, {Ba{\~n}ados}, {Becker}, {Bosman}, {Carniani}, {Cristiani}, {Cupani},
  {Davies}, {Eilers}, {Farina}, {Ferrara}, {Maiolino}, {Mazzucchelli},
  {Mesinger}, {Meyer}, {Onoue}, {Piconcelli}, {Ryan-Weber}, {Schindler},
  {Wang}, {Yang}, {Zhu}, \& {Fiore}}]{Bischetti2022Natur.605..244B}
{Bischetti}, M., {Feruglio}, C., {D'Odorico}, V., {et~al.} 2022, \nat, 605, 244

\bibitem[{{Bongiorno} {et~al.}(2007){Bongiorno}, {Zamorani}, {Gavignaud},
  {Marano}, {Paltani}, {Mathez}, {M{\o}ller}, {Picat}, {Cirasuolo},
  {Lamareille}, {Bottini}, {Garilli}, {Le Brun}, {Le F{\`e}vre}, {Maccagni},
  {Scaramella}, {Scodeggio}, {Tresse}, {Vettolani}, {Zanichelli}, {Adami},
  {Arnouts}, {Bardelli}, {Bolzonella}, {Cappi}, {Charlot}, {Ciliegi},
  {Contini}, {Foucaud}, {Franzetti}, {Guzzo}, {Ilbert}, {Iovino}, {McCracken},
  {Marinoni}, {Mazure}, {Meneux}, {Merighi}, {Pell{\`o}}, {Pollo}, {Pozzetti},
  {Radovich}, {Zucca}, {Hatziminaoglou}, {Polletta}, {Bondi}, {Brinchmann},
  {Cucciati}, {de la Torre}, {Gregorini}, {Mellier}, {Merluzzi}, {Temporin},
  {Vergani}, \& {Walcher}}]{Bongiorno2007A&A...472..443B}
{Bongiorno}, A., {Zamorani}, G., {Gavignaud}, I., {et~al.} 2007, \aap, 472, 443

\bibitem[{{Bonifacio} {et~al.}(2025){Bonifacio}, {Caffau}, {Fran{\c{c}}ois}, \&
  {Spite}}]{bonifacio2025}
{Bonifacio}, P., {Caffau}, E., {Fran{\c{c}}ois}, P., \& {Spite}, M. 2025,
  \aapr, 33, 2

\bibitem[{{Borguet} {et~al.}(2013){Borguet}, {Arav}, {Edmonds}, {Chamberlain},
  \& {Benn}}]{Borguet2013ApJ...762...49B}
{Borguet}, B. C.~J., {Arav}, N., {Edmonds}, D., {Chamberlain}, C., \& {Benn},
  C. 2013, \apj, 762, 49

\bibitem[{{Borucki} {et~al.}(2010){Borucki}, {Koch}, {Basri}, {Batalha},
  {Brown}, {Caldwell}, {Caldwell}, {Christensen-Dalsgaard}, {Cochran},
  {DeVore}, {Dunham}, {Dupree}, {Gautier}, {Geary}, {Gilliland}, {Gould},
  {Howell}, {Jenkins}, {Kondo}, {Latham}, {Marcy}, {Meibom}, {Kjeldsen},
  {Lissauer}, {Monet}, {Morrison}, {Sasselov}, {Tarter}, {Boss}, {Brownlee},
  {Owen}, {Buzasi}, {Charbonneau}, {Doyle}, {Fortney}, {Ford}, {Holman},
  {Seager}, {Steffen}, {Welsh}, {Rowe}, {Anderson}, {Buchhave}, {Ciardi},
  {Walkowicz}, {Sherry}, {Horch}, {Isaacson}, {Everett}, {Fischer}, {Torres},
  {Johnson}, {Endl}, {MacQueen}, {Bryson}, {Dotson}, {Haas}, {Kolodziejczak},
  {Van Cleve}, {Chandrasekaran}, {Twicken}, {Quintana}, {Clarke}, {Allen},
  {Li}, {Wu}, {Tenenbaum}, {Verner}, {Bruhweiler}, {Barnes}, \&
  {Prsa}}]{Borucki2010}
{Borucki}, W.~J., {Koch}, D., {Basri}, G., {et~al.} 2010, Science, 327, 977

\bibitem[{{Botelho} {et~al.}(2020){Botelho}, {Milone}, {Mel{\'e}ndez},
  {Alves-Brito}, {Spina}, \& {Bean}}]{Botelho2020}
{Botelho}, R.~B., {Milone}, A. d.~C., {Mel{\'e}ndez}, J., {et~al.} 2020,
  \mnras, 499, 2196

\bibitem[{{Bouchy} {et~al.}(2001){Bouchy}, {Pepe}, \& {Queloz}}]{bouchy2001}
{Bouchy}, F., {Pepe}, F., \& {Queloz}, D. 2001, \aap, 374, 733

\bibitem[{{Bouma} {et~al.}(2019){Bouma}, {Hartman}, {Bhatti}, {Winn}, \&
  {Bakos}}]{2019ApJS..245...13B}
{Bouma}, L.~G., {Hartman}, J.~D., {Bhatti}, W., {Winn}, J.~N., \& {Bakos},
  G.~{\'A}. 2019, \apjs, 245, 13

\bibitem[{{Bouma} {et~al.}(2020){Bouma}, {Hartman}, {Brahm}, {Evans},
  {Collins}, {Zhou}, {Sarkis}, {Quinn}, {de Leon}, {Livingsto n}, {Bergmann},
  {Stassun}, {Bhatti}, {Winn}, {Bakos}, {Abe}, {Crouzet}, {Dransfield},
  {Guillot}, {Marie-Sa inte}, {M{\'e}karnia}, {Triaud}, {Tinney}, {Henning},
  {Espinoza}, {Jord{\'a}n}, {Barbieri}, {Nandakumar}, {Trifono v}, {Vines},
  {Vuckovic}, {Ziegler}, {Law}, {Mann}, {Ricker}, {Vanderspek}, {Seager},
  {Jenkins}, {Burke}, {Dragomir}, {Levine}, {Quintana}, {Rodriguez}, {Smith},
  \& {Wohler}}]{2020AJ....160..239B}
{Bouma}, L.~G., {Hartman}, J.~D., {Brahm}, R., {et~al.} 2020, \aj, 160, 239

\bibitem[{{Bouvier}(2008)}]{bouvier2008}
{Bouvier}, J. 2008, \aap, 489, L53

\bibitem[{{Bouvier} {et~al.}(2007{\natexlab{a}}){Bouvier}, {Alencar},
  {Boutelier}, {Dougados}, {Balog}, {Grankin}, {Hodgkin}, {Ibrahimov}, {Kun},
  {Magakian}, \& {Pinte}}]{bouvier2007b}
{Bouvier}, J., {Alencar}, S.~H.~P., {Boutelier}, T., {et~al.}
  2007{\natexlab{a}}, \aap, 463, 1017

\bibitem[{{Bouvier} {et~al.}(2007{\natexlab{b}}){Bouvier}, {Alencar},
  {Harries}, {Johns-Krull}, \& {Romanova}}]{bouvier2007a}
{Bouvier}, J., {Alencar}, S.~H.~P., {Harries}, T.~J., {Johns-Krull}, C.~M., \&
  {Romanova}, M.~M. 2007{\natexlab{b}}, in Protostars and Planets V, ed.
  B.~{Reipurth}, D.~{Jewitt}, \& K.~{Keil}, 479

\bibitem[{{Bowen} {et~al.}(2005){Bowen}, {Jenkins}, {Pettini}, \&
  {Tripp}}]{bowen2005}
{Bowen}, D.~V., {Jenkins}, E.~B., {Pettini}, M., \& {Tripp}, T.~M. 2005, \apj,
  635, 880

\bibitem[{{Bragaglia} {et~al.}(2017){Bragaglia}, {Carretta}, {D'Orazi},
  {Sollima}, {Donati}, {Gratton}, \& {Lucatello}}]{Bragaglia2017}
{Bragaglia}, A., {Carretta}, E., {D'Orazi}, V., {et~al.} 2017, \aap, 607, A44

\bibitem[{{Bramich} \& {Horne}(2006)}]{Bramich2006MNRAS.367.1677B}
{Bramich}, D.~M. \& {Horne}, K. 2006, \mnras, 367, 1677

\bibitem[{{Brasseur} {et~al.}(2024){Brasseur}, {Jardine}, \&
  {Hussain}}]{brasseur2024}
{Brasseur}, C.~E., {Jardine}, M.~M., \& {Hussain}, G.~A.~J. 2024, \mnras, 530,
  2442

\bibitem[{{Brauer} {et~al.}(2025){Brauer}, {Mead}, {Wise}, {Bryan}, {Low},
  {Ji}, {Emerick}, {Andersson}, {Frebel}, \& {C{\^o}t{\'e}}}]{brauer2025}
{Brauer}, K., {Mead}, J., {Wise}, J.~H., {et~al.} 2025, \apj, 993, 2

\bibitem[{{Brazzini} {et~al.}(2025){Brazzini}, {D'Odorico}, {Bischetti},
  {Feruglio}, {Cupani}, {Becker}, \& {Tripodi}}]{Brazzini2025A&A...698A.145B}
{Brazzini}, M., {D'Odorico}, V., {Bischetti}, M., {et~al.} 2025, \aap, 698,
  A145

\bibitem[{{Bromm}(2013)}]{Bromm2013}
{Bromm}, V. 2013, Reports on Progress in Physics, 76, 112901

\bibitem[{{Brown}(2025)}]{Brown2025arXiv250301533B}
{Brown}, A. G.~A. 2025, arXiv e-prints, arXiv:2503.01533

\bibitem[{{Brucalassi} {et~al.}(2017){Brucalassi}, {Koppenhoefer}, {Saglia},
  {Pasquini}, {Ruiz}, {Bonifacio}, {Bedin}, {Libralato}, {Biazzo}, {Melo},
  {Lovis}, \& {Randich}}]{brucalassi2017}
{Brucalassi}, A., {Koppenhoefer}, J., {Saglia}, R., {et~al.} 2017, \aap, 603,
  A85

\bibitem[{{Brucalassi} {et~al.}(2014){Brucalassi}, {Pasquini}, {Saglia},
  {Ruiz}, {Bonifacio}, {Bedin}, {Biazzo}, {Melo}, {Lovis}, \&
  {Randich}}]{brucalassi2014}
{Brucalassi}, A., {Pasquini}, L., {Saglia}, R., {et~al.} 2014, \aap, 561, L9

\bibitem[{{Brucalassi} {et~al.}(2016){Brucalassi}, {Pasquini}, {Saglia},
  {Ruiz}, {Bonifacio}, {Le{\~a}o}, {Canto Martins}, {de Medeiros}, {Bedin},
  {Biazzo}, {Melo}, {Lovis}, \& {Randich}}]{brucalassi2016}
{Brucalassi}, A., {Pasquini}, L., {Saglia}, R., {et~al.} 2016, \aap, 592, L1

\bibitem[{{Brucalassi} {et~al.}(2022){Brucalassi}, {Tozzi}, {Oliva},
  {Gonzalez}, {Randich}, {de Silva}, {Tolstoy}, {Bianco}, {Genoni}, {Pariani},
  {Lawrence}, {Bensby}, {Hill}, {Jeffries}, {Lagarde}, {Magrini}, {Smiljanic},
  \& {Skulad{\"o}ttir}}]{Brucalassi2022}
{Brucalassi}, A., {Tozzi}, A., {Oliva}, E., {et~al.} 2022, in Society of
  Photo-Optical Instrumentation Engineers (SPIE) Conference Series, Vol. 12184,
  Ground-based and Airborne Instrumentation for Astronomy IX, ed. C.~J.
  {Evans}, J.~J. {Bryant}, \& K.~{Motohara}, 121841A

\bibitem[{{Bruntt} {et~al.}(2003){Bruntt}, {Grundahl}, {Tingley}, {Frandsen},
  {Stetson}, \& {Thomsen}}]{Bruntt2003A&A...410..323B}
{Bruntt}, H., {Grundahl}, F., {Tingley}, B., {et~al.} 2003, \aap, 410, 323

\bibitem[{{Buder} {et~al.}(2025){Buder}, {Buck}, {Sk{\'u}lad{\'o}ttir}, {Ness},
  {McKenzie}, \& {Monty}}]{buder2025}
{Buder}, S., {Buck}, T., {Sk{\'u}lad{\'o}ttir}, {\'A}., {et~al.} 2025, arXiv
  e-prints, arXiv:2510.11284

\bibitem[{{Burbidge} {et~al.}(1957){Burbidge}, {Burbidge}, {Fowler}, \&
  {Hoyle}}]{B2FH}
{Burbidge}, E.~M., {Burbidge}, G.~R., {Fowler}, W.~A., \& {Hoyle}, F. 1957,
  Reviews of Modern Physics, 29, 547

\bibitem[{{Burke} {et~al.}(2006){Burke}, {Gaudi}, {DePoy}, \&
  {Pogge}}]{Burke2006AJ....132..210B}
{Burke}, C.~J., {Gaudi}, B.~S., {DePoy}, D.~L., \& {Pogge}, R.~W. 2006, \aj,
  132, 210

\bibitem[{{Burn} \& {Mordasini}(2024)}]{Burn2025}
{Burn}, R. \& {Mordasini}, C. 2024, in Handbook of Exoplanets (Springer
  International Publishing), 143--2

\bibitem[{{Busso} {et~al.}(1999){Busso}, {Gallino}, \&
  {Wasserburg}}]{busso1999}
{Busso}, M., {Gallino}, R., \& {Wasserburg}, G.~J. 1999, \araa, 37, 239

\bibitem[{{Butcher}(1987)}]{Butcher1987}
{Butcher}, H.~R. 1987, \nat, 328, 127

\bibitem[{{Cadelano} {et~al.}(2022){Cadelano}, {Dalessandro}, {Salaris},
  {Bastian}, {Mucciarelli}, {Saracino}, {Martocchia}, \&
  {Cabrera-Ziri}}]{Cadelano2022}
{Cadelano}, M., {Dalessandro}, E., {Salaris}, M., {et~al.} 2022, \apjl, 924, L2

\bibitem[{{Caffau} {et~al.}(2005{\natexlab{a}}){Caffau}, {Bonifacio},
  {Faraggiana}, {Fran{\c{c}}ois}, {Gratton}, \& {Barbieri}}]{caffau2005a}
{Caffau}, E., {Bonifacio}, P., {Faraggiana}, R., {et~al.} 2005{\natexlab{a}},
  \aap, 441, 533

\bibitem[{{Caffau} {et~al.}(2005{\natexlab{b}}){Caffau}, {Bonifacio},
  {Faraggiana}, \& {Sbordone}}]{caffau2005b}
{Caffau}, E., {Bonifacio}, P., {Faraggiana}, R., \& {Sbordone}, L.
  2005{\natexlab{b}}, \aap, 436, L9

\bibitem[{{Caffau} {et~al.}(2014){Caffau}, {Monaco}, {Spite}, {Bonifacio},
  {Carraro}, {Ludwig}, {Villanova}, {Beletsky}, \& {Sbordone}}]{caffau2014}
{Caffau}, E., {Monaco}, L., {Spite}, M., {et~al.} 2014, \aap, 568, A29

\bibitem[{{Caffau} {et~al.}(2010){Caffau}, {Sbordone}, {Ludwig}, {Bonifacio},
  \& {Spite}}]{caffau2010}
{Caffau}, E., {Sbordone}, L., {Ludwig}, H.~G., {Bonifacio}, P., \& {Spite}, M.
  2010, Astronomische Nachrichten, 331, 725

\bibitem[{{Caffau} {et~al.}(2008){Caffau}, {Sbordone}, {Ludwig}, {Bonifacio},
  {Steffen}, \& {Behara}}]{Caffau2008A&A...483..591C}
{Caffau}, E., {Sbordone}, L., {Ludwig}, H.-G., {et~al.} 2008, \aap, 483, 591

\bibitem[{{Caliskan} {et~al.}(2026){Caliskan}, {Amarsi}, {J{\"o}nsson},
  {Grevesse}, \& {Sahoo}}]{Caliskan2026arXiv260505356C}
{Caliskan}, S., {Amarsi}, A.~M., {J{\"o}nsson}, P., {Grevesse}, N., \& {Sahoo},
  B.~K. 2026, arXiv e-prints, arXiv:2605.05356

\bibitem[{{Caliskan} {et~al.}(2025){Caliskan}, {Amarsi}, {Racca},
  {Koutsouridou}, {Barklem}, {Lind}, \& {Salvadori}}]{caliskan2025}
{Caliskan}, S., {Amarsi}, A.~M., {Racca}, M., {et~al.} 2025, \aap, 696, A210

\bibitem[{{Caliskan} {et~al.}(2024){Caliskan}, {Grumer}, \&
  {Amarsi}}]{caliskan2024}
{Caliskan}, S., {Grumer}, J., \& {Amarsi}, A.~M. 2024, Journal of Physics B
  Atomic Molecular Physics, 57, 055003

\bibitem[{{Cameron}(1957)}]{Cameron1957}
{Cameron}, A.~G.~W. 1957, \pasp, 69, 201

\bibitem[{{Cameron} \& {Fowler}(1971)}]{cameron71LSb}
{Cameron}, A.~G.~W. \& {Fowler}, W.~A. 1971, \apj, 164, 111

\bibitem[{{Canocchi} {et~al.}(2024){Canocchi}, {Lind}, {Lagae}, {Pietrow},
  {Amarsi}, {Kiselman}, {Andriienko}, \& {Hoeijmakers}}]{Canocchi2024_sun}
{Canocchi}, G., {Lind}, K., {Lagae}, C., {et~al.} 2024, \aap, 683, A242

\bibitem[{{Cantat-Gaudin} {et~al.}(2020){Cantat-Gaudin}, {Anders},
  {Castro-Ginard}, {Jordi}, {Romero-G{\'o}mez}, {Soubiran}, {Casamiquela},
  {Tarricq}, {Moitinho}, {Vallenari}, {Bragaglia}, {Krone-Martins}, \&
  {Kounkel}}]{cantatgaudin2020}
{Cantat-Gaudin}, T., {Anders}, F., {Castro-Ginard}, A., {et~al.} 2020, \aap,
  640, A1

\bibitem[{{Carleo} {et~al.}(2021){Carleo}, {Desidera}, {Nardiello},
  {Malavolta}, {Lanza}, {Livingston}, {Locci}, {Marzari}, {Messina}, {Turrini},
  {Baratella}, {Borsa}, {D'Orazi}, {Nascimbeni}, {Pinamonti}, {Rainer}, {Alei},
  {Bignamini}, {Gratton}, {Micela}, {Montalto}, {Sozzetti}, {Squicciarini},
  {Affer}, {Benatti}, {Biazzo}, {Bonomo}, {Claudi}, {Cosentino}, {Covino},
  {Damasso}, {Esposito}, {Fiorenzano}, {Frustagli}, {Giacobbe}, {Harutyunyan},
  {Leto}, {Magazz{\`u}}, {Maggio}, {Mainella}, {Maldonado}, {Mallonn},
  {Mancini}, {Molinari}, {Molinaro}, {Pagano}, {Pedani}, {Piotto}, {Poretti},
  {Redfield}, \& {Scandariato}}]{Carleo2021A&A...645A..71C}
{Carleo}, I., {Desidera}, S., {Nardiello}, D., {et~al.} 2021, \aap, 645, A71

\bibitem[{{Carretta} \& {Bragaglia}(2021)}]{CarrettaBragaglia2021}
{Carretta}, E. \& {Bragaglia}, A. 2021, \aap, 646, A9

\bibitem[{{Carretta} \& {Bragaglia}(2024)}]{Carretta2024}
{Carretta}, E. \& {Bragaglia}, A. 2024, \aap, 690, A158

\bibitem[{{Carretta} {et~al.}(2009{\natexlab{a}}){Carretta}, {Bragaglia},
  {Gratton}, \& {Lucatello}}]{carretta2009b}
{Carretta}, E., {Bragaglia}, A., {Gratton}, R., \& {Lucatello}, S.
  2009{\natexlab{a}}, \aap, 505, 139

\bibitem[{{Carretta} {et~al.}(2009{\natexlab{b}}){Carretta}, {Bragaglia},
  {Gratton}, {Lucatello}, {Catanzaro}, {Leone}, {Bellazzini}, {Claudi},
  {D'Orazi}, {Momany}, {Ortolani}, {Pancino}, {Piotto}, {Recio-Blanco}, \&
  {Sabbi}}]{carretta2009a}
{Carretta}, E., {Bragaglia}, A., {Gratton}, R.~G., {et~al.} 2009{\natexlab{b}},
  \aap, 505, 117

\bibitem[{{Cassisi} \& {Salaris}(2020)}]{CassisiSalaris2020}
{Cassisi}, S. \& {Salaris}, M. 2020, \aapr, 28, 5

\bibitem[{{Castro-Gonz{\'a}lez} {et~al.}(2025){Castro-Gonz{\'a}lez}, {Bouchy},
  {Correia}, {Sozzetti}, {Lillo-Box}, {Figueira}, {Lavie}, {Lovis}, {Hobson},
  {Sousa}, {Adibekyan}, {Standing}, {Hara}, {Barrado}, {Silva}, {Bourrier},
  {Korth}, {Santos}, {Damasso}, {Zapatero Osorio}, {Rodrigues}, {Alibert},
  {Barros}, {Cristiani}, {Di Marcantonio}, {Gonz{\'a}lez Hern{\'a}ndez}, {Lo
  Curto}, {Martins}, {Nunes}, {Palle}, {Pepe}, {Su{\'a}rez Mascare{\~n}o}, \&
  {Tabernero}}]{Castro-Gonz2025}
{Castro-Gonz{\'a}lez}, A., {Bouchy}, F., {Correia}, A.~C.~M., {et~al.} 2025,
  \aap, 699, A344

\bibitem[{{Castro-Tapia} {et~al.}(2024){Castro-Tapia}, {Aguilera-G{\'o}mez}, \&
  {Chanam{\'e}}}]{castro24LSb}
{Castro-Tapia}, M., {Aguilera-G{\'o}mez}, C., \& {Chanam{\'e}}, J. 2024, \aap,
  690, A367

\bibitem[{{Cayrel} {et~al.}(2001){Cayrel}, {Hill}, {Beers}, {Barbuy}, {Spite},
  {Spite}, {Plez}, {Andersen}, {Bonifacio}, {Fran{\c{c}}ois}, {Molaro},
  {Nordstr{\"o}m}, \& {Primas}}]{Cayrel2001}
{Cayrel}, R., {Hill}, V., {Beers}, T.~C., {et~al.} 2001, \nat, 409, 691

\bibitem[{{Cehula} {et~al.}(2024){Cehula}, {Thompson}, \&
  {Metzger}}]{Cehula2024}
{Cehula}, J., {Thompson}, T.~A., \& {Metzger}, B.~D. 2024, \mnras, 528, 5323

\bibitem[{{Cescutti} {et~al.}(2021){Cescutti}, {Morossi}, {Franchini}, {Di
  Marcantonio}, {Chiappini}, {Steffen}, {Valentini}, {Fran{\c{c}}ois},
  {Christlieb}, {Cort{\'e}s}, {Kobayashi}, \& {Depagne}}]{Cecutti2021}
{Cescutti}, G., {Morossi}, C., {Franchini}, M., {et~al.} 2021, \aap, 654, A164

\bibitem[{{Cescutti} {et~al.}(2015){Cescutti}, {Romano}, {Matteucci},
  {Chiappini}, \& {Hirschi}}]{Cescutti2015}
{Cescutti}, G., {Romano}, D., {Matteucci}, F., {Chiappini}, C., \& {Hirschi},
  R. 2015, \aap, 577, A139

\bibitem[{{Chen} {et~al.}(2025){Chen}, {Landry}, {Read}, \&
  {Siegel}}]{Chen2025}
{Chen}, H.-Y., {Landry}, P., {Read}, J.~S., \& {Siegel}, D.~M. 2025, \apj, 985,
  154

\bibitem[{{Choi} {et~al.}(2020){Choi}, {Leighly}, {Terndrup}, {Gallagher}, \&
  {Richards}}]{Choi2020ApJ...891...53C}
{Choi}, H., {Leighly}, K.~M., {Terndrup}, D.~M., {Gallagher}, S.~C., \&
  {Richards}, G.~T. 2020, \apj, 891, 53

\bibitem[{{Choi} {et~al.}(2016){Choi}, {Dotter}, {Conroy}, {Cantiello},
  {Paxton}, \& {Johnson}}]{2016ApJ...823..102C}
{Choi}, J., {Dotter}, A., {Conroy}, C., {et~al.} 2016, \apj, 823, 102

\bibitem[{{Choplin} {et~al.}(2024){Choplin}, {Siess}, {Goriely}, \&
  {Martinet}}]{Choplin2024}
{Choplin}, A., {Siess}, L., {Goriely}, S., \& {Martinet}, S. 2024, Galaxies,
  12, 66

\bibitem[{{Cimatti} \& {Moresco}(2023)}]{cimatti23}
{Cimatti}, A. \& {Moresco}, M. 2023, \apj, 953, 149

\bibitem[{{Cioni} {et~al.}(2019){Cioni}, {Storm}, {Bell}, {Lemasle},
  {Niederhofer}, {Bestenlehner}, {El Youssoufi}, {Feltzing},
  {Gonz{\'a}lez-Fern{\'a}ndez}, {Grebel}, {Hobbs}, {Irwin}, {Jablonka}, {Koch},
  {Schnurr}, {Schmidt}, \& {Steinmetz}}]{cioni2019}
{Cioni}, M. . R.~L., {Storm}, J., {Bell}, C.~P.~M., {et~al.} 2019, The
  Messenger, 175, 54

\bibitem[{{Cirasuolo} {et~al.}(2020){Cirasuolo}, {Fairley}, {Rees}, {Gonzalez},
  {Taylor}, {Maiolino}, {Afonso}, {Evans}, {Flores}, {Lilly}, {Oliva},
  {Paltani}, {Vanzi}, {Abreu}, {Accardo}, {Adams}, {{\'A}lvarez M{\'e}ndez},
  {Amans}, {Amarantidis}, {Atek}, {Atkinson}, {Banerji}, {Barrett},
  {Barrientos}, {Bauer}, {Beard}, {B{\'e}chet}, {Belfiore}, {Bellazzini},
  {Benoist}, {Best}, {Biazzo}, {Black}, {Boettger}, {Bonifacio}, {Bowler},
  {Bragaglia}, {Brierley}, {Brinchmann}, {Brinkmann}, {Buat}, {Buitrago},
  {Burgarella}, {Burningham}, {Buscher}, {Cabral}, {Caffau}, {Cardoso},
  {Carnall}, {Carollo}, {Castillo}, {Castignani}, {Catelan}, {Cicone},
  {Cimatti}, {Cioni}, {Clementini}, {Cochrane}, {Coelho}, {Colling}, {Contini},
  {Contreras}, {Conzelmann}, {Cresci}, {Cropper}, {Cucciati}, {Cullen},
  {Cumani}, {Curti}, {Da Silva}, {Daddi}, {Dalessandro}, {Dalessio}, {Dauvin},
  {Davidson}, {de Laverny}, {Delplancke-Str{\"o}bele}, {De Lucia}, {Del
  Vecchio}, {Dessauges-Zavadsky}, {Di Matteo}, {Dole}, {Drass}, {Dunlop},
  {D{\"u}nner}, {Eales}, {Ellis}, {Enriques}, {Fasola}, {Ferguson}, {Ferruzzi},
  {Fisher}, {Flores}, {Fontana}, {Forchi}, {Francois}, {Franzetti}, {Gargiulo},
  {Garilli}, {Gaudemard}, {Gieles}, {Gilmore}, {Ginolfi}, {Gomes}, {Guinouard},
  {Gutierrez}, {Haigron}, {Hammer}, {Hammersley}, {Haniff}, {Harrison},
  {Haywood}, {Hill}, {Hubin}, {Humphrey}, {Ibata}, {Infante}, {Ives}, {Ivison},
  {Iwert}, {Jablonka}, {Jakob}, {Jarvis}, {King}, {Kneib}, {Laporte},
  {Lawrence}, {Lee}, {Li Causi}, {Lorenzoni}, {Lucatello}, {Luco}, {Macleod},
  {Magliocchetti}, {Magrini}, {Mainieri}, {Maire}, {Mannucci}, {Martin},
  {Matute}, {Maurogordato}, {McGee}, {Mcleod}, {McLure}, {McMahon}, {Melse},
  {Messias}, {Mucciarelli}, {Nisini}, {Nix}, {Norberg}, {Oesch}, {Oliveira},
  {Origlia}, {Padilla}, {Palsa}, {Pancino}, {Papaderos}, {Pappalardo}, {Parry},
  {Pasquini}, {Peacock}, {Pedichini}, {Pello}, {Peng}, {Pentericci}, {Pfuhl},
  {Piazzesi}, {Popovic}, {Pozzetti}, {Puech}, {Puzia}, {Raichoor}, {Randich},
  {Recio-Blanco}, {Reis}, {Reix}, {Renzini}, {Rodrigues}, {Rojas},
  {Rojas-Arriagada}, {Rota}, {Royer}, {Sacco}, {Sanchez-Janssen}, {Sanna},
  {Santos}, {Sarzi}, {Schaerer}, {Schiavon}, {Schnell}, {Schultheis},
  {Scodeggio}, {Serjeant}, {Shen}, {Simmonds}, {Smoker}, {Sobral}, {Sordet},
  {Sp{\'e}rone}, {Strachan}, {Sun}, {Swinbank}, {Tait}, {Tereno}, {Tojeiro},
  {Torres}, {Tosi}, {Tozzi}, {Tresiter}, {Valenti}, {Valenzuela Navarro},
  {Vanzella}, {Vergani}, {Verhamme}, {Vernet}, {Vignali}, {Vinther}, {Von
  Dran}, {Waring}, {Watson}, {Wild}, {Willesme}, {Woodward}, {Wuyts}, {Yang},
  {Zamorani}, {Zoccali}, {Bluck}, \& {Trussler}}]{cirasuolo2020}
{Cirasuolo}, M., {Fairley}, A., {Rees}, P., {et~al.} 2020, The Messenger, 180,
  10

\bibitem[{{Clarkson} {et~al.}(2018){Clarkson}, {Herwig}, \&
  {Pignatari}}]{Clarkson2018}
{Clarkson}, O., {Herwig}, F., \& {Pignatari}, M. 2018, \mnras, 474, L37

\bibitem[{Clear {et~al.}(2022)Clear, Pickering, Nave, Uylings, \&
  Raassen}]{Clear2022}
Clear, C.~P., Pickering, J.~C., Nave, G., Uylings, P., \& Raassen, T. 2022, The
  Astrophysical Journal Supplement Series, 261, 35

\bibitem[{Clear {et~al.}(2023{\natexlab{a}})Clear, Pickering, Nave, Uylings, \&
  Raassen}]{Clear2023b}
Clear, C.~P., Pickering, J.~C., Nave, G., Uylings, P., \& Raassen, T.
  2023{\natexlab{a}}, The Astrophysical Journal Supplement Series, 269, 36

\bibitem[{Clear {et~al.}(2023{\natexlab{b}})Clear, Uylings, Raassen, Nave, \&
  Pickering}]{Clear2023a}
Clear, C.~P., Uylings, P., Raassen, T., Nave, G., \& Pickering, J.~C.
  2023{\natexlab{b}}, Monthly Notices of the Royal Astronomical Society, 519,
  4040

\bibitem[{{Colless} {et~al.}(2001){Colless}, {Dalton}, {Maddox}, {Sutherland},
  {Norberg}, {Cole}, {Bland-Hawthorn}, {Bridges}, {Cannon}, {Collins}, {Couch},
  {Cross}, {Deeley}, {De Propris}, {Driver}, {Efstathiou}, {Ellis}, {Frenk},
  {Glazebrook}, {Jackson}, {Lahav}, {Lewis}, {Lumsden}, {Madgwick}, {Peacock},
  {Peterson}, {Price}, {Seaborne}, \& {Taylor}}]{colless01}
{Colless}, M., {Dalton}, G., {Maddox}, S., {et~al.} 2001, \mnras, 328, 1039

\bibitem[{{Collier Cameron}(1995)}]{colliercameron1995}
{Collier Cameron}, A. 1995, \mnras, 275, 534

\bibitem[{{Collier Cameron} \& {Campbell}(1993)}]{colliercameron1993}
{Collier Cameron}, A. \& {Campbell}, C.~G. 1993, \aap, 274, 309

\bibitem[{{Collier Cameron} {et~al.}(1990){Collier Cameron}, {Duncan},
  {Ehrenfreund}, {Foing}, {Kuntz}, {Penston}, {Robinson}, \&
  {Soderblom}}]{colliercameron1990}
{Collier Cameron}, A., {Duncan}, D.~K., {Ehrenfreund}, P., {et~al.} 1990,
  \mnras, 247, 415

\bibitem[{{Collier Cameron} \& {Robinson}(1989)}]{colliercameron1989}
{Collier Cameron}, A. \& {Robinson}, R.~D. 1989, \mnras, 236, 57

\bibitem[{{Collier Cameron} \& {Unruh}(1994)}]{colliercameron1994}
{Collier Cameron}, A. \& {Unruh}, Y.~C. 1994, \mnras, 269, 814

\bibitem[{{Columba} {et~al.}(2023){Columba}, {Danielski}, {Dorozsmai},
  {Toonen}, \& {Lopez Puertas}}]{Columba2023}
{Columba}, G., {Danielski}, C., {Dorozsmai}, A., {Toonen}, S., \& {Lopez
  Puertas}, M. 2023, \aap, 675, A156

\bibitem[{{Comer{\'o}n} \& {Pasquali}(2005)}]{Comeron2005A&A...430..541C}
{Comer{\'o}n}, F. \& {Pasquali}, A. 2005, \aap, 430, 541

\bibitem[{Concepcion {et~al.}(2023)Concepcion, Clear, Ding, \&
  Pickering}]{Concepcion2023}
Concepcion, F., Clear, C.~P., Ding, M., \& Pickering, J.~C. 2023, European
  Physical Journal D, 77

\bibitem[{{Cooke}(2026)}]{cooke26LSb}
{Cooke}, R. 2026, in Encyclopedia of Astrophysics, Volume 5, Vol.~5, 159--183

\bibitem[{{Cooper} {et~al.}(2023){Cooper}, {Koposov}, {Allende Prieto},
  {Manser}, {Kizhuprakkat}, {Myers}, {Dey}, {G{\"a}nsicke}, {Li}, {Rockosi},
  {Valluri}, {Najita}, {Deason}, {Raichoor}, {Wang}, {Ting}, {Kim}, {Carrillo},
  {Wang}, {Beraldo e Silva}, {Han}, {Ding}, {S{\'a}nchez-Conde}, {Aguilar},
  {Ahlen}, {Bailey}, {Belokurov}, {Brooks}, {Cunha}, {Dawson}, {de la Macorra},
  {Doel}, {Eisenstein}, {Fagrelius}, {Fanning}, {Font-Ribera}, {Forero-Romero},
  {Gazta{\~n}aga}, {Gontcho a Gontcho}, {Guy}, {Honscheid}, {Kehoe}, {Kisner},
  {Kremin}, {Landriau}, {Levi}, {Martini}, {Meisner}, {Miquel}, {Moustakas},
  {Nie}, {Palanque-Delabrouille}, {Percival}, {Poppett}, {Prada}, {Rehemtulla},
  {Schlafly}, {Schlegel}, {Schubnell}, {Sharples}, {Tarl{\'e}}, {Wechsler},
  {Weinberg}, {Zhou}, \& {Zou}}]{desi23}
{Cooper}, A.~P., {Koposov}, S.~E., {Allende Prieto}, C., {et~al.} 2023, \apj,
  947, 37

\bibitem[{{Coria} {et~al.}(2023){Coria}, {Crossfield}, {Lothringer}, {Flores},
  {Prantzos}, \& {Freedman}}]{Coria2023}
{Coria}, D.~R., {Crossfield}, I. J.~M., {Lothringer}, J., {et~al.} 2023, \apj,
  954, 121

\bibitem[{{Cosentino} {et~al.}(2012){Cosentino}, {Lovis}, {Pepe}, {Collier
  Cameron}, {Latham}, {Molinari}, {Udry}, {Bezawada}, {Black}, {Born},
  {Buchschacher}, {Charbonneau}, {Figueira}, {Fleury}, {Galli}, {Gallie},
  {Gao}, {Ghedina}, {Gonzalez}, {Gonzalez}, {Guerra}, {Henry}, {Horne},
  {Hughes}, {Kelly}, {Lodi}, {Lunney}, {Maire}, {Mayor}, {Micela}, {Ordway},
  {Peacock}, {Phillips}, {Piotto}, {Pollacco}, {Queloz}, {Rice}, {Riverol},
  {Riverol}, {San Juan}, {Sasselov}, {Segransan}, {Sozzetti}, {Sosnowska},
  {Stobie}, {Szentgyorgyi}, {Vick}, \& {Weber}}]{Cosentino2012SPIE.8446E..1VC}
{Cosentino}, R., {Lovis}, C., {Pepe}, F., {et~al.} 2012, in Society of
  Photo-Optical Instrumentation Engineers (SPIE) Conference Series, Vol. 8446,
  Ground-based and Airborne Instrumentation for Astronomy IV, ed. I.~S.
  {McLean}, S.~K. {Ramsay}, \& H.~{Takami}, 84461V

\bibitem[{{C{\^o}t{\'e}} {et~al.}(2019){C{\^o}t{\'e}}, {Eichler}, {Arcones},
  {Hansen}, {Simonetti}, {Frebel}, {Fryer}, {Pignatari}, {Reichert},
  {Belczynski}, \& {Matteucci}}]{Cote2019}
{C{\^o}t{\'e}}, B., {Eichler}, M., {Arcones}, A., {et~al.} 2019, \apj, 875, 106

\bibitem[{{Cottaar} \& {H{\'e}nault-Brunet}(2014)}]{Cottaar2014}
{Cottaar}, M. \& {H{\'e}nault-Brunet}, V. 2014, \aap, 562, A20

\bibitem[{{Covino} {et~al.}(2013){Covino}, {Esposito}, {Barbieri}, {Mancini},
  {Nascimbeni}, {Claudi}, {Desidera}, {Gratton}, {Lanza}, {Sozzetti}, {Biazzo},
  {Affer}, {Gandolfi}, {Munari}, {Pagano}, {Bonomo}, {Collier Cameron},
  {H{\'e}brard}, {Maggio}, {Messina}, {Micela}, {Molinari}, {Pepe}, {Piotto},
  {Ribas}, {Santos}, {Southworth}, {Shkolnik}, {Triaud}, {Bedin}, {Benatti},
  {Boccato}, {Bonavita}, {Borsa}, {Borsato}, {Brown}, {Carolo}, {Ciceri},
  {Cosentino}, {Damasso}, {Faedi}, {Mart{\'\i}nez Fiorenzano}, {Latham},
  {Lovis}, {Mordasini}, {Nikolov}, {Poretti}, {Rainer}, {Rebolo L{\'o}pez},
  {Scandariato}, {Silvotti}, {Smareglia}, {Alcal{\'a}}, {Cunial}, {Di
  Fabrizio}, {Di Mauro}, {Giacobbe}, {Granata}, {Harutyunyan}, {Knapic},
  {Lattanzi}, {Leto}, {Lodato}, {Malavolta}, {Marzari}, {Molinaro},
  {Nardiello}, {Pedani}, {Prisinzano}, \&
  {Turrini}}]{Covino2013A&A...554A..28C}
{Covino}, E., {Esposito}, M., {Barbieri}, M., {et~al.} 2013, \aap, 554, A28

\bibitem[{{Cowan} \& {Rose}(1977)}]{Cowan-1977}
{Cowan}, J.~J. \& {Rose}, W.~K. 1977, \apj, 212, 149

\bibitem[{{Cowan} {et~al.}(2002){Cowan}, {Sneden}, {Burles}, {Ivans}, {Beers},
  {Truran}, {Lawler}, {Primas}, {Fuller}, {Pfeiffer}, \& {Kratz}}]{Cowan2002}
{Cowan}, J.~J., {Sneden}, C., {Burles}, S., {et~al.} 2002, \apj, 572, 861

\bibitem[{{Cowan} {et~al.}(2021){Cowan}, {Sneden}, {Lawler}, {Aprahamian},
  {Wiescher}, {Langanke}, {Mart{\'\i}nez-Pinedo}, \& {Thielemann}}]{Cowan2021}
{Cowan}, J.~J., {Sneden}, C., {Lawler}, J.~E., {et~al.} 2021, Reviews of Modern
  Physics, 93, 015002

\bibitem[{{Cowan} {et~al.}(1991){Cowan}, {Thielemann}, \& {Truran}}]{cowan1991}
{Cowan}, J.~J., {Thielemann}, F.-K., \& {Truran}, J.~W. 1991, \araa, 29, 447

\bibitem[{Cowan(2023)}]{cowan2023theory}
Cowan, R.~D. 2023, The theory of atomic structure and spectra, Vol.~3 (Univ of
  California Press)

\bibitem[{{Crenshaw} {et~al.}(1999){Crenshaw}, {Kraemer}, {Boggess}, {Maran},
  {Mushotzky}, \& {Wu}}]{Crenshaw1999ApJ...516..750C}
{Crenshaw}, D.~M., {Kraemer}, S.~B., {Boggess}, A., {et~al.} 1999, \apj, 516,
  750

\bibitem[{{Curtis} {et~al.}(2018){Curtis}, {Vanderburg}, {Torres}, {Kraus},
  {Huber}, {Mann}, {Rizzuto}, {Isaacson}, {Howard}, {Henze}, {Fulton}, \&
  {Wright}}]{curtis2018}
{Curtis}, J.~L., {Vanderburg}, A., {Torres}, G., {et~al.} 2018, \aj, 155, 173

\bibitem[{{Da Costa} {et~al.}(2013){Da Costa}, {Norris}, \&
  {Yong}}]{DaCosta2013}
{Da Costa}, G.~S., {Norris}, J.~E., \& {Yong}, D. 2013, \apj, 769, 8

\bibitem[{{da Silva} \& {Smiljanic}(2023)}]{daSilvaSmiljanic2023}
{da Silva}, A.~R. \& {Smiljanic}, R. 2023, \aap, 677, A74

\bibitem[{{Dai} {et~al.}(2025){Dai}, {Liu}, {Pang}, {Jiang}, {de Leon},
  {Zhong}, \& {Zhou}}]{Dai2025}
{Dai}, Y.-Z., {Liu}, H.-G., {Pang}, X., {et~al.} 2025, arXiv e-prints,
  arXiv:2512.07029

\bibitem[{{Daley-Yates} \& {Jardine}(2024)}]{daley-yates2024}
{Daley-Yates}, S. \& {Jardine}, M.~M. 2024, \mnras, 534, 621

\bibitem[{{Dalton} {et~al.}(2014){Dalton}, {Trager}, {Abrams}, {Bonifacio},
  {L{\'o}pez Aguerri}, {Middleton}, {Benn}, {Dee}, {Say{\`e}de}, {Lewis},
  {Pragt}, {Pico}, {Walton}, {Rey}, {Allende Prieto}, {Pe{\~n}ate}, {Lhome},
  {Ag{\'o}cs}, {Alonso}, {Terrett}, {Brock}, {Gilbert}, {Ridings}, {Guinouard},
  {Verheijen}, {Tosh}, {Rogers}, {Steele}, {Stuik}, {Tromp}, {Jasko}, {Kragt},
  {Lesman}, {Mottram}, {Bates}, {Gribbin}, {Rodriguez}, {Delgado}, {Martin},
  {Cano}, {Navarro}, {Irwin}, {Lewis}, {Gonzalez Solares}, {O'Mahony},
  {Bianco}, {Zurita}, {ter Horst}, {Molinari}, {Lodi}, {Guerra}, {Vallenari},
  \& {Baruffolo}}]{dalton2014}
{Dalton}, G., {Trager}, S., {Abrams}, D.~C., {et~al.} 2014, in SPIE Conf. Ser.,
  Vol. 9147, 91470L

\bibitem[{{Damasso} {et~al.}(2024){Damasso}, {Polychroni}, {Locci}, {Turrini},
  {Maggio}, {Cubillos}, {Baratella}, {Biazzo}, {Benatti}, {Mantovan},
  {Nardiello}, {Desidera}, {Bonomo}, {Pinamonti}, {Malavolta}, {Marzari},
  {Sozzetti}, \& {Spinelli}}]{Damasso2024A&A...688A..15D}
{Damasso}, M., {Polychroni}, D., {Locci}, D., {et~al.} 2024, \aap, 688, A15

\bibitem[{{Damiani} {et~al.}(2016){Damiani}, {Bonito}, {Magrini}, {Prisinzano},
  {Mapelli}, {Micela}, {Kalari}, {Ma{\'\i}z Apell{\'a}niz}, {Gilmore},
  {Randich}, {Alfaro}, {Flaccomio}, {Koposov}, {Klutsch}, {Lanzafame},
  {Pancino}, {Sacco}, {Bayo}, {Carraro}, {Casey}, {Costado}, {Franciosini},
  {Hourihane}, {Lardo}, {Lewis}, {Monaco}, {Morbidelli}, {Worley}, {Zaggia},
  {Zwitter}, \& {Dorda}}]{Damiani2016A&A...591A..74D}
{Damiani}, F., {Bonito}, R., {Magrini}, L., {et~al.} 2016, \aap, 591, A74

\bibitem[{{Damiani} {et~al.}(2017){Damiani}, {Bonito}, {Prisinzano}, {Zwitter},
  {Bayo}, {Kalari}, {Jim{\'e}nez-Esteban}, {Costado}, {Jofr{\'e}}, {Randich},
  {Flaccomio}, {Lanzafame}, {Lardo}, {Morbidelli}, \&
  {Zaggia}}]{Damiani2017A&A...604A.135D}
{Damiani}, F., {Bonito}, R., {Prisinzano}, L., {et~al.} 2017, \aap, 604, A135

\bibitem[{{Daniel} {et~al.}(2017){Daniel}, {Heggie}, \& {Varri}}]{Daniel2017}
{Daniel}, K.~J., {Heggie}, D.~C., \& {Varri}, A.~L. 2017, \mnras, 468, 1453

\bibitem[{Davidson {et~al.}(1992)Davidson, Snoek, Volten, \&
  Doenszelmann}]{Davidson1992}
Davidson, M.~D., Snoek, L.~C., Volten, H., \& Doenszelmann, A. 1992, Astronomy
  and Astrophysics (ISSN 0004-6361), 255

\bibitem[{{Davies} {et~al.}(2014){Davies}, {Adams}, {Armitage}, {Chambers},
  {Ford}, {Morbidelli}, {Raymond}, \& {Veras}}]{davies2014}
{Davies}, M.~B., {Adams}, F.~C., {Armitage}, P., {et~al.} 2014, in Protostars
  and Planets VI, ed. H.~{Beuther}, R.~S. {Klessen}, C.~P. {Dullemond}, \&
  T.~{Henning}, 787

\bibitem[{{de Blok}(2010)}]{blok2010}
{de Blok}, W.~J.~G. 2010, Advances in Astronomy, 2010, 789293

\bibitem[{{de Jong} {et~al.}(2019){de Jong}, {Agertz}, {Berbel}, {Aird},
  {Alexander}, {Amarsi}, {Anders}, {Andrae}, {Ansarinejad}, {Ansorge},
  {Antilogus}, {Anwand -Heerwart}, {Arentsen}, {Arnadottir}, {Asplund},
  {Auger}, {Azais}, {Baade}, {Baker}, {Baker}, {Balbinot}, {Baldry}, {Banerji},
  {Barden}, {Barklem}, {Barth{\'e}l{\'e}my-Mazot}, {Battistini}, {Bauer},
  {Bell}, {Bellido-Tirado}, {Bellstedt}, {Belokurov}, {Bensby}, {Bergemann},
  {Bestenlehner}, {Bielby}, {Bilicki}, {Blake}, {Bland-Hawthorn}, {Boeche},
  {Boland}, {Boller}, {Bongard}, {Bongiorno}, {Bonifacio}, {Boudon}, {Brooks},
  {Brown}, {Brown}, {Br{\"u}ggen}, {Brynnel}, {Brzeski}, {Buchert},
  {Buschkamp}, {Caffau}, {Caillier}, {Carrick}, {Casagrande}, {Case}, {Casey},
  {Cesarini}, {Cescutti}, {Chapuis}, {Chiappini}, {Childress}, {Christlieb},
  {Church}, {Cioni}, {Cluver}, {Colless}, {Collett}, {Comparat}, {Cooper},
  {Couch}, {Courbin}, {Croom}, {Croton}, {Daguis{\'e}}, {Dalton}, {Davies},
  {Davis}, {de Laverny}, {Deason}, {Dionies}, {Disseau}, {Doel}, {D{\"o}scher},
  {Driver}, {Dwelly}, {Eckert}, {Edge}, {Edvardsson}, {Youssoufi}, {Elhaddad},
  {Enke}, {Erfanianfar}, {Farrell}, {Fechner}, {Feiz}, {Feltzing}, {Ferreras},
  {Feuerstein}, {Feuillet}, {Finoguenov}, {Ford}, {Fotopoulou}, {Fouesneau},
  {Frenk}, {Frey}, {Gaessler}, {Geier}, {Fusillo}, {Gerhard}, {Giannantonio},
  {Giannone}, {Gibson}, {Gillingham}, {Gonz{\'a}lez-Fern{\'a}ndez},
  {Gonzalez-Solares}, {Gottloeber}, {Gould}, {Grebel}, {Gueguen}, {Guiglion},
  {Haehnelt}, {Hahn}, {Hansen}, {Hartman}, {Hauptner}, {Hawkins}, {Haynes},
  {Haynes}, {Heiter}, {Helmi}, {Aguayo}, {Hewett}, {Hinton}, {Hobbs}, {Hoenig},
  {Hofman}, {Hook}, {Hopgood}, {Hopkins}, {Hourihane}, {Howes}, {Howlett},
  {Huet}, {Irwin}, {Iwert}, {Jablonka}, {Jahn}, {Jahnke}, {Jarno}, {Jin},
  {Jofre}, {Johl}, {Jones}, {J{\"o}nsson}, {Jordan}, {Karovicova}, {Khalatyan},
  {Kelz}, {Kennicutt}, {King}, {Kitaura}, {Klar}, {Klauser}, {Kneib}, {Koch},
  {Koposov}, {Kordopatis}, {Korn}, {Kosmalski}, {Kotak}, {Kovalev}, {Kreckel},
  {Kripak}, {Krumpe}, {Kuijken}, {Kunder}, {Kushniruk}, {Lam}, {Lamer},
  {Laurent}, {Lawrence}, {Lehmitz}, {Lemasle}, {Lewis}, {Li}, {Lidman}, {Lind},
  {Liske}, {Lizon}, {Loveday}, {Ludwig}, {McDermid}, {Maguire}, {Mainieri},
  {Mali}, {Mandel}, {Mandel}, {Mannering}, {Martell}, {Martinez Delgado},
  {Matijevic}, {McGregor}, {McMahon}, {McMillan}, {Mena}, {Merloni}, {Meyer},
  {Michel}, {Micheva}, {Migniau}, {Minchev}, {Monari}, {Muller}, {Murphy},
  {Muthukrishna}, {Nandra}, {Navarro}, {Ness}, {Nichani}, {Nichol}, {Nicklas},
  {Niederhofer}, {Norberg}, {Obreschkow}, {Oliver}, {Owers}, {Pai},
  {Pankratow}, {Parkinson}, {Paschke}, {Paterson}, {Pecontal}, {Parry},
  {Phillips}, {Pillepich}, {Pinard}, {Pirard}, {Piskunov}, {Plank},
  {Pl{\"u}schke}, {Pons}, {Popesso}, {Power}, {Pragt}, {Pramskiy}, {Pryer},
  {Quattri}, {Queiroz}, {Quirrenbach}, {Rahurkar}, {Raichoor}, {Ramstedt},
  {Rau}, {Recio-Blanco}, {Reiss}, {Renaud}, {Revaz}, {Rhode}, {Richard},
  {Richter}, {Rix}, {Robotham}, {Roelfsema}, {Romaniello}, {Rosario},
  {Rothmaier}, {Roukema}, {Ruchti}, {Rupprecht}, {Rybizki}, {Ryde}, {Saar},
  {Sadler}, {Sahl{\'e}n}, {Salvato}, {Sassolas}, {Saunders}, {Saviauk},
  {Sbordone}, {Schmidt}, {Schnurr}, {Scholz}, {Schwope}, {Seifert}, {Shanks},
  {Sheinis}, {Sivov}, {Sk{\'u}lad{\'o}ttir}, {Smartt}, {Smedley}, {Smith},
  {Smith}, {Sorce}, {Spitler}, {Starkenburg}, {Steinmetz}, {Stilz}, {Storm},
  {Sullivan}, {Sutherland}, {Swann}, {Tamone}, {Taylor}, {Teillon}, {Tempel},
  {ter Horst}, {Thi}, {Tolstoy}, {Trager}, {Traven}, {Tremblay}, {Tresse},
  {Valentini}, {van de Weygaert}, {van den Ancker}, {Veljanoski}, {Venkatesan},
  {Wagner}, {Wagner}, {Walcher}, {Waller}, {Walton}, {Wang}, {Winkler},
  {Wisotzki}, {Worley}, {Worseck}, {Xiang}, {Xu}, {Yong}, {Zhao}, {Zheng},
  {Zscheyge}, \& {Zucker}}]{dejong2019}
{de Jong}, R.~S., {Agertz}, O., {Berbel}, A.~A., {et~al.} 2019, The Messenger,
  175, 3

\bibitem[{{de los Reyes} {et~al.}(2020){de los Reyes}, {Kirby}, {Seitenzahl},
  \& {Shen}}]{delosReyes2020}
{de los Reyes}, M. A.~C., {Kirby}, E.~N., {Seitenzahl}, I.~R., \& {Shen}, K.~J.
  2020, \apj, 891, 85

\bibitem[{{de Mink} {et~al.}(2009){de Mink}, {Pols}, {Langer}, \&
  {Izzard}}]{deMink2009}
{de Mink}, S.~E., {Pols}, O.~R., {Langer}, N., \& {Izzard}, R.~G. 2009, \aap,
  507, L1

\bibitem[{{De Silva} {et~al.}(2015){De Silva}, {Freeman}, {Bland-Hawthorn},
  {Martell}, {de Boer}, {Asplund}, {Keller}, {Sharma}, {Zucker}, {Zwitter},
  {Anguiano}, {Bacigalupo}, {Bayliss}, {Beavis}, {Bergemann}, {Campbell},
  {Cannon}, {Carollo}, {Casagrande}, {Casey}, {Da Costa}, {D'Orazi}, {Dotter},
  {Duong}, {Heger}, {Ireland}, {Kafle}, {Kos}, {Lattanzio}, {Lewis}, {Lin},
  {Lind}, {Munari}, {Nataf}, {O'Toole}, {Parker}, {Reid}, {Schlesinger},
  {Sheinis}, {Simpson}, {Stello}, {Ting}, {Traven}, {Watson}, {Wittenmyer},
  {Yong}, \& {{\v Z}erjal}}]{desilva2015}
{De Silva}, G.~M., {Freeman}, K.~C., {Bland-Hawthorn}, J., {et~al.} 2015,
  \mnras, 449, 2604

\bibitem[{{Deason} {et~al.}(2022){Deason}, {Bose}, {Fattahi}, {Amorisco},
  {Hellwing}, \& {Frenk}}]{deason2022}
{Deason}, A.~J., {Bose}, S., {Fattahi}, A., {et~al.} 2022, \mnras, 511, 4044

\bibitem[{{Debes} \& {Sigurdsson}(2002)}]{DebesSigurdsson2002}
{Debes}, J.~H. \& {Sigurdsson}, S. 2002, \apj, 572, 556

\bibitem[{{Decressin} {et~al.}(2007){Decressin}, {Charbonnel}, \&
  {Meynet}}]{Decressin2007}
{Decressin}, T., {Charbonnel}, C., \& {Meynet}, G. 2007, \aap, 475, 859

\bibitem[{{Deepak} \& {Lambert}(2021)}]{DeepakLambert2021}
{Deepak} \& {Lambert}, D.~L. 2021, \mnras, 507, 205

\bibitem[{{Dekker} {et~al.}(2000){Dekker}, {D'Odorico}, {Kaufer}, {Delabre}, \&
  {Kotzlowski}}]{dekker2000}
{Dekker}, H., {D'Odorico}, S., {Kaufer}, A., {Delabre}, B., \& {Kotzlowski}, H.
  2000, in Society of Photo-Optical Instrumentation Engineers (SPIE) Conference
  Series, Vol. 4008, Optical and IR Telescope Instrumentation and Detectors,
  ed. M.~{Iye} \& A.~F. {Moorwood}, 534--545

\bibitem[{{Delgado Mena} {et~al.}(2015){Delgado Mena}, {Bertr{\'a}n de Lis},
  {Adibekyan}, {Sousa}, {Figueira}, {Mortier}, {Gonz{\'a}lez Hern{\'a}ndez},
  {Tsantaki}, {Israelian}, \& {Santos}}]{delgadomena2015}
{Delgado Mena}, E., {Bertr{\'a}n de Lis}, S., {Adibekyan}, V.~Z., {et~al.}
  2015, \aap, 576, A69

\bibitem[{{Delgado Mena} {et~al.}(2023){Delgado Mena}, {Gomes da Silva},
  {Faria}, {Santos}, {Martins}, {Tsantaki}, {Mortier}, {Sousa}, \&
  {Lovis}}]{delgadomena23}
{Delgado Mena}, E., {Gomes da Silva}, J., {Faria}, J.~P., {et~al.} 2023, \aap,
  679, A94

\bibitem[{{Delgado Mena} {et~al.}(2018){Delgado Mena}, {Lovis}, {Santos},
  {Gomes da Silva}, {Mortier}, {Tsantaki}, {Sousa}, {Figueira}, {Cunha},
  {Campante}, {Adibekyan}, {Faria}, \& {Montalto}}]{delgadomena2018}
{Delgado Mena}, E., {Lovis}, C., {Santos}, N.~C., {et~al.} 2018, \aap, 619, A2

\bibitem[{{Deliyannis} {et~al.}(2019){Deliyannis}, {Anthony-Twarog},
  {Lee-Brown}, \& {Twarog}}]{deliyannis19LSb}
{Deliyannis}, C.~P., {Anthony-Twarog}, B.~J., {Lee-Brown}, D.~B., \& {Twarog},
  B.~A. 2019, \aj, 158, 163

\bibitem[{{Deliyannis} {et~al.}(1990){Deliyannis}, {Demarque}, \&
  {Kawaler}}]{delyannis1990}
{Deliyannis}, C.~P., {Demarque}, P., \& {Kawaler}, S.~D. 1990, \apjs, 73, 21

\bibitem[{{Denissenkov}(2010)}]{Denissenkov10}
{Denissenkov}, P.~A. 2010, \apj, 723, 563

\bibitem[{{Denissenkov} {et~al.}(2019){Denissenkov}, {Herwig}, {Woodward},
  {Andrassy}, {Pignatari}, \& {Jones}}]{Denissenkov2019}
{Denissenkov}, P.~A., {Herwig}, F., {Woodward}, P., {et~al.} 2019, \mnras, 488,
  4258

\bibitem[{{Denissenkov} {et~al.}(2015){Denissenkov}, {VandenBerg}, {Hartwick},
  {Herwig}, {Weiss}, \& {Paxton}}]{Denissenkov2015}
{Denissenkov}, P.~A., {VandenBerg}, D.~A., {Hartwick}, F.~D.~A., {et~al.} 2015,
  \mnras, 448, 3314

\bibitem[{{Deprince} {et~al.}(2025){Deprince}, {Wagle}, {Ben Nasr}, {Carvajal
  Gallego}, {Godefroid}, {Goriely}, {Just}, {Palmeri}, {Quinet}, \& {Van
  Eck}}]{deprince2025}
{Deprince}, J., {Wagle}, G., {Ben Nasr}, S., {et~al.} 2025, \aap, 696, A32

\bibitem[{{Dessauges-Zavadsky} {et~al.}(2007){Dessauges-Zavadsky}, {Calura},
  {Prochaska}, {D'Odorico}, \& {Matteucci}}]{Dessauges-Zavadsky2007}
{Dessauges-Zavadsky}, M., {Calura}, F., {Prochaska}, J.~X., {D'Odorico}, S., \&
  {Matteucci}, F. 2007, \aap, 470, 431

\bibitem[{{Dickinson} {et~al.}(2023){Dickinson}, {Smith}, {Andrews}, {Milne},
  {Kilpatrick}, \& {Milisavljevic}}]{Dickinson2023}
{Dickinson}, D., {Smith}, N., {Andrews}, J., {et~al.} 2023, in American
  Astronomical Society Meeting Abstracts, Vol. 241, American Astronomical
  Society Meeting Abstracts \#241, 207.16

\bibitem[{{Dickson} {et~al.}(2024){Dickson}, {Smith}, {H{\'e}nault-Brunet},
  {Gieles}, \& {Baumgardt}}]{dickson2024}
{Dickson}, N., {Smith}, P.~J., {H{\'e}nault-Brunet}, V., {Gieles}, M., \&
  {Baumgardt}, H. 2024, \mnras, 529, 331

\bibitem[{{Diehl} {et~al.}(2022){Diehl}, {Korn}, {Leibundgut}, {Lugaro}, \&
  {Wallner}}]{Diehl2022}
{Diehl}, R., {Korn}, A.~J., {Leibundgut}, B., {Lugaro}, M., \& {Wallner}, A.
  2022, Progress in Particle and Nuclear Physics, 127, 103983

\bibitem[{Ding(2025)}]{Ding25atoms}
Ding, M. 2025, Atoms, 13, 35

\bibitem[{Ding {et~al.}(2025)Ding, Kozuki, Concepcion, Nave, \&
  Pickering}]{Ding2025}
Ding, M., Kozuki, H., Concepcion, F., Nave, G., \& Pickering, J.~C. 2025,
  Monthly Notices of the Royal Astronomical Society, 536, 274

\bibitem[{Ding {et~al.}(2024{\natexlab{a}})Ding, Pickering, Ryabtsev, Kononov,
  \& Ryabchikova}]{Ding2024a}
Ding, M., Pickering, J., Ryabtsev, A., Kononov, E., \& Ryabchikova, T.
  2024{\natexlab{a}}, Astronomy \& Astrophysics, 149, 1

\bibitem[{Ding \& Pickering(2020)}]{Ding2020}
Ding, M. \& Pickering, J.~C. 2020, The Astrophysical Journal Supplement Series,
  251, 24

\bibitem[{Ding {et~al.}(2024{\natexlab{b}})Ding, Ryabtsev, Kononov,
  Ryabchikova, \& Pickering}]{Ding2024b}
Ding, M., Ryabtsev, A.~N., Kononov, E.~Y., Ryabchikova, T., \& Pickering, J.~C.
  2024{\natexlab{b}}, Astronomy and Astrophysics, 692

\bibitem[{{Dixon} {et~al.}(2025){Dixon}, {Ezzeddine}, {Li}, {Merle},
  {Bautista}, \& {Guo}}]{dixon25}
{Dixon}, J.~D., {Ezzeddine}, R., {Li}, Y., {et~al.} 2025, \apj, 994, 44

\bibitem[{{Donati} {et~al.}(2011){Donati}, {Gregory}, {Alencar}, {Bouvier},
  {Hussain}, {Skelly}, {Dougados}, {Jardine}, {M{\'e}nard}, {Romanova}, \&
  {Unruh}}]{donati2011}
{Donati}, J.~F., {Gregory}, S.~G., {Alencar}, S.~H.~P., {et~al.} 2011, \mnras,
  417, 472

\bibitem[{{Donati} {et~al.}(2013){Donati}, {Gregory}, {Alencar}, {Hussain},
  {Bouvier}, {Jardine}, {M{\'e}nard}, {Dougados}, {Romanova}, \& {MaPP
  Collaboration}}]{donati2013}
{Donati}, J.~F., {Gregory}, S.~G., {Alencar}, S.~H.~P., {et~al.} 2013, \mnras,
  436, 881

\bibitem[{{Donati} \& {Landstreet}(2009)}]{donati2009}
{Donati}, J.~F. \& {Landstreet}, J.~D. 2009, \araa, 47, 333

\bibitem[{{Donati} {et~al.}(2008){Donati}, {Morin}, {Petit}, {Delfosse},
  {Forveille}, {Auri{\`e}re}, {Cabanac}, {Dintrans}, {Fares}, {Gastine},
  {Jardine}, {Ligni{\`e}res}, {Paletou}, {Ramirez Velez}, \&
  {Th{\'e}ado}}]{donati2008b}
{Donati}, J.~F., {Morin}, J., {Petit}, P., {et~al.} 2008, \mnras, 390, 545

\bibitem[{{Donati} {et~al.}(2010){Donati}, {Skelly}, {Bouvier}, {Gregory},
  {Grankin}, {Jardine}, {Hussain}, {M{\'e}nard}, {Dougados}, {Unruh},
  {Mohanty}, {Auri{\`e}re}, {Morin}, {Far{\`e}s}, \& {MAPP
  Collaboration}}]{donati2010}
{Donati}, J.~F., {Skelly}, M.~B., {Bouvier}, J., {et~al.} 2010, \mnras, 409,
  1347

\bibitem[{{Dondoglio} {et~al.}(2025){Dondoglio}, {Marino}, {Milone}, {Jang},
  {Cordoni}, {D'Antona}, {Renzini}, {Tailo}, {Sanchez}, {Muratore}, {Ziliotto},
  {Barbieri}, {Bortolan}, {Lagioia}, {Legnardi}, {Lionetto}, \&
  {Mohandasan}}]{Dondoglio2025}
{Dondoglio}, E., {Marino}, A.~F., {Milone}, A.~P., {et~al.} 2025, \aap, 697,
  A135

\bibitem[{{D'Orazi} {et~al.}(2015){D'Orazi}, {Gratton}, {Angelou}, {Bragaglia},
  {Carretta}, {Lattanzio}, {Lucatello}, {Momany}, {Sollima}, \&
  {Beccari}}]{Dorazi2015}
{D'Orazi}, V., {Gratton}, R.~G., {Angelou}, G.~C., {et~al.} 2015, \mnras, 449,
  4038

\bibitem[{{D'Orazi} {et~al.}(2009){D'Orazi}, {Magrini}, {Randich}, {Galli},
  {Busso}, \& {Sestito}}]{Dorazi2009}
{D'Orazi}, V., {Magrini}, L., {Randich}, S., {et~al.} 2009, \apjl, 693, L31

\bibitem[{{Dorn} {et~al.}(2023){Dorn}, {Bristow}, {Smoker}, {Rodler}, {Lavail},
  {Accardo}, {van den Ancker}, {Baade}, {Baruffolo}, {Courtney-Barrer},
  {Blanco}, {Brucalassi}, {Cumani}, {Follert}, {Haimerl}, {Hatzes}, {Haug},
  {Heiter}, {Hinterschuster}, {Hubin}, {Ives}, {Jung}, {Jones}, {Kaeufl},
  {Kirchbauer}, {Klein}, {Kochukhov}, {Korhonen}, {K{\"o}hler}, {Lizon},
  {Moins}, {Molina-Conde}, {Marquart}, {Neeser}, {Oliva}, {Pallanca},
  {Pasquini}, {Paufique}, {Piskunov}, {Reiners}, {Schneller}, {Schmutzer},
  {Seemann}, {Slumstrup}, {Smette}, {Stegmeier}, {Stempels}, {Tordo},
  {Valenti}, {Valenzuela}, {Vernet}, {Vinther}, \& {Wehrhahn}}]{dorn23}
{Dorn}, R.~J., {Bristow}, P., {Smoker}, J.~V., {et~al.} 2023, \aap, 671, A24

\bibitem[{{Dotter}(2016)}]{2016ApJS..222....8D}
{Dotter}, A. 2016, \apjs, 222, 8

\bibitem[{{Dotter} {et~al.}(2017){Dotter}, {Conroy}, {Cargile}, \&
  {Asplund}}]{dotter2017}
{Dotter}, A., {Conroy}, C., {Cargile}, P., \& {Asplund}, M. 2017, \apj, 840, 99

\bibitem[{Dowd {et~al.}(2025)Dowd, Doyle, \& Dunne}]{Dowd2025}
Dowd, K., Doyle, E., \& Dunne, P. 2025, Experimental Astronomy, 60

\bibitem[{{Duffau} {et~al.}(2017){Duffau}, {Caffau}, {Sbordone}, {Bonifacio},
  {Andrievsky}, {Korotin}, {Babusiaux}, {Salvadori}, {Monaco},
  {Fran{\c{c}}ois}, {Sk{\'u}lad{\'o}ttir}, {Bragaglia}, {Donati}, {Spina},
  {Gallagher}, {Ludwig}, {Christlieb}, {Hansen}, {Mott}, {Steffen}, {Zaggia},
  {Blanco-Cuaresma}, {Calura}, {Friel}, {Jim{\'e}nez-Esteban}, {Koch},
  {Magrini}, {Pancino}, {Tang}, {Tautvai{\v{s}}ien{\.{e}}}, {Vallenari},
  {Hawkins}, {Gilmore}, {Randich}, {Feltzing}, {Bensby}, {Flaccomio},
  {Smiljanic}, {Bayo}, {Carraro}, {Casey}, {Costado}, {Damiani}, {Franciosini},
  {Hourihane}, {Jofr{\'e}}, {Lardo}, {Lewis}, {Morbidelli}, {Sousa}, \&
  {Worley}}]{duffau2017}
{Duffau}, S., {Caffau}, E., {Sbordone}, L., {et~al.} 2017, \aap, 604, A128

\bibitem[{{Duncan} {et~al.}(1991){Duncan}, {Vaughan}, {Wilson}, {Preston},
  {Frazer}, {Lanning}, {Misch}, {Mueller}, {Soyumer}, {Woodard}, {Baliunas},
  {Noyes}, {Hartmann}, {Porter}, {Zwaan}, {Middelkoop}, {Rutten}, \&
  {Mihalas}}]{Duncan1991}
{Duncan}, D.~K., {Vaughan}, A.~H., {Wilson}, O.~C., {et~al.} 1991, \apjs, 76,
  383

\bibitem[{{Dunn} {et~al.}(2010){Dunn}, {Bautista}, {Arav}, {Moe}, {Korista},
  {Costantini}, {Benn}, {Ellison}, \& {Edmonds}}]{Dunn2010ApJ...709..611D}
{Dunn}, J.~P., {Bautista}, M., {Arav}, N., {et~al.} 2010, \apj, 709, 611

\bibitem[{{Dunstone} {et~al.}(2006){Dunstone}, {Collier Cameron}, {Barnes}, \&
  {Jardine}}]{dunstone2006}
{Dunstone}, N.~J., {Collier Cameron}, A., {Barnes}, J.~R., \& {Jardine}, M.
  2006, \mnras, 373, 1308

\bibitem[{{Duquennoy} \& {Mayor}(1991)}]{duquennoy1991}
{Duquennoy}, A. \& {Mayor}, M. 1991, \aap, 248, 485

\bibitem[{{Durney} \& {Latour}(1978)}]{durney1978}
{Durney}, B.~R. \& {Latour}, J. 1978, Geophysical and Astrophysical Fluid
  Dynamics, 9, 241

\bibitem[{{Dzuba} {et~al.}(1999){Dzuba}, {Flambaum}, \&
  {Webb}}]{Dzuba1999PhRvL..82..888D}
{Dzuba}, V.~A., {Flambaum}, V.~V., \& {Webb}, J.~K. 1999, \prl, 82, 888

\bibitem[{{Eistrup} {et~al.}(2016){Eistrup}, {Walsh}, \& {van
  Dishoeck}}]{Eistrup2016}
{Eistrup}, C., {Walsh}, C., \& {van Dishoeck}, E.~F. 2016, \aap, 595, A83

\bibitem[{{Eistrup} {et~al.}(2018){Eistrup}, {Walsh}, \& {van
  Dishoeck}}]{Eistrup2018}
{Eistrup}, C., {Walsh}, C., \& {van Dishoeck}, E.~F. 2018, \aap, 613, A14

\bibitem[{{Eitner} {et~al.}(2024){Eitner}, {Bergemann}, {Hoppe}, {Nordlund},
  {Plez}, \& {Klevas}}]{eitner2024}
{Eitner}, P., {Bergemann}, M., {Hoppe}, R., {et~al.} 2024, \aap, 688, A52

\bibitem[{{Eitner} {et~al.}(2025){Eitner}, {Bergemann}, {Hoppe}, {Storm},
  {Lipatova}, {Glover}, {Klessen}, {Nordlund}, \& {Popovas}}]{eitner2025}
{Eitner}, P., {Bergemann}, M., {Hoppe}, R., {et~al.} 2025, arXiv e-prints,
  arXiv:2509.24555

\bibitem[{{Eitner} {et~al.}(2023){Eitner}, {Bergemann}, {Ruiter}, {Avril},
  {Seitenzahl}, {Gent}, \& {C{\^o}t{\'e}}}]{eitner2023}
{Eitner}, P., {Bergemann}, M., {Ruiter}, A.~J., {et~al.} 2023, \aap, 677, A151

\bibitem[{{El-Badry} {et~al.}(2021){El-Badry}, {Rix}, \&
  {Heintz}}]{el-badry2021}
{El-Badry}, K., {Rix}, H.-W., \& {Heintz}, T.~M. 2021, \mnras, 506, 2269

\bibitem[{{Ernandes} {et~al.}(2024){Ernandes}, {Feuillet}, {Feltzing}, \&
  {Sk{\'u}lad{\'o}ttir}}]{ernandes2024}
{Ernandes}, H., {Feuillet}, D., {Feltzing}, S., \& {Sk{\'u}lad{\'o}ttir},
  {\'A}. 2024, \aap, 691, A333

\bibitem[{{Ernandes} {et~al.}(2025){Ernandes}, {Sk{\'u}lad{\'o}ttir},
  {Feltzing}, \& {Feuillet}}]{ernandes2025}
{Ernandes}, H., {Sk{\'u}lad{\'o}ttir}, {\'A}., {Feltzing}, S., \& {Feuillet},
  D. 2025, \aap, 703, A256

\bibitem[{Even {et~al.}(2020)Even, Korobkin, Fryer, Fontes, Wollaeger,
  Hungerford, Lippuner, Miller, Mumpower, \& Misch}]{Even2020}
Even, W., Korobkin, O., Fryer, C.~L., {et~al.} 2020, The Astrophysical Journal,
  899, 24

\bibitem[{{Fakhouri} {et~al.}(2010){Fakhouri}, {Ma}, \&
  {Boylan-Kolchin}}]{fakhouri2010}
{Fakhouri}, O., {Ma}, C.-P., \& {Boylan-Kolchin}, M. 2010, \mnras, 406, 2267

\bibitem[{{Faria} {et~al.}(2018){Faria}, {Santos}, {Figueira}, \&
  {Brewer}}]{Faria2018}
{Faria}, J.~P., {Santos}, N.~C., {Figueira}, P., \& {Brewer}, B.~J. 2018, The
  Journal of Open Source Software, 3, 487

\bibitem[{{Feltzing} {et~al.}(2017){Feltzing}, {Howes}, {McMillan}, \&
  {Stonkut{\.e}}}]{feltzing2017}
{Feltzing}, S., {Howes}, L.~M., {McMillan}, P.~J., \& {Stonkut{\.e}}, E. 2017,
  \mnras, 465, L109

\bibitem[{{Ferraro} {et~al.}(2018){Ferraro}, {Mucciarelli}, {Lanzoni},
  {Pallanca}, {Lapenna}, {Origlia}, {Dalessandro}, {Valenti}, {Beccari},
  {Bellazzini}, {Vesperini}, {Varri}, \& {Sollima}}]{Ferraro2018}
{Ferraro}, F.~R., {Mucciarelli}, A., {Lanzoni}, B., {et~al.} 2018, \apj, 860,
  50

\bibitem[{{Fields} \& {Olive}(2022)}]{fields22LSb}
{Fields}, B.~D. \& {Olive}, K.~A. 2022, \jcap, 2022, 078

\bibitem[{{Fields} {et~al.}(2020){Fields}, {Olive}, {Yeh}, \&
  {Young}}]{Fields2020}
{Fields}, B.~D., {Olive}, K.~A., {Yeh}, T.-H., \& {Young}, C. 2020, \jcap,
  2020, 010

\bibitem[{{Filomeno} {et~al.}(2024){Filomeno}, {Biazzo}, {Baratella},
  {Benatti}, {D'Orazi}, {Desidera}, {Mancini}, {Messina}, {Polychroni},
  {Turrini}, {Cabona}, {Carleo}, {Damasso}, {Malavolta}, {Mantovan},
  {Nardiello}, {Scandariato}, {Sozzetti}, {Zingales}, {Andreuzzi},
  {Antoniucci}, {Bignamini}, {Bonomo}, {Claudi}, {Cosentino}, {Fiorenzano},
  {Fonte}, {Harutyunyan}, \& {Knapic}}]{Filomeno2024}
{Filomeno}, S., {Biazzo}, K., {Baratella}, M., {et~al.} 2024, \aap, 690, A370

\bibitem[{Fischer {et~al.}(2016)Fischer, Godefroid, Brage, JÃ¶nsson, \&
  Gaigalas}]{Fischer_2016}
Fischer, C.~F., Godefroid, M., Brage, T., JÃ¶nsson, P., \& Gaigalas, G. 2016,
  Journal of Physics B: Atomic, Molecular and Optical Physics, 49, 182004

\bibitem[{{Fischer} \& {Valenti}(2005)}]{fischer2005}
{Fischer}, D.~A. \& {Valenti}, J. 2005, \apj, 622, 1102

\bibitem[{{Fishlock} {et~al.}(2017){Fishlock}, {Yong}, {Karakas},
  {Alves-Brito}, {Mel{\'e}ndez}, {Nissen}, {Kobayashi}, \&
  {Casey}}]{fishlock2017}
{Fishlock}, C.~K., {Yong}, D., {Karakas}, A.~I., {et~al.} 2017, \mnras, 466,
  4672

\bibitem[{{Fitzpatrick} {et~al.}(2024){Fitzpatrick}, {Placco}, {Bolton},
  {Merino}, {Ridgway}, \& {Stanghellini}}]{Fitzpatrick2024}
{Fitzpatrick}, M., {Placco}, V., {Bolton}, A., {et~al.} 2024, arXiv e-prints,
  arXiv:2401.01982

\bibitem[{{Folsom} {et~al.}(2018){Folsom}, {Bouvier}, {Petit}, {L{\`e}bre},
  {Amard}, {Palacios}, {Morin}, {Donati}, \& {Vidotto}}]{folsom2018}
{Folsom}, C.~P., {Bouvier}, J., {Petit}, P., {et~al.} 2018, \mnras, 474, 4956

\bibitem[{Fontes {et~al.}(2017)Fontes, Fryer, \& Hungerford}]{2017fontes}
Fontes, C.~J., Fryer, C.~L., \& Hungerford, A.~L. 2017, in AIP Conference
  Proceedings, Vol. 1811 (American Institute of Physics Inc.)

\bibitem[{Fontes {et~al.}(2020)Fontes, Fryer, Hungerford, Wollaeger, \&
  Korobkin}]{Fontes2020}
Fontes, C.~J., Fryer, C.~L., Hungerford, A.~L., Wollaeger, R.~T., \& Korobkin,
  O. 2020, Monthly Notices of the Royal Astronomical Society, 493, 4143

\bibitem[{{Forbes} \& {Bridges}(2010)}]{Forbes2010}
{Forbes}, D.~A. \& {Bridges}, T. 2010, \mnras, 404, 1203

\bibitem[{{Fortney} {et~al.}(2007){Fortney}, {Marley}, \&
  {Barnes}}]{fortney2007}
{Fortney}, J.~J., {Marley}, M.~S., \& {Barnes}, J.~W. 2007, \apj, 659, 1661

\bibitem[{{Frebel} {et~al.}(2007){Frebel}, {Christlieb}, {Norris}, {Thom},
  {Beers}, \& {Rhee}}]{Frebel2007}
{Frebel}, A., {Christlieb}, N., {Norris}, J.~E., {et~al.} 2007, \apjl, 660,
  L117

\bibitem[{{Freeman} \& {Bland-Hawthorn}(2002)}]{freeman2002}
{Freeman}, K. \& {Bland-Hawthorn}, J. 2002, \araa, 40, 487

\bibitem[{{Freytag} {et~al.}(2012){Freytag}, {Steffen}, {Ludwig},
  {Wedemeyer-B{\"o}hm}, {Schaffenberger}, \& {Steiner}}]{freytag2012}
{Freytag}, B., {Steffen}, M., {Ludwig}, H.~G., {et~al.} 2012, Journal of
  Computational Physics, 231, 919

\bibitem[{{Fr{\"o}hlich} {et~al.}(2006){Fr{\"o}hlich}, {Mart{\'\i}nez-Pinedo},
  {Liebend{\"o}rfer}, {Thielemann}, {Bravo}, {Hix}, {Langanke}, \&
  {Zinner}}]{Frohlich2006}
{Fr{\"o}hlich}, C., {Mart{\'\i}nez-Pinedo}, G., {Liebend{\"o}rfer}, M.,
  {et~al.} 2006, \prl, 96, 142502

\bibitem[{{Gaia Collaboration} {et~al.}(2023){Gaia Collaboration}, {Vallenari},
  {Brown}, {Prusti}, {de Bruijne}, {Arenou}, {Babusiaux}, {Biermann},
  {Creevey}, {Ducourant}, {Evans}, {Eyer}, {Guerra}, {Hutton}, {Jordi},
  {Klioner}, {Lammers}, {Lindegren}, {Luri}, {Mignard}, {Panem}, {Pourbaix},
  {Randich}, {Sartoretti}, {Soubiran}, {Tanga}, {Walton}, {Bailer-Jones},
  {Bastian}, {Drimmel}, {Jansen}, {Katz}, {Lattanzi}, {van Leeuwen}, {Bakker},
  {Cacciari}, {Casta{\~n}eda}, {De Angeli}, {Fabricius}, {Fouesneau},
  {Fr{\'e}mat}, {Galluccio}, {Guerrier}, {Heiter}, {Masana}, {Messineo},
  {Mowlavi}, {Nicolas}, {Nienartowicz}, {Pailler}, {Panuzzo}, {Riclet}, {Roux},
  {Seabroke}, {Sordo}, {Th{\'e}venin}, {Gracia-Abril}, {Portell}, {Teyssier},
  {Altmann}, {Andrae}, {Audard}, {Bellas-Velidis}, {Benson}, {Berthier},
  {Blomme}, {Burgess}, {Busonero}, {Busso}, {C{\'a}novas}, {Carry}, {Cellino},
  {Cheek}, {Clementini}, {Damerdji}, {Davidson}, {de Teodoro}, {Nu{\~n}ez
  Campos}, {Delchambre}, {Dell'Oro}, {Esquej}, {Fern{\'a}ndez-Hern{\'a}ndez},
  {Fraile}, {Garabato}, {Garc{\'\i}a-Lario}, {Gosset}, {Haigron}, {Halbwachs},
  {Hambly}, {Harrison}, {Hern{\'a}ndez}, {Hestroffer}, {Hodgkin}, {Holl},
  {Jan{\ss}en}, {Jevardat de Fombelle}, {Jordan}, {Krone-Martins}, {Lanzafame},
  {L{\"o}ffler}, {Marchal}, {Marrese}, {Moitinho}, {Muinonen}, {Osborne},
  {Pancino}, {Pauwels}, {Recio-Blanco}, {Reyl{\'e}}, {Riello}, {Rimoldini},
  {Roegiers}, {Rybizki}, {Sarro}, {Siopis}, {Smith}, {Sozzetti}, {Utrilla},
  {van Leeuwen}, {Abbas}, {{\'A}brah{\'a}m}, {Abreu Aramburu}, {Aerts},
  {Aguado}, {Ajaj}, {Aldea-Montero}, {Altavilla}, {{\'A}lvarez}, {Alves},
  {Anders}, {Anderson}, {Anglada Varela}, {Antoja}, {Baines}, {Baker},
  {Balaguer-N{\'u}{\~n}ez}, {Balbinot}, {Balog}, {Barache}, {Barbato},
  {Barros}, {Barstow}, {Bartolom{\'e}}, {Bassilana}, {Bauchet}, {Becciani},
  {Bellazzini}, {Berihuete}, {Bernet}, {Bertone}, {Bianchi}, {Binnenfeld},
  {Blanco-Cuaresma}, {Blazere}, {Boch}, {Bombrun}, {Bossini}, {Bouquillon},
  {Bragaglia}, {Bramante}, {Breedt}, {Bressan}, {Brouillet}, {Brugaletta},
  {Bucciarelli}, {Burlacu}, {Butkevich}, {Buzzi}, {Caffau}, {Cancelliere},
  {Cantat-Gaudin}, {Carballo}, {Carlucci}, {Carnerero}, {Carrasco},
  {Casamiquela}, {Castellani}, {Castro-Ginard}, {Chaoul}, {Charlot}, {Chemin},
  {Chiaramida}, {Chiavassa}, {Chornay}, {Comoretto}, {Contursi}, {Cooper},
  {Cornez}, {Cowell}, {Crifo}, {Cropper}, {Crosta}, {Crowley}, {Dafonte},
  {Dapergolas}, {David}, {David}, {de Laverny}, {De Luise}, {De March}, {De
  Ridder}, {de Souza}, {de Torres}, {del Peloso}, {del Pozo}, {Delbo},
  {Delgado}, {Delisle}, {Demouchy}, {Dharmawardena}, {Di Matteo}, {Diakite},
  {Diener}, {Distefano}, {Dolding}, {Edvardsson}, {Enke}, {Fabre}, {Fabrizio},
  {Faigler}, {Fedorets}, {Fernique}, {Fienga}, {Figueras}, {Fournier},
  {Fouron}, {Fragkoudi}, {Gai}, {Garcia-Gutierrez}, {Garcia-Reinaldos},
  {Garc{\'\i}a-Torres}, {Garofalo}, {Gavel}, {Gavras}, {Gerlach}, {Geyer},
  {Giacobbe}, {Gilmore}, {Girona}, {Giuffrida}, {Gomel}, {Gomez},
  {Gonz{\'a}lez-N{\'u}{\~n}ez}, {Gonz{\'a}lez-Santamar{\'\i}a},
  {Gonz{\'a}lez-Vidal}, {Granvik}, {Guillout}, {Guiraud},
  {Guti{\'e}rrez-S{\'a}nchez}, {Guy}, {Hatzidimitriou}, {Hauser}, {Haywood},
  {Helmer}, {Helmi}, {Sarmiento}, {Hidalgo}, {Hilger}, {H{\l}adczuk}, {Hobbs},
  {Holland}, {Huckle}, {Jardine}, {Jasniewicz}, {Jean-Antoine Piccolo},
  {Jim{\'e}nez-Arranz}, {Jorissen}, {Juaristi Campillo}, {Julbe}, {Karbevska},
  {Kervella}, {Khanna}, {Kontizas}, {Kordopatis}, {Korn}, {K{\'o}sp{\'a}l},
  {Kostrzewa-Rutkowska}, {Kruszy{\'n}ska}, {Kun}, {Laizeau}, {Lambert},
  {Lanza}, {Lasne}, {Le Campion}, {Lebreton}, {Lebzelter}, {Leccia}, {Leclerc},
  {Lecoeur-Taibi}, {Liao}, {Licata}, {Lindstr{\o}m}, {Lister}, {Livanou},
  {Lobel}, {Lorca}, {Loup}, {Madrero Pardo}, {Magdaleno Romeo}, {Managau},
  {Mann}, {Manteiga}, {Marchant}, {Marconi}, {Marcos}, {Marcos Santos},
  {Mar{\'\i}n Pina}, {Marinoni}, {Marocco}, {Marshall}, {Martin Polo},
  {Mart{\'\i}n-Fleitas}, {Marton}, {Mary}, {Masip}, {Massari},
  {Mastrobuono-Battisti}, {Mazeh}, {McMillan}, {Messina}, {Michalik}, {Millar},
  {Mints}, {Molina}, {Molinaro}, {Moln{\'a}r}, {Monari}, {Mongui{\'o}},
  {Montegriffo}, {Montero}, {Mor}, {Mora}, {Morbidelli}, {Morel}, {Morris},
  {Muraveva}, {Murphy}, {Musella}, {Nagy}, {Noval}, {Oca{\~n}a}, {Ogden},
  {Ordenovic}, {Osinde}, {Pagani}, {Pagano}, {Palaversa}, {Palicio},
  {Pallas-Quintela}, {Panahi}, {Payne-Wardenaar}, {Pe{\~n}alosa Esteller},
  {Penttil{\"a}}, {Pichon}, {Piersimoni}, {Pineau}, {Plachy}, {Plum}, {Poggio},
  {Pr{\v{s}}a}, {Pulone}, {Racero}, {Ragaini}, {Rainer}, {Raiteri}, {Rambaux},
  {Ramos}, {Ramos-Lerate}, {Re Fiorentin}, {Regibo}, {Richards}, {Rios Diaz},
  {Ripepi}, {Riva}, {Rix}, {Rixon}, {Robichon}, {Robin}, {Robin}, {Roelens},
  {Rogues}, {Rohrbasser}, {Romero-G{\'o}mez}, {Rowell}, {Royer}, {Ruz Mieres},
  {Rybicki}, {Sadowski}, {S{\'a}ez N{\'u}{\~n}ez}, {Sagrist{\`a} Sell{\'e}s},
  {Sahlmann}, {Salguero}, {Samaras}, {Sanchez Gimenez}, {Sanna},
  {Santove{\~n}a}, {Sarasso}, {Schultheis}, {Sciacca}, {Segol}, {Segovia},
  {S{\'e}gransan}, {Semeux}, {Shahaf}, {Siddiqui}, {Siebert}, {Siltala},
  {Silvelo}, {Slezak}, {Slezak}, {Smart}, {Snaith}, {Solano}, {Solitro},
  {Souami}, {Souchay}, {Spagna}, {Spina}, {Spoto}, {Steele},
  {Steidelm{\"u}ller}, {Stephenson}, {S{\"u}veges}, {Surdej}, {Szabados},
  {Szegedi-Elek}, {Taris}, {Taylor}, {Teixeira}, {Tolomei}, {Tonello}, {Torra},
  {Torra}, {Torralba Elipe}, {Trabucchi}, {Tsounis}, {Turon}, {Ulla}, {Unger},
  {Vaillant}, {van Dillen}, {van Reeven}, {Vanel}, {Vecchiato}, {Viala},
  {Vicente}, {Voutsinas}, {Weiler}, {Wevers}, {Wyrzykowski}, {Yoldas}, {Yvard},
  {Zhao}, {Zorec}, {Zucker}, \& {Zwitter}}]{gaiadr3_a}
{Gaia Collaboration}, {Vallenari}, A., {Brown}, A.~G.~A., {et~al.} 2023, \aap,
  674, A1

\bibitem[{{Gaidos} {et~al.}(2017){Gaidos}, {Mann}, {Rizzuto}, {Nofi}, {Mace},
  {Vanderburg}, {Feiden}, {Narita}, {Takeda}, {Esposito}, {De Rosa}, {Ansdell},
  {Hirano}, {Graham}, {Kraus}, \& {Jaffe}}]{Gaidos2017MNRAS.464..850G}
{Gaidos}, E., {Mann}, A.~W., {Rizzuto}, A., {et~al.} 2017, \mnras, 464, 850

\bibitem[{Gaigalas {et~al.}(2019)Gaigalas, Kato, Rynkun, Rada—, \&
  Tanaka}]{Gaigalas2019}
Gaigalas, G., Kato, D., Rynkun, P., Rada—, L., \& Tanaka, M. 2019, The
  Astrophysical Journal Supplement Series, 240, 29

\bibitem[{{Gallagher} {et~al.}(2012){Gallagher}, {Ryan}, {Hosford},
  {Garc{\'\i}a P{\'e}rez}, {Aoki}, \& {Honda}}]{Gallagher2012}
{Gallagher}, A.~J., {Ryan}, S.~G., {Hosford}, A., {et~al.} 2012, \aap, 538,
  A118

\bibitem[{{Gallet} \& {Bouvier}(2013)}]{gallet2013}
{Gallet}, F. \& {Bouvier}, J. 2013, \aap, 556, A36

\bibitem[{{Gallet} \& {Bouvier}(2015)}]{gallet2015}
{Gallet}, F. \& {Bouvier}, J. 2015, \aap, 577, A98

\bibitem[{{Gallet} {et~al.}(2019){Gallet}, {Zanni}, \& {Amard}}]{gallet2019}
{Gallet}, F., {Zanni}, C., \& {Amard}, L. 2019, \aap, 632, A6

\bibitem[{{Gao} {et~al.}(2018){Gao}, {Lind}, {Amarsi}, {Buder}, {Dotter},
  {Nordlander}, {Asplund}, {Bland-Hawthorn}, {de Silva}, {D'Orazi}, {Freeman},
  {Kos}, {Lewis}, {Lin}, {Martell}, {Schlesinger}, {Sharma}, {Simpson},
  {Zucker}, {Zwitter}, {da Costa}, {Anguiano}, {Horner}, {Hyde}, {Kafle},
  {Nataf}, {Reid}, {Stello}, {Ting}, \& {Galah Collaboration}}]{gao2018}
{Gao}, X., {Lind}, K., {Amarsi}, A.~M., {et~al.} 2018, \mnras, 481, 2666

\bibitem[{{Gerber} {et~al.}(2023){Gerber}, {Magg}, {Plez}, {Bergemann},
  {Heiter}, {Olander}, \& {Hoppe}}]{gerber2023}
{Gerber}, J.~M., {Magg}, E., {Plez}, B., {et~al.} 2023, \aap, 669, A43

\bibitem[{Ghosh \& Sharma(2026)}]{GHOSH2026113273}
Ghosh, N. \& Sharma, L. 2026, Radiation Physics and Chemistry, 239, 113273

\bibitem[{{Gibson} {et~al.}(2009){Gibson}, {Jiang}, {Brandt}, {Hall}, {Shen},
  {Wu}, {Anderson}, {Schneider}, {Vanden Berk}, {Gallagher}, {Fan}, \&
  {York}}]{Gibson2009ApJ...692..758G}
{Gibson}, R.~R., {Jiang}, L., {Brandt}, W.~N., {et~al.} 2009, \apj, 692, 758

\bibitem[{{Gieles} {et~al.}(2025){Gieles}, {Padoan}, {Charbonnel}, {Vink}, \&
  {Ram{\'\i}rez-Galeano}}]{Gieles2025}
{Gieles}, M., {Padoan}, P., {Charbonnel}, C., {Vink}, J.~S., \&
  {Ram{\'\i}rez-Galeano}, L. 2025, \mnras, 544, 483

\bibitem[{{Gieles} \& {Zocchi}(2015)}]{GielesZocchi2015}
{Gieles}, M. \& {Zocchi}, A. 2015, \mnras, 454, 576

\bibitem[{{Gilliland} {et~al.}(2000){Gilliland}, {Brown}, {Guhathakurta},
  {Sarajedini}, {Milone}, {Albrow}, {Baliber}, {Bruntt}, {Burrows},
  {Charbonneau}, {Choi}, {Cochran}, {Edmonds}, {Frandsen}, {Howell}, {Lin},
  {Marcy}, {Mayor}, {Naef}, {Sigurdsson}, {Stagg}, {Vandenberg}, {Vogt}, \&
  {Williams}}]{gilliland2000}
{Gilliland}, R.~L., {Brown}, T.~M., {Guhathakurta}, P., {et~al.} 2000, \apjl,
  545, L47

\bibitem[{{Gilmore} {et~al.}(2022){Gilmore}, {Randich}, {Worley}, {Hourihane},
  {Gonneau}, {Sacco}, {Lewis}, {Magrini}, {Fran{\c{c}}ois}, {Jeffries},
  {Koposov}, {Bragaglia}, {Alfaro}, {Allende Prieto}, {Blomme}, {Korn},
  {Lanzafame}, {Pancino}, {Recio-Blanco}, {Smiljanic}, {Van Eck}, {Zwitter},
  {Bensby}, {Flaccomio}, {Irwin}, {Franciosini}, {Morbidelli}, {Damiani},
  {Bonito}, {Friel}, {Vink}, {Prisinzano}, {Abbas}, {Hatzidimitriou}, {Held},
  {Jordi}, {Paunzen}, {Spagna}, {Jackson}, {Ma{\'\i}z Apell{\'a}niz},
  {Asplund}, {Bonifacio}, {Feltzing}, {Binney}, {Drew}, {Ferguson}, {Micela},
  {Negueruela}, {Prusti}, {Rix}, {Vallenari}, {Bergemann}, {Casey}, {de
  Laverny}, {Frasca}, {Hill}, {Lind}, {Sbordone}, {Sousa}, {Adibekyan},
  {Caffau}, {Daflon}, {Feuillet}, {Gebran}, {Gonzalez Hernandez}, {Guiglion},
  {Herrero}, {Lobel}, {Merle}, {Mikolaitis}, {Montes}, {Morel}, {Ruchti},
  {Soubiran}, {Tabernero}, {Tautvai{\v{s}}ien{\.{e}}}, {Traven}, {Valentini},
  {Van der Swaelmen}, {Villanova}, {Viscasillas V{\'a}zquez}, {Bayo}, {Biazzo},
  {Carraro}, {Edvardsson}, {Heiter}, {Jofr{\'e}}, {Marconi}, {Martayan},
  {Masseron}, {Monaco}, {Walton}, {Zaggia}, {Aguirre B{\o}rsen-Koch}, {Alves},
  {Balaguer-Nunez}, {Barklem}, {Barrado}, {Bellazzini}, {Berlanas}, {Binks},
  {Bressan}, {Capuzzo-Dolcetta}, {Casagrande}, {Casamiquela}, {Collins},
  {D'Orazi}, {Dantas}, {Debattista}, {Delgado-Mena}, {Di Marcantonio},
  {Drazdauskas}, {Evans}, {Famaey}, {Franchini}, {Fr{\'e}mat}, {Fu}, {Geisler},
  {Gerhard}, {Gonz{\'a}lez Solares}, {Grebel}, {Guti{\'e}rrez Albarr{\'a}n},
  {Jim{\'e}nez-Esteban}, {J{\"o}nsson}, {Khachaturyants}, {Kordopatis}, {Kos},
  {Lagarde}, {Ludwig}, {Mahy}, {Mapelli}, {Marfil}, {Martell}, {Messina},
  {Miglio}, {Minchev}, {Moitinho}, {Montalban}, {Monteiro}, {Morossi},
  {Mowlavi}, {Mucciarelli}, {Murphy}, {Nardetto}, {Ortolani}, {Paletou},
  {Palou{\v{s}}}, {Pickering}, {Quirrenbach}, {Re Fiorentin}, {Read}, {Romano},
  {Ryde}, {Sanna}, {Santos}, {Seabroke}, {Spina}, {Steinmetz}, {Stonkut{\'e}},
  {Sutorius}, {Th{\'e}venin}, {Tosi}, {Tsantaki}, {Wright}, {Wyse}, {Zoccali},
  {Zorec}, \& {Zucker}}]{Gilmore2022}
{Gilmore}, G., {Randich}, S., {Worley}, C.~C., {et~al.} 2022, \aap, 666, A120

\bibitem[{{Giribaldi} {et~al.}(2021){Giribaldi}, {da Silva}, {Smiljanic}, \&
  {Cornejo Espinoza}}]{Giribaldi2021}
{Giribaldi}, R.~E., {da Silva}, A.~R., {Smiljanic}, R., \& {Cornejo Espinoza},
  D. 2021, \aap, 650, A194

\bibitem[{{Giribaldi} {et~al.}(2025){Giribaldi}, {Magrini},
  {Schiappacasse-Ulloa}, {Randich}, \& {Merle}}]{Giribaldi2025}
{Giribaldi}, R.~E., {Magrini}, L., {Schiappacasse-Ulloa}, J., {Randich}, S., \&
  {Merle}, T. 2025, \aap, 702, A65

\bibitem[{{Giribaldi} \& {Smiljanic}(2023)}]{GiribaldiSmiljanic2023}
{Giribaldi}, R.~E. \& {Smiljanic}, R. 2023, Experimental Astronomy, 55, 117

\bibitem[{{Giribaldi} {et~al.}(2019){Giribaldi}, {Ubaldo-Melo}, {Porto de
  Mello}, {Pasquini}, {Ludwig}, {Ulmer-Moll}, \&
  {Lorenzo-Oliveira}}]{giribaldi2019A&A...624A..10G}
{Giribaldi}, R.~E., {Ubaldo-Melo}, M.~L., {Porto de Mello}, G.~F., {et~al.}
  2019, \aap, 624, A10

\bibitem[{{Giribaldi} {et~al.}(2023){Giribaldi}, {Van Eck}, {Merle},
  {Jorissen}, {Krynski}, {Planquart}, {Valentini}, {Chiappini}, \& {Van
  Winckel}}]{giribaldi2023A&A...679A.110G}
{Giribaldi}, R.~E., {Van Eck}, S., {Merle}, T., {et~al.} 2023, \aap, 679, A110

\bibitem[{{Giribaldi} {et~al.}(2026){Giribaldi}, {Vescovi}, {Magrini},
  {Cristallo}, {D'Orazi}, {Piersanti}, {Cornejo Espinoza}, {Randich}, \&
  {Baratella}}]{Giribaldi2026arXiv260511074G}
{Giribaldi}, R.~E., {Vescovi}, D., {Magrini}, L., {et~al.} 2026, arXiv
  e-prints, arXiv:2605.11074

\bibitem[{{Gomes da Silva} {et~al.}(2022){Gomes da Silva}, {Bensabat},
  {Monteiro}, \& {Santos}}]{GomesdaSilva2022}
{Gomes da Silva}, J., {Bensabat}, A., {Monteiro}, T., \& {Santos}, N.~C. 2022,
  \aap, 668, A174

\bibitem[{{Gomes da Silva} {et~al.}(2021){Gomes da Silva}, {Santos},
  {Adibekyan}, {Sousa}, {Campante}, {Figueira}, {Bossini}, {Delgado-Mena},
  {Monteiro}, {de Laverny}, {Recio-Blanco}, \& {Lovis}}]{GomesdaSilva2021}
{Gomes da Silva}, J., {Santos}, N.~C., {Adibekyan}, V., {et~al.} 2021, \aap,
  646, A77

\bibitem[{{Gonzalez}(2015)}]{ggonzalez2015}
{Gonzalez}, G. 2015, \mnras, 446, 1020

\bibitem[{{Gonzalez} {et~al.}(2020){Gonzalez}, {Mucciarelli}, {Origlia},
  {Schultheis}, {Caffau}, {Di Matteo}, {Randich}, {Recio-Blanco}, {Zoccali},
  {Bonifacio}, {Dalessandro}, {Schiavon}, {Pancino}, {Taylor}, {Valenti},
  {Rojas-Arriagada}, {Sacco}, {Biazzo}, {Bellazzini}, {Cioni}, {Clementini},
  {Contreras Ramos}, {de Laverny}, {Evans}, {Haywood}, {Hill}, {Ibata},
  {Lucatello}, {Magrini}, {Martin}, {Nisini}, {Sanna}, {Cirasuolo}, {Maiolino},
  {Afonso}, {Lilly}, {Flores}, {Oliva}, {Paltani}, \& {Vanzi}}]{gonzalez2020}
{Gonzalez}, O.~A., {Mucciarelli}, A., {Origlia}, L., {et~al.} 2020, The
  Messenger, 180, 18

\bibitem[{{Gonz{\'a}lez Picos} {et~al.}(2025){Gonz{\'a}lez Picos}, {Snellen},
  \& {de Regt}}]{GonzalezPicos2025}
{Gonz{\'a}lez Picos}, D., {Snellen}, I., \& {de Regt}, S. 2025, Nature
  Astronomy, arXiv:2508.18424

\bibitem[{{Goriely} {et~al.}(2011){Goriely}, {Bauswein}, \&
  {Janka}}]{Goriely2011}
{Goriely}, S., {Bauswein}, A., \& {Janka}, H.-T. 2011, \apjl, 738, L32

\bibitem[{{Goupil} {et~al.}(2024){Goupil}, {Catala}, {Samadi}, {Belkacem},
  {Ouazzani}, {Reese}, {Appourchaux}, {Mathur}, {Cabrera}, {B{\"o}rner},
  {Paproth}, {Moedas}, {Verma}, {Lebreton}, {Deal}, {Ballot}, {Chaplin},
  {Christensen-Dalsgaard}, {Cunha}, {Lanza}, {Miglio}, {Morel}, {Serenelli},
  {Mosser}, {Creevey}, {Moya}, {Garcia}, {Nielsen}, \&
  {Hatt}}]{Goupil2024A&A...683A..78G}
{Goupil}, M.~J., {Catala}, C., {Samadi}, R., {et~al.} 2024, \aap, 683, A78

\bibitem[{{Gratton} {et~al.}(2019){Gratton}, {Bragaglia}, {Carretta},
  {D'Orazi}, {Lucatello}, \& {Sollima}}]{gratton2019}
{Gratton}, R., {Bragaglia}, A., {Carretta}, E., {et~al.} 2019, \aapr, 27, 8

\bibitem[{{Gratton} {et~al.}(2004){Gratton}, {Sneden}, \&
  {Carretta}}]{Gratton2004}
{Gratton}, R., {Sneden}, C., \& {Carretta}, E. 2004, \araa, 42, 385

\bibitem[{{Gratton} {et~al.}(2001){Gratton}, {Bonifacio}, {Bragaglia},
  {Carretta}, {Castellani}, {Centurion}, {Chieffi}, {Claudi}, {Clementini},
  {D'Antona}, {Desidera}, {Fran{\c{c}}ois}, {Grundahl}, {Lucatello}, {Molaro},
  {Pasquini}, {Sneden}, {Spite}, \& {Straniero}}]{gratton2001}
{Gratton}, R.~G., {Bonifacio}, P., {Bragaglia}, A., {et~al.} 2001, \aap, 369,
  87

\bibitem[{{Gratton} {et~al.}(2012){Gratton}, {Carretta}, \&
  {Bragaglia}}]{gratton2012}
{Gratton}, R.~G., {Carretta}, E., \& {Bragaglia}, A. 2012, \aapr, 20, 50

\bibitem[{{Gravity Collaboration} {et~al.}(2020){Gravity Collaboration},
  {Garcia Lopez}, {Natta}, {Caratti o Garatti}, {Ray}, {Fedriani},
  {Koutoulaki}, {Klarmann}, {Perraut}, {Sanchez-Bermudez}, {Benisty},
  {Dougados}, {Labadie}, {Brandner}, {Garcia}, {Henning}, {Caselli}, {Duvert},
  {de Zeeuw}, {Grellmann}, {Abuter}, {Amorim}, {Baub{\"o}ck}, {Berger},
  {Bonnet}, {Buron}, {Cl{\'e}net}, {Coud{\'e} Du Foresto}, {de Wit}, {Eckart},
  {Eisenhauer}, {Filho}, {Gao}, {Garcia Dabo}, {Gendron}, {Genzel},
  {Gillessen}, {Habibi}, {Haubois}, {Haussmann}, {Hippler}, {Hubert},
  {Horrobin}, {Jimenez Rosales}, {Jocou}, {Kervella}, {Kolb}, {Lacour}, {Le
  Bouquin}, {L{\'e}na}, {Ott}, {Paumard}, {Perrin}, {Pfuhl}, {Ramirez}, {Rau},
  {Rousset}, {Scheithauer}, {Shangguan}, {Stadler}, {Straub}, {Straubmeier},
  {Sturm}, {van Dishoeck}, {Vincent}, {von Fellenberg}, {Widmann}, {Wieprecht},
  {Wiest}, {Wiezorrek}, {Woillez}, {Yazici}, \& {Zins}}]{garcialopez2020}
{Gravity Collaboration}, {Garcia Lopez}, R., {Natta}, A., {et~al.} 2020, \nat,
  584, 547

\bibitem[{{Gray}(1977)}]{Gray1977}
{Gray}, D.~F. 1977, \apj, 218, 530

\bibitem[{{Green} {et~al.}(2019){Green}, {Schlafly}, {Zucker}, {Speagle}, \&
  {Finkbeiner}}]{Green2019}
{Green}, G.~M., {Schlafly}, E., {Zucker}, C., {Speagle}, J.~S., \&
  {Finkbeiner}, D. 2019, \apj, 887, 93

\bibitem[{{Gregory} {et~al.}(2016){Gregory}, {Adams}, \&
  {Davies}}]{gregory2016}
{Gregory}, S.~G., {Adams}, F.~C., \& {Davies}, C.~L. 2016, \mnras, 457, 3836

\bibitem[{{Griffith} {et~al.}(2021){Griffith}, {Weinberg}, {Johnson}, {Beaton},
  {Garc{\'\i}a-Hern{\'a}ndez}, {Hasselquist}, {Holtzman}, {Johnson},
  {J{\"o}nsson}, {Lane}, {Nataf}, \& {Roman-Lopes}}]{griffith2021}
{Griffith}, E., {Weinberg}, D.~H., {Johnson}, J.~A., {et~al.} 2021, \apj, 909,
  77

\bibitem[{{Gruyters} {et~al.}(2014){Gruyters}, {Nordlander}, \&
  {Korn}}]{gruyters2014}
{Gruyters}, P., {Nordlander}, T., \& {Korn}, A.~J. 2014, \aap, 567, A72

\bibitem[{Gu(2008)}]{Gu2008FAC}
Gu, M.~F. 2008, Canadian Journal of Physics, 86, 675

\bibitem[{{Gustafsson}(2025)}]{gustafsson2025}
{Gustafsson}, B. 2025, \aapr, 33, 3

\bibitem[{{Gustafsson} {et~al.}(2008){Gustafsson}, {Edvardsson}, {Eriksson},
  {J{\o}rgensen}, {Nordlund}, \& {Plez}}]{gustafsson2008}
{Gustafsson}, B., {Edvardsson}, B., {Eriksson}, K., {et~al.} 2008, \aap, 486,
  951

\bibitem[{{H{\"a}berle} {et~al.}(2024){H{\"a}berle}, {Neumayer}, {Seth},
  {Bellini}, {Libralato}, {Baumgardt}, {Whitaker}, {Dumont}, {Alfaro-Cuello},
  {Anderson}, {Clontz}, {Kacharov}, {Kamann}, {Feldmeier-Krause}, {Milone},
  {Nitschai}, {Pechetti}, \& {van de Ven}}]{Haberle2024}
{H{\"a}berle}, M., {Neumayer}, N., {Seth}, A., {et~al.} 2024, \nat, 631, 285

\bibitem[{{Hamann} {et~al.}(2001){Hamann}, {Barlow}, {Chaffee}, {Foltz}, \&
  {Weymann}}]{Hamann2001ApJ...550..142H}
{Hamann}, F.~W., {Barlow}, T.~A., {Chaffee}, F.~C., {Foltz}, C.~B., \&
  {Weymann}, R.~J. 2001, \apj, 550, 142

\bibitem[{{Hamers} \& {Tremaine}(2017)}]{hamers2017}
{Hamers}, A.~S. \& {Tremaine}, S. 2017, \aj, 154, 272

\bibitem[{{Hammer} {et~al.}(2014){Hammer}, {Barbuy}, {Cuby}, {Kaper}, {Morris},
  {Evans}, {Jagourel}, {Dalton}, {Rees}, {Puech}, {Rodrigues}, {Pearson}, \&
  {Disseau}}]{Hammer2014SPIE.9147E..27H}
{Hammer}, F., {Barbuy}, B., {Cuby}, J.~G., {et~al.} 2014, in Society of
  Photo-Optical Instrumentation Engineers (SPIE) Conference Series, Vol. 9147,
  Ground-based and Airborne Instrumentation for Astronomy V, ed. S.~K.
  {Ramsay}, I.~S. {McLean}, \& H.~{Takami}, 914727

\bibitem[{{Hammer} {et~al.}(2021){Hammer}, {Morris}, {Cuby}, {Kaper},
  {Steinmetz}, {Afonso}, {Barbuy}, {Bergin}, {Finogenov}, {Gallego}, {Kassin},
  {Miller}, {{\"O}stlin}, {Pentericci}, {Schaerer}, {Ziegler}, {Chemla},
  {Dalton}, {De Frondat}, {Evans}, {Le Mignant}, {Puech}, {Rodrigues},
  {Sanchez-Janssen}, {Taburet}, {Tasca}, {Yang}, {Zanchetta}, {Dohlen},
  {Dubbeldam}, {El Hadi}, {Janssen}, {Kelz}, {Larrieu}, {Lewis}, {MacIntosh},
  {Morris}, {Navarro}, \& {Seifert}}]{hammer2021}
{Hammer}, F., {Morris}, S., {Cuby}, J.-G., {et~al.} 2021, The Messenger, 182,
  33

\bibitem[{{Hansen} {et~al.}(2018){Hansen}, {El-Souri}, {Monaco}, {Villanova},
  {Bonifacio}, {Caffau}, \& {Sbordone}}]{Hansen2018ApJ...855...83H}
{Hansen}, C.~J., {El-Souri}, M., {Monaco}, L., {et~al.} 2018, \apj, 855, 83

\bibitem[{{Hansen} {et~al.}(2014){Hansen}, {Montes}, \& {Arcones}}]{Hansen2014}
{Hansen}, C.~J., {Montes}, F., \& {Arcones}, A. 2014, \apj, 797, 123

\bibitem[{{Harris}(1996)}]{harris1996}
{Harris}, W.~E. 1996, \aj, 112, 1487

\bibitem[{{Harris}(2010)}]{Harris2010}
{Harris}, W.~E. 2010, arXiv e-prints, arXiv:1012.3224

\bibitem[{Hartman {et~al.}(2017)Hartman, Engstrom, Lundberg, Nilsson, Quinet,
  Fivet, Palmeri, Malcheva, \& Blagoev}]{Hartman2017}
Hartman, H., Engstrom, L., Lundberg, H., {et~al.} 2017, Astronomy and
  Astrophysics, 600

\bibitem[{Hartman {et~al.}(2015)Hartman, Nilsson, Engstrom, \&
  Lundberg}]{Hartman2015}
Hartman, H., Nilsson, H., Engstrom, L., \& Lundberg, H. 2015, Astronomy and
  Astrophysics, 584

\bibitem[{{Hartman} {et~al.}(2009){Hartman}, {Bakos}, {Torres}, {Kov{\'a}cs},
  {Noyes}, {P{\'a}l}, {Latham}, {Sip{\H{o}}cz}, {Fischer}, {Johnson}, {Marcy},
  {Butler}, {Howard}, {Esquerdo}, {Sasselov}, {Kov{\'a}cs}, {Stefanik},
  {Fernandez}, {L{\'a}z{\'a}r}, {Papp}, \&
  {S{\'a}ri}}]{Hartman2009ApJ...706..785H}
{Hartman}, J.~D., {Bakos}, G.~{\'A}., {Torres}, G., {et~al.} 2009, \apj, 706,
  785

\bibitem[{{Hartmann} {et~al.}(2016){Hartmann}, {Herczeg}, \&
  {Calvet}}]{hartmann2016}
{Hartmann}, L., {Herczeg}, G., \& {Calvet}, N. 2016, \araa, 54, 135

\bibitem[{Hartog {et~al.}(2015)Hartog, Palmer, \& Lawler}]{Hartog2015}
Hartog, E.~A., Palmer, A.~J., \& Lawler, J.~E. 2015, Journal of Physics B:
  Atomic, Molecular and Optical Physics, 48

\bibitem[{Hartog {et~al.}(2020)Hartog, Lawler, \& Roederer}]{DenHartog2020}
Hartog, E. A.~D., Lawler, J.~E., \& Roederer, I.~U. 2020, The Astrophysical
  Journal Supplement Series, 248, 10

\bibitem[{Hartog {et~al.}(2021)Hartog, Lawler, \& Roederer}]{DenHartog2021}
Hartog, E. A.~D., Lawler, J.~E., \& Roederer, I.~U. 2021, The Astrophysical
  Journal Supplement Series, 254, 5

\bibitem[{{Haywood} {et~al.}(2013){Haywood}, {Di Matteo}, {Lehnert}, {Katz}, \&
  {G{\'o}mez}}]{Haywood13}
{Haywood}, M., {Di Matteo}, P., {Lehnert}, M.~D., {Katz}, D., \& {G{\'o}mez},
  A. 2013, \aap, 560, A109

\bibitem[{{Heger} \& {Woosley}(2002)}]{heger2002}
{Heger}, A. \& {Woosley}, S.~E. 2002, \apj, 567, 532

\bibitem[{{Heggie}(2014)}]{Heggie14}
{Heggie}, D.~C. 2014, \mnras, 445, 3435

\bibitem[{Heise {et~al.}(1994)Heise, Hollandt, Kling, Kock, \&
  KÃ¼hne}]{Heise1994}
Heise, C., Hollandt, J., Kling, R., Kock, M., \& KÃ¼hne, M. 1994, Applied
  Optics, 33, 5111

\bibitem[{{Helled} {et~al.}(2014){Helled}, {Bodenheimer}, {Podolak}, {Boley},
  {Meru}, {Nayakshin}, {Fortney}, {Mayer}, {Alibert}, \& {Boss}}]{Helled2014}
{Helled}, R., {Bodenheimer}, P., {Podolak}, M., {et~al.} 2014, in Protostars
  and Planets VI, ed. H.~{Beuther}, R.~S. {Klessen}, C.~P. {Dullemond}, \&
  T.~{Henning}, 643--665

\bibitem[{{Helmi} {et~al.}(2018){Helmi}, {Babusiaux}, {Koppelman}, {Massari},
  {Veljanoski}, \& {Brown}}]{helmi2018}
{Helmi}, A., {Babusiaux}, C., {Koppelman}, H.~H., {et~al.} 2018, \nat, 563, 85

\bibitem[{{Helmi} {et~al.}(1999){Helmi}, {White}, {de Zeeuw}, \&
  {Zhao}}]{helmi1999}
{Helmi}, A., {White}, S.~D.~M., {de Zeeuw}, P.~T., \& {Zhao}, H. 1999, \nat,
  402, 53

\bibitem[{{H{\'e}nault-Brunet} {et~al.}(2015){H{\'e}nault-Brunet}, {Gieles},
  {Agertz}, \& {Read}}]{HenaultBrunet2015}
{H{\'e}nault-Brunet}, V., {Gieles}, M., {Agertz}, O., \& {Read}, J.~I. 2015,
  \mnras, 450, 1164

\bibitem[{{Hill} {et~al.}(2017){Hill}, {Christlieb}, {Beers}, {Barklem},
  {Kratz}, {Nordstr{\"o}m}, {Pfeiffer}, \& {Farouqi}}]{Hill2017}
{Hill}, V., {Christlieb}, N., {Beers}, T.~C., {et~al.} 2017, \aap, 607, A91

\bibitem[{{Hill} {et~al.}(2011){Hill}, {Lecureur}, {G{\'o}mez}, {Zoccali},
  {Schultheis}, {Babusiaux}, {Royer}, {Barbuy}, {Arenou}, {Minniti}, \&
  {Ortolani}}]{hill2011}
{Hill}, V., {Lecureur}, A., {G{\'o}mez}, A., {et~al.} 2011, \aap, 534, A80

\bibitem[{{Hill} {et~al.}(2002){Hill}, {Plez}, {Cayrel}, {Beers},
  {Nordstr{\"o}m}, {Andersen}, {Spite}, {Spite}, {Barbuy}, {Bonifacio},
  {Depagne}, {Fran{\c c}ois}, \& {Primas}}]{Hill2002}
{Hill}, V., {Plez}, B., {Cayrel}, R., {et~al.} 2002, \aap, 387, 560

\bibitem[{{Hill} {et~al.}(2019){Hill}, {Sk{\'u}lad{\'o}ttir}, {Tolstoy},
  {Venn}, {Shetrone}, {Jablonka}, {Primas}, {Battaglia}, {de Boer},
  {Fran{\c{c}}ois}, {Helmi}, {Kaufer}, {Letarte}, {Starkenburg}, \&
  {Spite}}]{hill2019}
{Hill}, V., {Sk{\'u}lad{\'o}ttir}, {\'A}., {Tolstoy}, E., {et~al.} 2019, \aap,
  626, A15

\bibitem[{{Hirano} {et~al.}(2014){Hirano}, {Hosokawa}, {Yoshida}, {Umeda},
  {Omukai}, {Chiaki}, \& {Yorke}}]{hirano2014}
{Hirano}, S., {Hosokawa}, T., {Yoshida}, N., {et~al.} 2014, \apj, 781, 60

\bibitem[{{Holmbeck} {et~al.}(2018){Holmbeck}, {Beers}, {Roederer}, {Placco},
  {Hansen}, {Sakari}, {Sneden}, {Liu}, {Lee}, {Cowan}, \&
  {Frebel}}]{Holmbeck-2018}
{Holmbeck}, E.~M., {Beers}, T.~C., {Roederer}, I.~U., {et~al.} 2018, \apjl,
  859, L24

\bibitem[{Holmes {et~al.}(2016)Holmes, Pickering, Ruffoni, Blackwell-Whitehead,
  Nilsson, Engstrom, Hartman, Lundberg, \& Belmonte}]{Holmes2016}
Holmes, C.~E., Pickering, J.~C., Ruffoni, M.~P., {et~al.} 2016, The
  Astrophysical Journal Supplement Series, 224, 35

\bibitem[{{Homma} {et~al.}(2024){Homma}, {Chiba}, {Komiyama}, {Tanaka},
  {Okamoto}, {Tanaka}, {Ishigaki}, {Hayashi}, {Arimoto}, {Lupton}, {Strauss},
  {Miyazaki}, {Wang}, \& {Murayama}}]{homma2024}
{Homma}, D., {Chiba}, M., {Komiyama}, Y., {et~al.} 2024, \pasj, 76, 733

\bibitem[{{Hoppe} {et~al.}(2026){Hoppe}, {Bergemann}, {Eitner}, {Ellwarth},
  {Nordlund}, {Leenaarts}, {Plez}, \& {Serenelli}}]{hoppe25}
{Hoppe}, R., {Bergemann}, M., {Eitner}, P., {et~al.} 2026, \mnras, 546,
  staf2085

\bibitem[{{Horta} {et~al.}(2023){Horta}, {Schiavon}, {Mackereth}, {Weinberg},
  {Hasselquist}, {Feuillet}, {O'Connell}, {Anguiano}, {Allende-Prieto},
  {Beaton}, {Bizyaev}, {Cunha}, {Geisler}, {Garc{\'\i}a-Hern{\'a}ndez},
  {Holtzman}, {J{\"o}nsson}, {Lane}, {Majewski}, {M{\'e}sz{\'a}ros}, {Minniti},
  {Nitschelm}, {Shetrone}, {Smith}, \& {Zasowski}}]{Horta2023}
{Horta}, D., {Schiavon}, R.~P., {Mackereth}, J.~T., {et~al.} 2023, \mnras, 520,
  5671

\bibitem[{{Hoyle}(1946)}]{Hoyle1946}
{Hoyle}, F. 1946, \mnras, 106, 343

\bibitem[{{Hunt} \& {Reffert}(2023)}]{hunt23}
{Hunt}, E.~L. \& {Reffert}, S. 2023, \aap, 673, A114

\bibitem[{{Ibata} {et~al.}(2021){Ibata}, {Malhan}, {Martin}, {Aubert},
  {Famaey}, {Bianchini}, {Monari}, {Siebert}, {Thomas}, {Bellazzini},
  {Bonifacio}, {Caffau}, \& {Renaud}}]{ibata2021}
{Ibata}, R., {Malhan}, K., {Martin}, N., {et~al.} 2021, \apj, 914, 123

\bibitem[{{Ibata} {et~al.}(1994){Ibata}, {Gilmore}, \& {Irwin}}]{ibata1994}
{Ibata}, R.~A., {Gilmore}, G., \& {Irwin}, M.~J. 1994, \nat, 370, 194

\bibitem[{{Ibata} {et~al.}(1995){Ibata}, {Gilmore}, \& {Irwin}}]{ibata1995}
{Ibata}, R.~A., {Gilmore}, G., \& {Irwin}, M.~J. 1995, \mnras, 277, 781

\bibitem[{{Ida} \& {Lin}(2004)}]{ida2004}
{Ida}, S. \& {Lin}, D.~N.~C. 2004, \apj, 616, 567

\bibitem[{{Inoue} {et~al.}(2023){Inoue}, {Maehara}, {Notsu}, {Namekata},
  {Honda}, {Namizaki}, {Nogami}, \& {Shibata}}]{inoue2023}
{Inoue}, S., {Maehara}, H., {Notsu}, Y., {et~al.} 2023, \apj, 948, 9

\bibitem[{{Israelian} {et~al.}(2009){Israelian}, {Delgado Mena}, {Santos},
  {Sousa}, {Mayor}, {Udry}, {Dom{\'{\i}}nguez Cerde{\~n}a}, {Rebolo}, \&
  {Randich}}]{israelian2009}
{Israelian}, G., {Delgado Mena}, E., {Santos}, N.~C., {et~al.} 2009, \nat, 462,
  189

\bibitem[{{Israelian} \& {Rebolo}(2001)}]{israelian2001sulphur}
{Israelian}, G. \& {Rebolo}, R. 2001, \apjl, 557, L43

\bibitem[{{Israelian} {et~al.}(2004){Israelian}, {Santos}, {Mayor}, \&
  {Rebolo}}]{israelian2004}
{Israelian}, G., {Santos}, N.~C., {Mayor}, M., \& {Rebolo}, R. 2004, \aap, 414,
  601

\bibitem[{{Ivanova} {et~al.}(2005){Ivanova}, {Belczynski}, {Fregeau}, \&
  {Rasio}}]{Ivanova2005}
{Ivanova}, N., {Belczynski}, K., {Fregeau}, J.~M., \& {Rasio}, F.~A. 2005,
  \mnras, 358, 572

\bibitem[{{Jardine} \& {Collier Cameron}(2019)}]{jardine2019}
{Jardine}, M. \& {Collier Cameron}, A. 2019, \mnras, 482, 2853

\bibitem[{{Jardine} {et~al.}(2020){Jardine}, {Collier Cameron}, {Donati}, \&
  {Hussain}}]{jardine2020}
{Jardine}, M., {Collier Cameron}, A., {Donati}, J.~F., \& {Hussain}, G.~A.~J.
  2020, \mnras, 491, 4076

\bibitem[{{Jeffers} {et~al.}(2007){Jeffers}, {Donati}, \& {Collier
  Cameron}}]{jeffers2007}
{Jeffers}, S.~V., {Donati}, J.~F., \& {Collier Cameron}, A. 2007, \mnras, 375,
  567

\bibitem[{{Jeffries}(1993)}]{jeffries1993}
{Jeffries}, R.~D. 1993, \mnras, 262, 369

\bibitem[{{Jeffries} {et~al.}(2011){Jeffries}, {Jackson}, {Briggs}, {Evans}, \&
  {Pye}}]{jeffries2011}
{Jeffries}, R.~D., {Jackson}, R.~J., {Briggs}, K.~R., {Evans}, P.~A., \& {Pye},
  J.~P. 2011, \mnras, 411, 2099

\bibitem[{{Jin} {et~al.}(2022){Jin}, {Trager}, {Dalton}, {Aguerri}, {Drew},
  {Falc{\'o}n-Barroso}, {G{\"a}nsicke}, {Hill}, {Iovino}, {Pieri}, {Poggianti},
  {Smith}, {Vallenari}, {Abrams}, {Aguado}, {Antoja}, {Arag{\'o}n-Salamanca},
  {Ascasibar}, {Babusiaux}, {Balcells}, {Barrena}, {Battaglia}, {Belokurov},
  {Bensby}, {Bonifacio}, {Bragaglia}, {Carrasco}, {Carrera}, {Cornwell},
  {Dom{\'\i}nguez-Palmero}, {Duncan}, {Famaey}, {Fari{\~n}a}, {Gonzalez},
  {Guest}, {Hatch}, {Hess}, {Hoskin}, {Irwin}, {Knapen}, {Koposov}, {Kuchner},
  {Laigle}, {Lewis}, {Longhett}, {Lucatello}, {M{\'e}ndez-Abreu}, {Mercurio},
  {Molaeinezhad}, {Mongui{\'o}}, {Morrison}, {Murphy}, {Peralta de Arriba},
  {P{\'e}rez}, {P{\'e}rez-R{\`a}fols}, {Pic{\'o}}, {Raddi}, {Romero-G{\'o}mez},
  {Royer}, {Siebert}, {Seabroke}, {Som}, {Terrett}, {Thomas}, {Wesson},
  {Worley}, {Alfaro}, {Allende Prieto}, {Alonso-Santiago}, {Amos}, {Ashley},
  {Balaguer-N{\'u}{\~n}ez}, {Balbinot}, {Bellazzini}, {Benn}, {Berlanas},
  {Bernard}, {Best}, {Bettoni}, {Bianco}, {Bishop}, {Blomqvist}, {Boeche},
  {Bolzonella}, {Bonoli}, {Bosma}, {Britavskiy}, {Busarello}, {Caffau},
  {Cantat-Gaudin}, {Castro-Ginard}, {Couto}, {Carbajo-Hijarrubia}, {Carter},
  {Casamiquela}, {Conrado}, {Corcho-Caballero}, {Costantin}, {Deason}, {de
  Burgos}, {De Grandi}, {Di Matteo}, {Dom{\'\i}nguez-G{\'o}mez}, {Dorda},
  {Drake}, {Dutta}, {Erkal}, {Feltzing}, {Ferr{\'e}-Mateu}, {Feuillet},
  {Figueras}, {Fossat}, {Franciosin}, {Frasca}, {Fumagalli}, {Gallazzi},
  {Garc{\'\i}a-Benito}, {Gentile Fusillo}, {Gebran}, {Gilbert}, {Gledhill},
  {Gonz{\'a}lez Delgado}, {Greimel}, {Guarcello}, {Guerra}, {Gullieuszik},
  {Haines}, {Hardcastle}, {Harris}, {Haywood}, {Helmi}, {Hernandez}, {Herrero},
  {Hughes}, {Irsic}, {Jablonka}, {Jarvis}, {Jordi}, {Kondapally}, {Kordopatis},
  {Krogager}, {La Barbera}, {Lam}, {Larsen}, {Lemasle}, {Lewis}, {Lhom{\'e}},
  {Lind}, {Lodi}, {Longobardi}, {Lonoce}, {Magrin}, {Ma{\'\i}z Apell{\'a}niz},
  {Marchal}, {Marco}, {Martin}, {Matsuno}, {Maurogordato}, {Merluzzi},
  {Miralda-Escud{\'e}}, {Molinari}, {Monari}, {Morelli}, {Mottram}, {Naylor},
  {Negueruela}, {O{\~n}orbe}, {Pancino}, {Peirani}, {Peletier}, {Pozzetti},
  {Rainer}, {Ramos}, {Read}, {Rossi}, {R{\"o}ttgering},
  {Rubi{\~n}o-Mart{\'\i}n}, {Sabater Montes}, {San Juan}, {Sanna}, {Schallig},
  {Schiavon}, {Schultheis}, {Serra}, {Shimwell}, {Sim{\'o}n-D{\'\i}az},
  {Smith}, {Sordo}, {Sorini}, {Soubiran}, {Starkenburg}, {Steele}, {Stott},
  {Stuik}, {Tolstoy}, {Tortora}, {Tsantaki}, {Van der Swaelmen}, {van Weeren},
  {Vergani}, {Verheijen}, {Verro}, {Vink}, {Vioque}, {Walcher}, {Walton},
  {Wegg}, {Weijmans}, {Williams}, {Wilson}, {Wright}, {Xylakis-Dornbusch},
  {Youakim}, {Zibetti}, \& {Zurita}}]{jin2022}
{Jin}, S., {Trager}, S.~C., {Dalton}, G.~B., {et~al.} 2022, arXiv e-prints,
  arXiv:2212.03981

\bibitem[{{Jofr{\'e}} {et~al.}(2020){Jofr{\'e}}, {Jackson}, \& {Tucci
  Maia}}]{jofre2020}
{Jofr{\'e}}, P., {Jackson}, H., \& {Tucci Maia}, M. 2020, \aap, 633, L9

\bibitem[{Johansson {et~al.}(2002)Johansson, Derkatch, Donnelly, Hartman,
  Hibbert, Karlsson, Kock, Li, Leckrone, LitzeÂ¨n, Lundberg, Mannervik,
  Norlin, Nilsson, Pickering, Raassen, Rostohar, Royen, Schmitt, Johanning,
  Sikstro, Smith, Svanberg, \& Wahlgren}]{Johansson2002}
Johansson, S., Derkatch, A., Donnelly, M.~P., {et~al.} 2002, Phys. Scr, 71

\bibitem[{{Johnson} {et~al.}(2010){Johnson}, {Dong}, \& {Gould}}]{Johnson2010}
{Johnson}, J.~A., {Dong}, S., \& {Gould}, A. 2010, \apj, 713, 713

\bibitem[{{Jones} {et~al.}(2018){Jones}, {Brahm}, {Espinoza}, {Jord{\'a}n},
  {Rojas}, {Rabus}, {Drass}, {Zapata}, {Soto}, {Jenkins}, {Vu{\v{c}}kovi{\'c}},
  {Ciceri}, \& {Sarkis}}]{jones2018}
{Jones}, M.~I., {Brahm}, R., {Espinoza}, N., {et~al.} 2018, \aap, 613, A76

\bibitem[{{Jones} {et~al.}(2016{\natexlab{a}}){Jones}, {Jenkins}, {Brahm},
  {Wittenmyer}, {Olivares E.}, {Melo}, {Rojo}, {Jord{\'a}n}, {Drass}, {Butler},
  \& {Wang}}]{Jones2016}
{Jones}, M.~I., {Jenkins}, J.~S., {Brahm}, R., {et~al.} 2016{\natexlab{a}},
  \aap, 590, A38

\bibitem[{{Jones} {et~al.}(2021){Jones}, {Wittenmyer}, {Aguilera-G{\'o}mez},
  {Soto}, {Torres}, {Trifonov}, {Jenkins}, {Zapata}, {Sarkis}, {Zakhozhay},
  {Brahm}, {Ram{\'\i}rez}, {Santana}, {Vines}, {D{\'\i}az},
  {Vu{\v{c}}kovi{\'c}}, \& {Pantoja}}]{Jones2021}
{Jones}, M.~I., {Wittenmyer}, R., {Aguilera-G{\'o}mez}, C., {et~al.} 2021,
  \aap, 646, A131

\bibitem[{{Jones} {et~al.}(2016{\natexlab{b}}){Jones}, {Ritter}, {Herwig},
  {Fryer}, {Pignatari}, {Bertolli}, \& {Paxton}}]{Jones2016nucleo}
{Jones}, S., {Ritter}, C., {Herwig}, F., {et~al.} 2016{\natexlab{b}}, \mnras,
  455, 3848

\bibitem[{J{\"o}nsson {et~al.}(2013)J{\"o}nsson, Gaigalas, Biero{\'n},
  Froese~Fischer, \& Grant}]{Jonsson2013GRASP2K}
J{\"o}nsson, P., Gaigalas, G., Biero{\'n}, J., Froese~Fischer, C., \& Grant,
  I.~P. 2013, Computer Physics Communications, 184, 2197

\bibitem[{{Jorissen} {et~al.}(2020){Jorissen}, {Van Winckel}, {Siess},
  {Escorza}, {Pourbaix}, \& {Van Eck}}]{Jorissen2020}
{Jorissen}, A., {Van Winckel}, H., {Siess}, L., {et~al.} 2020, \aap, 639, A7

\bibitem[{{Jung} {et~al.}(2024){Jung}, {Roca-F{\`a}brega}, {Kim}, {Genina},
  {Hausammann}, {Kim}, {Lupi}, {Nagamine}, {Powell}, {Revaz}, {Shimizu},
  {Vel{\'a}zquez}, {Ceverino}, {Primack}, {Quinn}, {Strawn}, {Abel}, {Dekel},
  {Dong}, {Oh}, {Teyssier}, \& {AGORA Collaboration}}]{jung2024}
{Jung}, M., {Roca-F{\`a}brega}, S., {Kim}, J.-H., {et~al.} 2024, \apj, 964, 123

\bibitem[{{Kacharov} {et~al.}(2015){Kacharov}, {Koch}, {Caffau}, \&
  {Sbordone}}]{kacharov2015}
{Kacharov}, N., {Koch}, A., {Caffau}, E., \& {Sbordone}, L. 2015, \aap, 577,
  A18

\bibitem[{{Kamann} {et~al.}(2018){Kamann}, {Husser}, {Dreizler}, {Emsellem},
  {Weilbacher}, {Martens}, {Bacon}, {den Brok}, {Giesers}, {Krajnovi{\'c}},
  {Roth}, {Wendt}, \& {Wisotzki}}]{Kamann2018}
{Kamann}, S., {Husser}, T.~O., {Dreizler}, S., {et~al.} 2018, \mnras, 473, 5591

\bibitem[{{Kanehisa} {et~al.}(2024){Kanehisa}, {Pawlowski}, {Heesters}, \&
  {M{\"u}ller}}]{kanehisa2024}
{Kanehisa}, K.~J., {Pawlowski}, M.~S., {Heesters}, N., \& {M{\"u}ller}, O.
  2024, \aap, 686, A280

\bibitem[{{K{\"a}ppeler} {et~al.}(2011){K{\"a}ppeler}, {Gallino}, {Bisterzo},
  \& {Aoki}}]{Kappeler2011}
{K{\"a}ppeler}, F., {Gallino}, R., {Bisterzo}, S., \& {Aoki}, W. 2011, Reviews
  of Modern Physics, 83, 157

\bibitem[{{Karakas} \& {Lattanzio}(2014)}]{KarakasLattanzio2014}
{Karakas}, A.~I. \& {Lattanzio}, J.~C. 2014, \pasa, 31, e030

\bibitem[{{Kasen} {et~al.}(2017){Kasen}, {Metzger}, {Barnes}, {Quataert}, \&
  {Ramirez-Ruiz}}]{Kasen2017}
{Kasen}, D., {Metzger}, B., {Barnes}, J., {Quataert}, E., \& {Ramirez-Ruiz}, E.
  2017, \nat, 551, 80

\bibitem[{{Katz} {et~al.}(2023){Katz}, {Sartoretti}, {Guerrier}, {Panuzzo},
  {Seabroke}, {Th{\'e}venin}, {Cropper}, {Benson}, {Blomme}, {Haigron},
  {Marchal}, {Smith}, {Baker}, {Chemin}, {Damerdji}, {David}, {Dolding},
  {Fr{\'e}mat}, {Gosset}, {Jan{\ss}en}, {Jasniewicz}, {Lobel}, {Plum},
  {Samaras}, {Snaith}, {Soubiran}, {Vanel}, {Zwitter}, {Antoja}, {Arenou},
  {Babusiaux}, {Brouillet}, {Caffau}, {Di Matteo}, {Fabre}, {Fabricius},
  {Fragkoudi}, {Haywood}, {Huckle}, {Hottier}, {Lasne}, {Leclerc},
  {Mastrobuono-Battisti}, {Royer}, {Teyssier}, {Zorec}, {Crifo}, {Jean-Antoine
  Piccolo}, {Turon}, \& {Viala}}]{Katz2023}
{Katz}, D., {Sartoretti}, P., {Guerrier}, A., {et~al.} 2023, \aap, 674, A5

\bibitem[{{Kawaler}(1988)}]{kawaler1988}
{Kawaler}, S.~D. 1988, \apj, 333, 236

\bibitem[{{King} \& {Wheatley}(2021)}]{king2021}
{King}, G.~W. \& {Wheatley}, P.~J. 2021, \mnras, 501, L28

\bibitem[{{Kirby} {et~al.}(2011){Kirby}, {Lanfranchi}, {Simon}, {Cohen}, \&
  {Guhathakurta}}]{kirby2011}
{Kirby}, E.~N., {Lanfranchi}, G.~A., {Simon}, J.~D., {Cohen}, J.~G., \&
  {Guhathakurta}, P. 2011, \apj, 727, 78

\bibitem[{{Kirby} {et~al.}(2019){Kirby}, {Xie}, {Guo}, {de los Reyes},
  {Bergemann}, {Kovalev}, {Shen}, {Piro}, \& {McWilliam}}]{kirby2019}
{Kirby}, E.~N., {Xie}, J.~L., {Guo}, R., {et~al.} 2019, \apj, 881, 45

\bibitem[{{Klessen} \& {Glover}(2023)}]{klessen2023}
{Klessen}, R.~S. \& {Glover}, S. C.~O. 2023, \araa, 61, 65

\bibitem[{Klose {et~al.}(2002)Klose, Fuhr, \& Wiese}]{Klose2002}
Klose, J.~Z., Fuhr, J.~R., \& Wiese, W.~L. 2002, Journal of Physical and
  Chemical Reference Data, 31, 217â€“230

\bibitem[{{Kobayashi} {et~al.}(2020){Kobayashi}, {Karakas}, \&
  {Lugaro}}]{kobayashi2020}
{Kobayashi}, C., {Karakas}, A.~I., \& {Lugaro}, M. 2020, \apj, 900, 179

\bibitem[{{Kobayashi} {et~al.}(2023){Kobayashi}, {Mandel}, {Belczynski},
  {Goriely}, {Janka}, {Just}, {Ruiter}, {Vanbeveren}, {Kruckow}, {Briel},
  {Eldridge}, \& {Stanway}}]{Kobayashi2023}
{Kobayashi}, C., {Mandel}, I., {Belczynski}, K., {et~al.} 2023, \apjl, 943, L12

\bibitem[{{Koch} \& {Caffau}(2011)}]{koch&caffau2011}
{Koch}, A. \& {Caffau}, E. 2011, \aap, 534, A52

\bibitem[{Kodangil {et~al.}(2024)Kodangil, Domoto, Tanaka, Kato, Gaigalas,
  Tanuma, \& Nakamura}]{Kodangil2024}
Kodangil, S., Domoto, N., Tanaka, M., {et~al.} 2024, Journal of Quantitative
  Spectroscopy and Radiative Transfer, 322, 109011

\bibitem[{Kodangil {et~al.}(2025)Kodangil, Tanaka, Kato, Gaigalas, Tanuma, \&
  Nakamura}]{Kodangil2025}
Kodangil, S., Tanaka, M., Kato, D., {et~al.} 2025, European Physical Journal D,
  79

\bibitem[{Kodangil {et~al.}(2026)Kodangil, Tanaka, Kato, Gaigalas, Tanuma, \&
  Nakamura}]{Kodangil2026}
Kodangil, S., Tanaka, M., Kato, D., {et~al.} 2026, Journal of Quantitative
  Spectroscopy and Radiative Transfer, 348, 109692

\bibitem[{{Koenigl}(1991)}]{koenigl1991}
{Koenigl}, A. 1991, \apjl, 370, L39

\bibitem[{{Koester} {et~al.}(2014){Koester}, {G{\"a}nsicke}, \&
  {Farihi}}]{Koester2014}
{Koester}, D., {G{\"a}nsicke}, B.~T., \& {Farihi}, J. 2014, \aap, 566, A34

\bibitem[{{Konacki} {et~al.}(2009){Konacki}, {Muterspaugh}, {Kulkarni}, \&
  {He{\l}miniak}}]{Konacki2009}
{Konacki}, M., {Muterspaugh}, M.~W., {Kulkarni}, S.~R., \& {He{\l}miniak},
  K.~G. 2009, \apj, 704, 513

\bibitem[{{Kordopatis} {et~al.}(2025){Kordopatis}, {Feuillet}, {Lehmann},
  {Feltzing}, {Minchev}, {Hill}, \& {Ernandes}}]{Kordopatis2025}
{Kordopatis}, G., {Feuillet}, D., {Lehmann}, C., {et~al.} 2025, \aap, 703, A151

\bibitem[{{Kordopatis} {et~al.}(2023{\natexlab{a}}){Kordopatis}, {Hill}, \&
  {Lind}}]{Kordopatis2023a}
{Kordopatis}, G., {Hill}, V., \& {Lind}, K. 2023{\natexlab{a}}, \aap, 674, A104

\bibitem[{{Kordopatis} {et~al.}(2023{\natexlab{b}}){Kordopatis}, {Schultheis},
  {McMillan}, {Palicio}, {de Laverny}, {Recio-Blanco}, {Creevey},
  {{\'A}lvarez}, {Andrae}, {Poggio}, {Spitoni}, {Contursi}, {Zhao},
  {Oreshina-Slezak}, {Ordenovic}, \& {Bijaoui}}]{Kordopatis2023b}
{Kordopatis}, G., {Schultheis}, M., {McMillan}, P.~J., {et~al.}
  2023{\natexlab{b}}, \aap, 669, A104

\bibitem[{{Kordopatis} {et~al.}(2015){Kordopatis}, {Wyse}, {Gilmore},
  {Recio-Blanco}, {de Laverny}, {Hill}, {Adibekyan}, {Heiter}, {Minchev},
  {Famaey}, {Bensby}, {Feltzing}, {Guiglion}, {Korn}, {Mikolaitis},
  {Schultheis}, {Vallenari}, {Bayo}, {Carraro}, {Flaccomio}, {Franciosini},
  {Hourihane}, {Jofr{\'e}}, {Koposov}, {Lardo}, {Lewis}, {Lind}, {Magrini},
  {Morbidelli}, {Pancino}, {Randich}, {Sacco}, {Worley}, \&
  {Zaggia}}]{Kordopatis2015a}
{Kordopatis}, G., {Wyse}, R.~F.~G., {Gilmore}, G., {et~al.} 2015, \aap, 582,
  A122

\bibitem[{{Korn} {et~al.}(2006){Korn}, {Grundahl}, {Richard}, {Barklem},
  {Mashonkina}, {Collet}, {Piskunov}, \& {Gustafsson}}]{korn2006}
{Korn}, A.~J., {Grundahl}, F., {Richard}, O., {et~al.} 2006, \nat, 442, 657

\bibitem[{{Korn} {et~al.}(2007){Korn}, {Grundahl}, {Richard}, {Mashonkina},
  {Barklem}, {Collet}, {Gustafsson}, \& {Piskunov}}]{korn2007}
{Korn}, A.~J., {Grundahl}, F., {Richard}, O., {et~al.} 2007, \apj, 671, 402

\bibitem[{{Korotin} \& {Ku{\v{c}}inskas}(2022)}]{korotin2022}
{Korotin}, S. \& {Ku{\v{c}}inskas}, A. 2022, \aap, 657, L11

\bibitem[{{Korotin}(2020)}]{korotin2020}
{Korotin}, S.~A. 2020, Astronomy Letters, 46, 541

\bibitem[{{Korotin} \& {Kiselev}(2024)}]{korotin2024}
{Korotin}, S.~A. \& {Kiselev}, K.~O. 2024, Astronomy Reports, 68, 1159

\bibitem[{{Kostov} {et~al.}(2016){Kostov}, {Moore}, {Tamayo}, {Jayawardhana},
  \& {Rinehart}}]{Kostov16:CE}
{Kostov}, V.~B., {Moore}, K., {Tamayo}, D., {Jayawardhana}, R., \& {Rinehart},
  S.~A. 2016, \apj, 832, 183

\bibitem[{{Koutsouridou} {et~al.}(2024){Koutsouridou}, {Salvadori}, \&
  {Sk{\'u}lad{\'o}ttir}}]{koutsouridou2024}
{Koutsouridou}, I., {Salvadori}, S., \& {Sk{\'u}lad{\'o}ttir}, {\'A}. 2024,
  \apjl, 962, L26

\bibitem[{{Koutsouridou} {et~al.}(2025){Koutsouridou}, {Sk{\'u}lad{\'o}ttir},
  \& {Salvadori}}]{koutsouridou2025}
{Koutsouridou}, I., {Sk{\'u}lad{\'o}ttir}, {\'A}., \& {Salvadori}, S. 2025,
  \aap, 699, A32

\bibitem[{Kramida {et~al.}(2024)Kramida, {Yu.~Ralchenko}, Reader, \& {and NIST
  ASD Team}}]{NIST_ASD}
Kramida, A., {Yu.~Ralchenko}, Reader, J., \& {and NIST ASD Team}. 2024, {NIST
  Atomic Spectra Database (ver. 5.12), [Online]. Available:
  {\tt{https://physics.nist.gov/asd}} [2025, November 17]. National Institute
  of Standards and Technology, Gaithersburg, MD.}

\bibitem[{{Kratz} {et~al.}(2014){Kratz}, {Farouqi}, \&
  {M{\"o}ller}}]{Kratz2014}
{Kratz}, K.-L., {Farouqi}, K., \& {M{\"o}ller}, P. 2014, \apj, 792, 6

\bibitem[{{Kuerster} {et~al.}(1994){Kuerster}, {Schmitt}, \&
  {Cutispoto}}]{kuerster1994}
{Kuerster}, M., {Schmitt}, J.~H.~M.~M., \& {Cutispoto}, G. 1994, \aap, 289, 899

\bibitem[{{Kullmann} {et~al.}(2023){Kullmann}, {Goriely}, {Just}, {Bauswein},
  \& {Janka}}]{Kullmann-2023}
{Kullmann}, I., {Goriely}, S., {Just}, O., {Bauswein}, A., \& {Janka}, H.-T.
  2023, \mnras, 523, 2551

\bibitem[{{Kurucz}(2005)}]{kurucz05LSb}
{Kurucz}, R.~L. 2005, Memorie della Societa Astronomica Italiana Supplementi,
  8, 14

\bibitem[{{Kuske} {et~al.}(2025){Kuske}, {Arcones}, \& {Reichert}}]{Kuske2025}
{Kuske}, J., {Arcones}, A., \& {Reichert}, M. 2025, \apj, 990, 37

\bibitem[{{Lagae} {et~al.}(2025){Lagae}, {Amarsi}, \& {Lind}}]{lagae2025}
{Lagae}, C., {Amarsi}, A.~M., \& {Lind}, K. 2025, \aap, 697, A60

\bibitem[{{Lagarde} {et~al.}(2024){Lagarde}, {Minkevi{\v{c}}i{\={u}}t{\.{e}}},
  {Drazdauskas}, {Tautvai{\v{s}}ien{\.{e}}}, {Charbonnel}, {Reyl{\'e}},
  {Miglio}, {Kushwahaa}, \& {Bale}}]{Lagarde2024}
{Lagarde}, N., {Minkevi{\v{c}}i{\={u}}t{\.{e}}}, R., {Drazdauskas}, A.,
  {et~al.} 2024, \aap, 684, A70

\bibitem[{{Lagrange} {et~al.}(2010){Lagrange}, {Bonnefoy}, {Chauvin}, {Apai},
  {Ehrenreich}, {Boccaletti}, {Gratadour}, {Rouan}, {Mouillet}, {Lacour}, \&
  {Kasper}}]{lagrange2010}
{Lagrange}, A.~M., {Bonnefoy}, M., {Chauvin}, G., {et~al.} 2010, Science, 329,
  57

\bibitem[{{Lammers} \& {Winn}(2026)}]{Lammers2026}
{Lammers}, C. \& {Winn}, J.~N. 2026, \aj, 171, 18

\bibitem[{{Lanza}(2009)}]{lanza2009}
{Lanza}, A.~F. 2009, \aap, 505, 339

\bibitem[{{Lardo} {et~al.}(2023){Lardo}, {Salaris}, {Cassisi}, {Bastian},
  {Mucciarelli}, {Cabrera-Ziri}, \& {Dalessandro}}]{Lardo2023}
{Lardo}, C., {Salaris}, M., {Cassisi}, S., {et~al.} 2023, \aap, 669, A19

\bibitem[{{Latour} {et~al.}(2025){Latour}, {Kamann}, {Martocchia}, {Husser},
  {Saracino}, \& {Dreizler}}]{Latour2025}
{Latour}, M., {Kamann}, S., {Martocchia}, S., {et~al.} 2025, \aap, 694, A248

\bibitem[{Lawler \& Den\;Hartog(2019)}]{2019lawlercalibration}
Lawler, J.~E. \& Den\;Hartog, E.~A. 2019, Journal of Quantitative Spectroscopy
  and Radiative Transfer, 237, 106620

\bibitem[{Lawler {et~al.}(2022)Lawler, Schmidt, \& Hartog}]{Lawler2022a}
Lawler, J.~E., Schmidt, J.~R., \& Hartog, E. A.~D. 2022, The Astrophysical
  Journal Supplement Series, 258, 27

\bibitem[{{Ledda} {et~al.}(2023){Ledda}, {Danielski}, \& {Turrini}}]{Ledda2023}
{Ledda}, S., {Danielski}, C., \& {Turrini}, D. 2023, \aap, 675, A184

\bibitem[{{Lee} {et~al.}(2021{\natexlab{a}}){Lee}, {Webb}, {Carswell}, \&
  {Milakovi{\'c}}}]{Lee2021MNRAS.504.1787L}
{Lee}, C.-C., {Webb}, J.~K., {Carswell}, R.~F., \& {Milakovi{\'c}}, D.
  2021{\natexlab{a}}, \mnras, 504, 1787

\bibitem[{{Lee} {et~al.}(2021{\natexlab{b}}){Lee}, {Webb}, {Milakovi{\'c}}, \&
  {Carswell}}]{Lee2021MNRAS.507...27L_b}
{Lee}, C.-C., {Webb}, J.~K., {Milakovi{\'c}}, D., \& {Carswell}, R.~F.
  2021{\natexlab{b}}, \mnras, 507, 27

\bibitem[{{Leenaarts} \& {Carlsson}(2009)}]{leenaarts2009}
{Leenaarts}, J. \& {Carlsson}, M. 2009, in Astronomical Society of the Pacific
  Conference Series, Vol. 415, The Second Hinode Science Meeting: Beyond
  Discovery-Toward Understanding, ed. B.~{Lites}, M.~{Cheung}, T.~{Magara},
  J.~{Mariska}, \& K.~{Reeves}, 87

\bibitem[{{Legnardi} {et~al.}(2022){Legnardi}, {Milone}, {Armillotta},
  {Marino}, {Cordoni}, {Renzini}, {Vesperini}, {D'Antona}, {McKenzie}, {Yong},
  {Dondoglio}, {Lagioia}, {Carlos}, {Tailo}, {Jang}, \&
  {Mohandasan}}]{Legnardi2022}
{Legnardi}, M.~V., {Milone}, A.~P., {Armillotta}, L., {et~al.} 2022, \mnras,
  513, 735

\bibitem[{{Li} {et~al.}(2026){Li}, {Amarsi}, \&
  {J{\"o}nsson}}]{2026A&A...707A.141L}
{Li}, W., {Amarsi}, A.~M., \& {J{\"o}nsson}, P. 2026, \aap, 707, A141

\bibitem[{{Li} {et~al.}(2021){Li}, {Amarsi}, {Papoulia}, {Ekman}, \&
  {J{\"o}nsson}}]{li2021}
{Li}, W., {Amarsi}, A.~M., {Papoulia}, A., {Ekman}, J., \& {J{\"o}nsson}, P.
  2021, \mnras, 502, 3780

\bibitem[{{Libralato} {et~al.}(2016{\natexlab{a}}){Libralato}, {Bedin},
  {Nardiello}, \& {Piotto}}]{Libralato2016MNRAS.456.1137L}
{Libralato}, M., {Bedin}, L.~R., {Nardiello}, D., \& {Piotto}, G.
  2016{\natexlab{a}}, \mnras, 456, 1137

\bibitem[{{Libralato} {et~al.}(2022){Libralato}, {Bellini}, {Vesperini},
  {Piotto}, {Milone}, {van der Marel}, {Anderson}, {Aparicio}, {Barbuy},
  {Bedin}, {Borsato}, {Cassisi}, {Dalessandro}, {Ferraro}, {King}, {Lanzoni},
  {Nardiello}, {Ortolani}, {Sarajedini}, \& {Sohn}}]{Libralato2022}
{Libralato}, M., {Bellini}, A., {Vesperini}, E., {et~al.} 2022, \apj, 934, 150

\bibitem[{{Libralato} {et~al.}(2016{\natexlab{b}}){Libralato}, {Nardiello},
  {Bedin}, {Borsato}, {Granata}, {Malavolta}, {Piotto}, {Ochner}, {Cunial}, \&
  {Nascimbeni}}]{Libralato2016MNRAS.463.1780L}
{Libralato}, M., {Nardiello}, D., {Bedin}, L.~R., {et~al.} 2016{\natexlab{b}},
  \mnras, 463, 1780

\bibitem[{Liggins {et~al.}(2021{\natexlab{a}})Liggins, Pickering, Nave,
  Kramida, Gamrath, \& Quinet}]{Liggins2021b}
Liggins, F.~S., Pickering, J.~C., Nave, G., {et~al.} 2021{\natexlab{a}}, The
  Astrophysical Journal, 907, 69

\bibitem[{Liggins {et~al.}(2021{\natexlab{b}})Liggins, Pickering, Nave, Ward,
  \& Tchang-Brillet}]{Liggins2021a}
Liggins, F.~S., Pickering, J.~C., Nave, G., Ward, J.~W., \& Tchang-Brillet,
  W.~L. 2021{\natexlab{b}}, The Astrophysical Journal Supplement Series, 252,
  10

\bibitem[{{Lillo-Box} {et~al.}(2014){Lillo-Box}, {Barrado}, {Moya},
  {Montesinos}, {Montalb{\'a}n}, {Bayo}, {Barbieri}, {R{\'e}gulo}, {Mancini},
  {Bouy}, \& {Henning}}]{LilloBox2014}
{Lillo-Box}, J., {Barrado}, D., {Moya}, A., {et~al.} 2014, \aap, 562, A109

\bibitem[{{Limongi} \& {Chieffi}(2003)}]{limongi2003}
{Limongi}, M. \& {Chieffi}, A. 2003, \apj, 592, 404

\bibitem[{{Lind} \& {Amarsi}(2024)}]{LindAmarsi2024}
{Lind}, K. \& {Amarsi}, A.~M. 2024, \araa, 62, 475

\bibitem[{{Lind} {et~al.}(2009){Lind}, {Asplund}, \& {Barklem}}]{lind2009}
{Lind}, K., {Asplund}, M., \& {Barklem}, P.~S. 2009, \aap, 503, 541

\bibitem[{{Lind} {et~al.}(2022){Lind}, {Nordlander}, {Wehrhahn}, {Montelius},
  {Osorio}, {Barklem}, {Af{\c{s}}ar}, {Sneden}, \& {Kobayashi}}]{lind22}
{Lind}, K., {Nordlander}, T., {Wehrhahn}, A., {et~al.} 2022, \aap, 665, A33

\bibitem[{{Liu} {et~al.}(2018){Liu}, {Yong}, {Asplund}, {Feltzing}, {Mustill},
  {Mel{\'e}ndez}, {Ram{\'\i}rez}, \& {Lin}}]{liu2018}
{Liu}, F., {Yong}, D., {Asplund}, M., {et~al.} 2018, \aap, 614, A138

\bibitem[{{Liu} {et~al.}(2024){Liu}, {Alexander}, {Meyer}, {Nittler}, {Wang},
  \& {Stroud}}]{Liu2024}
{Liu}, N., {Alexander}, C. M.~O., {Meyer}, B.~S., {et~al.} 2024, \apjl, 961,
  L22

\bibitem[{{Lodders}(2003)}]{lodders03LSb}
{Lodders}, K. 2003, \apj, 591, 1220

\bibitem[{{Lodders}(2010)}]{Lodders2010}
{Lodders}, K. 2010, in Astrophysics and Space Science Proceedings, Vol.~16,
  Principles and Perspectives in Cosmochemistry, ed. A.~{Goswami} \& B.~E.
  {Reddy}, 379

\bibitem[{{Lovis} \& {Mayor}(2007)}]{lovis2007}
{Lovis}, C. \& {Mayor}, M. 2007, \aap, 472, 657

\bibitem[{{Lucertini} {et~al.}(2022){Lucertini}, {Monaco}, {Caffau},
  {Bonifacio}, \& {Mucciarelli}}]{lucertini2022}
{Lucertini}, F., {Monaco}, L., {Caffau}, E., {Bonifacio}, P., \& {Mucciarelli},
  A. 2022, \aap, 657, A29

\bibitem[{{Ludwig} {et~al.}(2009){Ludwig}, {Behara}, {Steffen}, \&
  {Bonifacio}}]{ludwig2009}
{Ludwig}, H.~G., {Behara}, N.~T., {Steffen}, M., \& {Bonifacio}, P. 2009, \aap,
  502, L1

\bibitem[{{Ludwig} {et~al.}(2010){Ludwig}, {Caffau}, {Steffen}, {Bonifacio}, \&
  {Sbordone}}]{Ludwig2010A&A...509A..84L}
{Ludwig}, H.-G., {Caffau}, E., {Steffen}, M., {Bonifacio}, P., \& {Sbordone},
  L. 2010, \aap, 509, A84

\bibitem[{{Lugaro} {et~al.}(2023){Lugaro}, {Pignatari}, {Reifarth}, \&
  {Wiescher}}]{Lugaro2023}
{Lugaro}, M., {Pignatari}, M., {Reifarth}, R., \& {Wiescher}, M. 2023, Annual
  Review of Nuclear and Particle Science, 73, 315

\bibitem[{Lundberg {et~al.}(2016)Lundberg, Hartman, EngstrÃ¶m, Nilsson,
  Persson, Palmeri, Quinet, Fivet, Malcheva, \& Blagoev}]{Lundberg2016}
Lundberg, H., Hartman, H., EngstrÃ¶m, L., {et~al.} 2016, Monthly Notices of
  the Royal Astronomical Society, 460, 356

\bibitem[{{Lundqvist, M.} {et~al.}(2007){Lundqvist, M.}, {Wahlgren, G. M.}, \&
  {Hill, V.}}]{Lundqvist2007}
{Lundqvist, M.}, {Wahlgren, G. M.}, \& {Hill, V.} 2007, A\&A, 463, 693

\bibitem[{{Maehara} {et~al.}(2021){Maehara}, {Notsu}, {Namekata}, {Honda},
  {Kowalski}, {Katoh}, {Ohshima}, {Iida}, {Oeda}, {Murata}, {Yamanaka},
  {Takagi}, {Sasada}, {Akitaya}, {Ikuta}, {Okamoto}, {Nogami}, \&
  {Shibata}}]{maehara2021}
{Maehara}, H., {Notsu}, Y., {Namekata}, K., {et~al.} 2021, \pasj, 73, 44

\bibitem[{{Magic} {et~al.}(2013){Magic}, {Collet}, {Asplund}, {Trampedach},
  {Hayek}, {Chiavassa}, {Stein}, \& {Nordlund}}]{magic2013a}
{Magic}, Z., {Collet}, R., {Asplund}, M., {et~al.} 2013, \aap, 557, A26

\bibitem[{{Magrini} {et~al.}(2023){Magrini}, {Bensby}, {Brucalassi}, {Randich},
  {Jeffries}, {de Silva}, {Skuladottir}, {Smiljanic}, {Gonzalez}, {Hill},
  {Lagarde}, {Tolstoy}, {Arroyo-Polonio}, {Baratella}, {Barnes}, {Battaglia},
  {Baumgardt}, {Bellazzini}, {Biazzo}, {Bragaglia}, {Carter}, {Casali},
  {Cescutti}, {Danielski}, {Delgado Mena}, {Drazdauskas}, {Gieles},
  {Giribaldi}, {Hawkins}, {Hoeijmakers}, {Jablonka}, {Kamath}, {Louth},
  {Fabiola Marino}, {Martell}, {Merle}, {Montet}, {Murphy}, {Nisini},
  {Nordlander}, {D'Orazi}, {Pino}, {Romano}, {Sacco}, {Sandford}, {Sollima},
  {Spina}, {Tautvaisiene}, {Ting}, {Tozzi}, {Van der Swaelmen}, {Van Eck},
  {Watson}, {Worley}, \& {Zocchi}}]{magrini2023}
{Magrini}, L., {Bensby}, T., {Brucalassi}, A., {et~al.} 2023, arXiv e-prints,
  arXiv:2312.08270

\bibitem[{{Magrini} {et~al.}(2022){Magrini}, {Danielski}, {Bossini}, {Rainer},
  {Turrini}, {Benatti}, {Brucalassi}, {Tsantaki}, {Delgado Mena}, {Sanna},
  {Biazzo}, {Campante}, {Van der Swaelmen}, {Sousa}, {He{\l}miniak}, {Neitzel},
  {Adibekyan}, {Bruno}, \& {Casali}}]{Magrini2022}
{Magrini}, L., {Danielski}, C., {Bossini}, D., {et~al.} 2022, \aap, 663, A161

\bibitem[{{Magrini} {et~al.}(2021){Magrini}, {Vescovi}, {Casali}, {Cristallo},
  {Viscasillas V{\'a}zquez}, {Cescutti}, {Spina}, {Van Der Swaelmen}, \&
  {Randich}}]{Magrini2021ymg}
{Magrini}, L., {Vescovi}, D., {Casali}, G., {et~al.} 2021, \aap, 646, L2

\bibitem[{{Majewski} {et~al.}(2017){Majewski}, {Schiavon}, {Frinchaboy},
  {Allende Prieto}, {Barkhouser}, {Bizyaev}, {Blank}, {Brunner}, {Burton},
  {Carrera}, {Chojnowski}, {Cunha}, {Epstein}, {Fitzgerald}, {Garc{\'\i}a
  P{\'e}rez}, {Hearty}, {Henderson}, {Holtzman}, {Johnson}, {Lam}, {Lawler},
  {Maseman}, {M{\'e}sz{\'a}ros}, {Nelson}, {Nguyen}, {Nidever}, {Pinsonneault},
  {Shetrone}, {Smee}, {Smith}, {Stolberg}, {Skrutskie}, {Walker}, {Wilson},
  {Zasowski}, {Anders}, {Basu}, {Beland}, {Blanton}, {Bovy}, {Brownstein},
  {Carlberg}, {Chaplin}, {Chiappini}, {Eisenstein}, {Elsworth}, {Feuillet},
  {Fleming}, {Galbraith-Frew}, {Garc{\'\i}a}, {Garc{\'\i}a-Hern{\'a}ndez},
  {Gillespie}, {Girardi}, {Gunn}, {Hasselquist}, {Hayden}, {Hekker}, {Ivans},
  {Kinemuchi}, {Klaene}, {Mahadevan}, {Mathur}, {Mosser}, {Muna}, {Munn},
  {Nichol}, {O'Connell}, {Parejko}, {Robin}, {Rocha-Pinto}, {Schultheis},
  {Serenelli}, {Shane}, {Silva Aguirre}, {Sobeck}, {Thompson}, {Troup},
  {Weinberg}, \& {Zamora}}]{majewski2017}
{Majewski}, S.~R., {Schiavon}, R.~P., {Frinchaboy}, P.~M., {et~al.} 2017, \aj,
  154, 94

\bibitem[{{Malavolta} {et~al.}(2018){Malavolta}, {Mayo}, {Louden}, {Rajpaul},
  {Bonomo}, {Buchhave}, {Kreidberg}, {Kristiansen}, {Lopez-Morales}, {Mortier},
  {Vanderburg}, {Coffinet}, {Ehrenreich}, {Lovis}, {Bouchy}, {Charbonneau},
  {Ciardi}, {Collier Cameron}, {Cosentino}, {Crossfield}, {Damasso},
  {Dressing}, {Dumusque}, {Everett}, {Figueira}, {Fiorenzano}, {Gonzales},
  {Haywood}, {Harutyunyan}, {Hirsch}, {Howell}, {Johnson}, {Latham}, {Lopez},
  {Mayor}, {Micela}, {Molinari}, {Nascimbeni}, {Pepe}, {Phillips}, {Piotto},
  {Rice}, {Sasselov}, {S{\'e}gransan}, {Sozzetti}, {Udry}, \&
  {Watson}}]{Malavolta2018}
{Malavolta}, L., {Mayo}, A.~W., {Louden}, T., {et~al.} 2018, \aj, 155, 107

\bibitem[{{Malavolta} {et~al.}(2016){Malavolta}, {Nascimbeni}, {Piotto},
  {Quinn}, {Borsato}, {Granata}, {Bonomo}, {Marzari}, {Bedin}, {Rainer},
  {Desidera}, {Lanza}, {Poretti}, {Sozzetti}, {White}, {Latham}, {Cunial},
  {Libralato}, {Nardiello}, {Boccato}, {Claudi}, {Cosentino}, {Covino},
  {Gratton}, {Maggio}, {Micela}, {Molinari}, {Pagano}, {Smareglia}, {Affer},
  {Andreuzzi}, {Aparicio}, {Benatti}, {Bignamini}, {Borsa}, {Damasso}, {Di
  Fabrizio}, {Harutyunyan}, {Esposito}, {Fiorenzano}, {Gandolfi}, {Giacobbe},
  {Gonz{\'a}lez Hern{\'a}ndez}, {Maldonado}, {Masiero}, {Molinaro}, {Pedani},
  \& {Scandariato}}]{malavolta2016}
{Malavolta}, L., {Nascimbeni}, V., {Piotto}, G., {et~al.} 2016, \aap, 588, A118

\bibitem[{{Mallinson} {et~al.}(2022){Mallinson}, {Lind}, {Amarsi}, {Barklem},
  {Grumer}, {Belyaev}, \& {Youakim}}]{mallinson2022}
{Mallinson}, J.~W.~E., {Lind}, K., {Amarsi}, A.~M., {et~al.} 2022, \aap, 668,
  A103

\bibitem[{{Mallinson} {et~al.}(2024){Mallinson}, {Lind}, {Amarsi}, \&
  {Youakim}}]{mallinson2024}
{Mallinson}, J.~W.~E., {Lind}, K., {Amarsi}, A.~M., \& {Youakim}, K. 2024,
  \aap, 687, A5

\bibitem[{{Manara} {et~al.}(2020){Manara}, {Natta}, {Rosotti}, {Alcal{\'a}},
  {Nisini}, {Lodato}, {Testi}, {Pascucci}, {Hillenbrand}, {Carpenter},
  {Scholz}, {Fedele}, {Frasca}, {Mulders}, {Rigliaco}, {Scardoni}, \&
  {Zari}}]{manara2020}
{Manara}, C.~F., {Natta}, A., {Rosotti}, G.~P., {et~al.} 2020, \aap, 639, A58

\bibitem[{{Mann} {et~al.}(2017){Mann}, {Gaidos}, {Vanderburg}, {Rizzuto},
  {Ansdell}, {Medina}, {Mace}, {Kraus}, \& {Sokal}}]{mann2017}
{Mann}, A.~W., {Gaidos}, E., {Vanderburg}, A., {et~al.} 2017, \aj, 153, 64

\bibitem[{{Mann} {et~al.}(2018){Mann}, {Vanderburg}, {Rizzuto}, {Kraus},
  {Berlind}, {Bieryla}, {Calkins}, {Esquerdo}, {Latham}, {Mace}, {Morris},
  {Quinn}, {Sokal}, \& {Stefanik}}]{Mann2018}
{Mann}, A.~W., {Vanderburg}, A., {Rizzuto}, A.~C., {et~al.} 2018, \aj, 155, 4

\bibitem[{Manrique {et~al.}(2025)Manrique, Pace, Aguilera, \&
  AragÃ³n}]{Manrique2025}
Manrique, J., Pace, D. M.~D., Aguilera, J.~A., \& AragÃ³n, C. 2025,
  Spectrochimica Acta - Part B Atomic Spectroscopy, 234

\bibitem[{{Mantovan} {et~al.}(2024){Mantovan}, {Malavolta}, {Desidera},
  {Zingales}, {Borsato}, {Piotto}, {Maggio}, {Locci}, {Polychroni}, {Turrini},
  {Baratella}, {Biazzo}, {Nardiello}, {Stassun}, {Nascimbeni}, {Benatti},
  {John}, {Watkins}, {Bieryla}, {Lissauer}, {Twicken}, {Lanza}, {Winn},
  {Messina}, {Montalto}, {Sozzetti}, {Boffin}, {Cheryasov}, {Strakhov},
  {Murgas}, {D'Arpa}, {Barkaoui}, {Benni}, {Bignamini}, {Bonomo}, {Borsa},
  {Cabona}, {Cameron}, {Claudi}, {Cochran}, {Collins}, {Damasso}, {Dong},
  {Endl}, {Fukui}, {F{\H{u}}r{\'e}sz}, {Gandolfi}, {Ghedina}, {Jenkins},
  {Kab{\'a}th}, {Latham}, {Lorenzi}, {Luque}, {Maldonado}, {McLeod},
  {Molinaro}, {Narita}, {Nowak}, {Orell-Miquel}, {Pall{\'e}}, {Parviainen},
  {Pedani}, {Quinn}, {Relles}, {Rowden}, {Scandariato}, {Schwarz}, {Seager},
  {Shporer}, {Vanderburg}, \& {Wilson}}]{Mantovan2024}
{Mantovan}, G., {Malavolta}, L., {Desidera}, S., {et~al.} 2024, \aap, 682, A129

\bibitem[{{Marconi} {et~al.}(2022{\natexlab{a}}){Marconi}, {Abreu},
  {Adibekyan}, {Alberti}, {Albrecht}, {Alcaniz}, {Aliverti}, {Allende Prieto},
  {Alvarado G{\'o}mez}, {Amado}, {Amate}, {Andersen}, {Artigau}, {Baker},
  {Baldini}, {Balestra}, {Barnes}, {Baron}, {Barros}, {Bauer}, {Beaulieu},
  {Bellido-Tirado}, {Benneke}, {Bensby}, {Bergin}, {Biazzo}, {Bik}, {Birkby},
  {Blind}, {Boisse}, {Bolmont}, {Bonaglia}, {Bonfils}, {Borsa}, {Brandeker},
  {Brandner}, {Broeg}, {Brogi}, {Brousseau}, {Brucalassi}, {Brynnel},
  {Buchhave}, {Buscher}, {Cabral}, {Calderone}, {Calvo-Ortega}, {Canto
  Martins}, {Cantalloube}, {Carbonaro}, {Chauvin}, {Chazelas}, {Cheffot},
  {Cheng}, {Chiavassa}, {Christensen}, {Cirami}, {Cook}, {Cooke}, {Coretti},
  {Covino}, {Cowan}, {Cresci}, {Cristiani}, {Cunha Parro}, {Cupani},
  {D'Odorico}, {de Castro Le{\~a}o}, {De Cia}, {De Medeiros}, {Debras},
  {Debus}, {Demangeon}, {Dessauges-Zavadsky}, {Di Marcantonio}, {Dionies},
  {Doyon}, {Dunn}, {Ehrenreich}, {Faria}, {Feruglio}, {Fisher}, {Fontana},
  {Fumagalli}, {Fusco}, {Fynbo}, {Gabella}, {Gaessler}, {Gallo}, {Gao},
  {Genolet}, {Genoni}, {Giacobbe}, {Giro}, {Gon{\c{c}}alves}, {Gonzalez},
  {Gonz{\'a}lez Hern{\'a}ndez}, {Gracia T{\'e}mich}, {Haehnelt}, {Haniff},
  {Hatzes}, {Helled}, {Hoeijmakers}, {Huke}, {J{\"a}rvinen}, {J{\"a}rvinen},
  {Kaminski}, {Korn}, {Kouach}, {Kowzan}, {Kreidberg}, {Landoni}, {Lanotte},
  {Lavail}, {Li}, {Liske}, {Lovis}, {Lucatello}, {Lunney}, {MacIntosh},
  {Madhusudhan}, {Magrini}, {Maiolino}, {Malo}, {Man}, {Marquart}, {Marques},
  {Martins}, {Martins}, {Maslowski}, {Mason}, {Mason}, {McCracken}, {Mergo},
  {Micela}, {Mitchell}, {Molli{\`e}re}, {Monteiro}, {Montgomery}, {Mordasini},
  {Morin}, {Mucciarelli}, {Murphy}, {N'Diaye}, {Neichel}, {Niedzielski},
  {Niemczura}, {Nortmann}, {Noterdaeme}, {Nunes}, {Oggioni}, {Oliva},
  {{\"O}nel}, {Origlia}, {{\"O}stlin}, {Palle}, {Papaderos}, {Pariani},
  {Pe{\~n}ate Castro}, {Pepe}, {Perreault Levasseur}, {Petit}, {Pino},
  {Piqueras}, {Pollo}, {Poppenhaeger}, {Quirrenbach}, {Rauscher}, {Rebolo},
  {Redaelli}, {Reffert}, {Reid}, {Reiners}, {Richter}, {Riva}, {Rivoire},
  {Rodr{\'\i}guez-L{\'o}pez}, {Roederer}, {Romano}, {Rousseau}, {Rowe},
  {Salvadori}, {Santos}, {Santos Diaz}, {Sanz-Forcada}, {Sarajlic}, {Sauvage},
  {Sch{\"a}fer}, {Schiavon}, {Schmidt}, {Selmi}, {Sivanandam}, {Sordet},
  {Sordo}, {Sortino}, {Sosnowska}, {Sousa}, {Stempels}, {Strassmeier},
  {Su{\'a}rez Mascare{\~n}o}, {Sulich}, {Sun}, {Tanvir}, {Tenegi-Sangin{\'e}s},
  {Thibault}, {Thompson}, {Tozzi}, {Turbet}, {Vall{\'e}e}, {Varas}, {Venn},
  {V{\'e}ran}, {Verma}, {Viel}, {Wade}, {Waring}, {Weber}, {Weder}, {Wehbe},
  {Weingrill}, {Woche}, {Xompero}, {Zackrisson}, {Zanutta}, {Zapatero Osorio},
  {Zechmeister}, \& {Zimara}}]{ANDES22}
{Marconi}, A., {Abreu}, M., {Adibekyan}, V., {et~al.} 2022{\natexlab{a}}, in
  Society of Photo-Optical Instrumentation Engineers (SPIE) Conference Series,
  Vol. 12184, Ground-based and Airborne Instrumentation for Astronomy IX, ed.
  C.~J. {Evans}, J.~J. {Bryant}, \& K.~{Motohara}, 1218424

\bibitem[{{Marconi} {et~al.}(2022{\natexlab{b}}){Marconi}, {Abreu},
  {Adibekyan}, {Alberti}, {Albrecht}, {Alcaniz}, {Aliverti}, {Allende Prieto},
  {Alvarado G{\'o}mez}, {Amado}, {Amate}, {Andersen}, {Artigau}, {Baker},
  {Baldini}, {Balestra}, {Barnes}, {Baron}, {Barros}, {Bauer}, {Beaulieu},
  {Bellido-Tirado}, {Benneke}, {Bensby}, {Bergin}, {Biazzo}, {Bik}, {Birkby},
  {Blind}, {Boisse}, {Bolmont}, {Bonaglia}, {Bonfils}, {Borsa}, {Brandeker},
  {Brandner}, {Broeg}, {Brogi}, {Brousseau}, {Brucalassi}, {Brynnel},
  {Buchhave}, {Buscher}, {Cabral}, {Calderone}, {Calvo-Ortega}, {Canto
  Martins}, {Cantalloube}, {Carbonaro}, {Chauvin}, {Chazelas}, {Cheffot},
  {Cheng}, {Chiavassa}, {Christensen}, {Cirami}, {Cook}, {Cooke}, {Coretti},
  {Covino}, {Cowan}, {Cresci}, {Cristiani}, {Cunha Parro}, {Cupani},
  {D'Odorico}, {de Castro Le{\~a}o}, {De Cia}, {De Medeiros}, {Debras},
  {Debus}, {Demangeon}, {Dessauges-Zavadsky}, {Di Marcantonio}, {Dionies},
  {Doyon}, {Dunn}, {Ehrenreich}, {Faria}, {Feruglio}, {Fisher}, {Fontana},
  {Fumagalli}, {Fusco}, {Fynbo}, {Gabella}, {Gaessler}, {Gallo}, {Gao},
  {Genolet}, {Genoni}, {Giacobbe}, {Giro}, {Gon{\c{c}}alves}, {Gonzalez},
  {Gonz{\'a}lez Hern{\'a}ndez}, {Gracia T{\'e}mich}, {Haehnelt}, {Haniff},
  {Hatzes}, {Helled}, {Hoeijmakers}, {Huke}, {J{\"a}rvinen}, {J{\"a}rvinen},
  {Kaminski}, {Korn}, {Kouach}, {Kowzan}, {Kreidberg}, {Landoni}, {Lanotte},
  {Lavail}, {Li}, {Liske}, {Lovis}, {Lucatello}, {Lunney}, {MacIntosh},
  {Madhusudhan}, {Magrini}, {Maiolino}, {Malo}, {Man}, {Marquart}, {Marques},
  {Martins}, {Martins}, {Maslowski}, {Mason}, {Mason}, {McCracken}, {Mergo},
  {Micela}, {Mitchell}, {Molli{\`e}re}, {Monteiro}, {Montgomery}, {Mordasini},
  {Morin}, {Mucciarelli}, {Murphy}, {N'Diaye}, {Neichel}, {Niedzielski},
  {Niemczura}, {Nortmann}, {Noterdaeme}, {Nunes}, {Oggioni}, {Oliva},
  {{\"O}nel}, {Origlia}, {{\"O}stlin}, {Palle}, {Papaderos}, {Pariani},
  {Pe{\~n}ate Castro}, {Pepe}, {Perreault Levasseur}, {Petit}, {Pino},
  {Piqueras}, {Pollo}, {Poppenhaeger}, {Quirrenbach}, {Rauscher}, {Rebolo},
  {Redaelli}, {Reffert}, {Reid}, {Reiners}, {Richter}, {Riva}, {Rivoire},
  {Rodr{\'\i}guez-L{\'o}pez}, {Roederer}, {Romano}, {Rousseau}, {Rowe},
  {Salvadori}, {Santos}, {Santos Diaz}, {Sanz-Forcada}, {Sarajlic}, {Sauvage},
  {Sch{\"a}fer}, {Schiavon}, {Schmidt}, {Selmi}, {Sivanandam}, {Sordet},
  {Sordo}, {Sortino}, {Sosnowska}, {Sousa}, {Stempels}, {Strassmeier},
  {Su{\'a}rez Mascare{\~n}o}, \& {Sulich}}]{Marconi2022SPIE12184E..24M}
{Marconi}, A., {Abreu}, M., {Adibekyan}, V., {et~al.} 2022{\natexlab{b}}, in
  Society of Photo-Optical Instrumentation Engineers (SPIE) Conference Series,
  Vol. 12184, Ground-based and Airborne Instrumentation for Astronomy IX, ed.
  C.~J. {Evans}, J.~J. {Bryant}, \& K.~{Motohara}, 1218424

\bibitem[{{Marino} {et~al.}(2015){Marino}, {Milone}, {Karakas}, {Casagrande},
  {Yong}, {Shingles}, {Da Costa}, {Norris}, {Stetson}, {Lind}, {Asplund},
  {Collet}, {Jerjen}, {Sbordone}, {Aparicio}, \& {Cassisi}}]{Marino2015}
{Marino}, A.~F., {Milone}, A.~P., {Karakas}, A.~I., {et~al.} 2015, \mnras, 450,
  815

\bibitem[{{Marino} {et~al.}(2009){Marino}, {Milone}, {Piotto}, {Villanova},
  {Bedin}, {Bellini}, \& {Renzini}}]{Marino2009}
{Marino}, A.~F., {Milone}, A.~P., {Piotto}, G., {et~al.} 2009, \aap, 505, 1099

\bibitem[{{Marino} {et~al.}(2019{\natexlab{a}}){Marino}, {Milone}, {Renzini},
  {D'Antona}, {Anderson}, {Bedin}, {Bellini}, {Cordoni}, {Lagioia}, {Piotto},
  \& {Tailo}}]{Marino2019a}
{Marino}, A.~F., {Milone}, A.~P., {Renzini}, A., {et~al.} 2019{\natexlab{a}},
  \mnras, 487, 3815

\bibitem[{{Marino} {et~al.}(2019{\natexlab{b}}){Marino}, {Milone}, {Sills},
  {Yong}, {Renzini}, {Bedin}, {Cordoni}, {D'Antona}, {Jerjen}, {Karakas},
  {Lagioia}, {Piotto}, \& {Tailo}}]{Marino2019b}
{Marino}, A.~F., {Milone}, A.~P., {Sills}, A., {et~al.} 2019{\natexlab{b}},
  \apj, 887, 91

\bibitem[{{Marsh}(2001)}]{marsh2001}
{Marsh}, T.~R. 2001, in Astrotomography, Indirect Imaging Methods in
  Observational Astronomy, ed. H.~M.~J. {Boffin}, D.~{Steeghs}, \&
  J.~{Cuypers}, Vol. 573 (Springer Berlin, Heidelberg), 1

\bibitem[{{Martell} {et~al.}(2021){Martell}, {Simpson}, {Balasubramaniam},
  {Buder}, {Sharma}, {Hon}, {Stello}, {Ting}, {Asplund}, {Bland-Hawthorn}, {De
  Silva}, {Freeman}, {Hayden}, {Kos}, {Lewis}, {Lind}, {Zucker}, {Zwitter},
  {Campbell}, {{\v{C}}otar}, {Horner}, {Montet}, \&
  {Wittenmyer}}]{martell21LSb}
{Martell}, S.~L., {Simpson}, J.~D., {Balasubramaniam}, A.~G., {et~al.} 2021,
  \mnras, 505, 5340

\bibitem[{{Martin}(2018)}]{Martin2018}
{Martin}, D.~V. 2018, in Handbook of Exoplanets, ed. H.~J. {Deeg} \& J.~A.
  {Belmonte} (Springer International Publishing), 156

\bibitem[{{Martin} {et~al.}(2019){Martin}, {Triaud}, {Udry}, {Marmier},
  {Maxted}, {Collier Cameron}, {Hellier}, {Pepe}, {Pollacco}, {S{\'e}gransan},
  \& {West}}]{Martin2019}
{Martin}, D.~V., {Triaud}, A. H.~M.~J., {Udry}, S., {et~al.} 2019, \aap, 624,
  A68

\bibitem[{{Martocchia} {et~al.}(2021){Martocchia}, {Lardo}, {Rejkuba},
  {Kamann}, {Bastian}, {Larsen}, {Cabrera-Ziri}, {Chantereau}, {Dalessandro},
  {Kacharov}, \& {Salaris}}]{Martocchia2021}
{Martocchia}, S., {Lardo}, C., {Rejkuba}, M., {et~al.} 2021, \mnras, 505, 5389

\bibitem[{{Mashonkina} {et~al.}(2014){Mashonkina}, {Christlieb}, \&
  {Eriksson}}]{Mashonkina2014}
{Mashonkina}, L., {Christlieb}, N., \& {Eriksson}, K. 2014, \aap, 569, A43

\bibitem[{{Mashonkina} {et~al.}(2023){Mashonkina}, {Pakhomov}, {Sitnova},
  {Smogorzhevskii}, {Jablonka}, \& {Hill}}]{mashonkina2023}
{Mashonkina}, L., {Pakhomov}, Y., {Sitnova}, T., {et~al.} 2023, \mnras, 524,
  3526

\bibitem[{{Mashonkina} {et~al.}(2012){Mashonkina}, {Ryabtsev}, \&
  {Frebel}}]{Mashonkina2012A&A...540A..98M}
{Mashonkina}, L., {Ryabtsev}, A., \& {Frebel}, A. 2012, \aap, 540, A98

\bibitem[{{Mashonkina} {et~al.}(2022){Mashonkina}, {Sitnova}, \&
  {Korotin}}]{mashonkina22}
{Mashonkina}, L., {Sitnova}, T., \& {Korotin}, S. 2022, Astronomy Letters, 48,
  303

\bibitem[{{Massari} {et~al.}(2019){Massari}, {Koppelman}, \&
  {Helmi}}]{Massari2019}
{Massari}, D., {Koppelman}, H.~H., \& {Helmi}, A. 2019, \aap, 630, L4

\bibitem[{{Masseron} {et~al.}(2014){Masseron}, {Plez}, {Van Eck}, {Colin},
  {Daoutidis}, {Godefroid}, {Coheur}, {Bernath}, {Jorissen}, \&
  {Christlieb}}]{Masseron2014}
{Masseron}, T., {Plez}, B., {Van Eck}, S., {et~al.} 2014, \aap, 571, A47

\bibitem[{{Matsuno} {et~al.}(2024){Matsuno}, {Amarsi}, {Carlos}, \&
  {Nissen}}]{matsuno2024}
{Matsuno}, T., {Amarsi}, A.~M., {Carlos}, M., \& {Nissen}, P.~E. 2024, \aap,
  688, A72

\bibitem[{{Matt} {et~al.}(2012){Matt}, {MacGregor}, {Pinsonneault}, \&
  {Greene}}]{matt2012}
{Matt}, S.~P., {MacGregor}, K.~B., {Pinsonneault}, M.~H., \& {Greene}, T.~P.
  2012, \apjl, 754, L26

\bibitem[{{Matteucci}(2021)}]{Matteucci2021}
{Matteucci}, F. 2021, \aapr, 29, 5

\bibitem[{{Matteucci} {et~al.}(1996){Matteucci}, {Molaro}, \&
  {Vladilo}}]{matteucci1996}
{Matteucci}, F., {Molaro}, P., \& {Vladilo}, G. 1996, arXiv e-prints, astro

\bibitem[{{Matteucci} {et~al.}(2014){Matteucci}, {Romano}, {Arcones},
  {Korobkin}, \& {Rosswog}}]{Matteucci2014}
{Matteucci}, F., {Romano}, D., {Arcones}, A., {Korobkin}, O., \& {Rosswog}, S.
  2014, \mnras, 438, 2177

\bibitem[{{Mayo} {et~al.}(2023){Mayo}, {Dressing}, {Vanderburg}, {Fortenbach},
  {Lienhard}, {Malavolta}, {Mortier}, {N{\'u}{\~n}ez}, {Richey-Yowell},
  {Turtelboom}, {Bonomo}, {Latham}, {L{\'o}pez-Morales}, {Shkolnik},
  {Sozzetti}, {Ag{\"u}eros}, {Borsato}, {Charbonneau}, {Cosentino}, {Douglas},
  {Dumusque}, {Ghedina}, {Gibson}, {Granata}, {Harutyunyan}, {Haywood},
  {Lacedelli}, {Lorenzi}, {Magazz{\`u}}, {Martinez Fiorenzano}, {Micela},
  {Molinari}, {Montalto}, {Nardiello}, {Nascimbeni}, {Pagano}, {Piotto},
  {Pino}, {Poretti}, {Scandariato}, {Udry}, \& {Buchhave}}]{Mayo2023}
{Mayo}, A.~W., {Dressing}, C.~D., {Vanderburg}, A., {et~al.} 2023, \aj, 165,
  235

\bibitem[{{Mayor} {et~al.}(2003){Mayor}, {Pepe}, {Queloz}, {Bouchy},
  {Rupprecht}, {Lo Curto}, {Avila}, {Benz}, {Bertaux}, {Bonfils}, {Dall},
  {Dekker}, {Delabre}, {Eckert}, {Fleury}, {Gilliotte}, {Gojak}, {Guzman},
  {Kohler}, {Lizon}, {Longinotti}, {Lovis}, {Megevand}, {Pasquini}, {Reyes},
  {Sivan}, {Sosnowska}, {Soto}, {Udry}, {van Kesteren}, {Weber}, \&
  {Weilenmann}}]{Mayor2003Msngr.114...20M}
{Mayor}, M., {Pepe}, F., {Queloz}, D., {et~al.} 2003, The Messenger, 114, 20

\bibitem[{{Mayor} \& {Queloz}(1995)}]{mayor95}
{Mayor}, M. \& {Queloz}, D. 1995, \nat, 378, 355

\bibitem[{{McCann} {et~al.}(2025){McCann}, {Ballance}, {McNeill}, {Sim}, \&
  {Ramsbottom}}]{mccann2025}
{McCann}, M., {Ballance}, C.~P., {McNeill}, F., {Sim}, S.~A., \& {Ramsbottom},
  C.~A. 2025, \mnras, 540, 2923

\bibitem[{{McConnachie}(2012)}]{mcconnachie2012}
{McConnachie}, A.~W. 2012, \aj, 144, 4

\bibitem[{{McGinnis} {et~al.}(2020){McGinnis}, {Bouvier}, \&
  {Gallet}}]{mcginnis2020}
{McGinnis}, P., {Bouvier}, J., \& {Gallet}, F. 2020, \mnras, 497, 2142

\bibitem[{{McKenzie} {et~al.}(2024){McKenzie}, {Yong}, {Karakas}, {Wang},
  {Monty}, {Marino}, {Milone}, {Nordlander}, {Mura-Guzm{\'a}n}, {Martell}, \&
  {Carlos}}]{McKenzie2024}
{McKenzie}, M., {Yong}, D., {Karakas}, A.~I., {et~al.} 2024, \mnras, 527, 7940

\bibitem[{{M{\'e}rand} {et~al.}(2021){M{\'e}rand}, {Andreani}, {Cirasuolo},
  {Comer{\'o}n}, {De Gregorio Monsalvo}, {Dessauges-Zavadsky}, {Emsellem},
  {Ivison}, {Kemper}, {Kerschbaum}, {Leibundgut}, {Liske}, {McLure},
  {Mroczkowski}, {Origlia}, {Philips}, \& {Sana}}]{merand21}
{M{\'e}rand}, A., {Andreani}, P., {Cirasuolo}, M., {et~al.} 2021, The
  Messenger, 184, 8

\bibitem[{{Merrill}(1952)}]{Merrill1952}
{Merrill}, P.~W. 1952, \apj, 116, 21

\bibitem[{{M{\'e}sz{\'a}ros} {et~al.}(2015){M{\'e}sz{\'a}ros}, {Martell},
  {Shetrone}, {Lucatello}, {Troup}, {Bovy}, {Cunha},
  {Garc{\'\i}a-Hern{\'a}ndez}, {Overbeek}, {Allende Prieto}, {Beers},
  {Frinchaboy}, {Garc{\'\i}a P{\'e}rez}, {Hearty}, {Holtzman}, {Majewski},
  {Nidever}, {Schiavon}, {Schneider}, {Sobeck}, {Smith}, {Zamora}, \&
  {Zasowski}}]{Meszaros2015}
{M{\'e}sz{\'a}ros}, S., {Martell}, S.~L., {Shetrone}, M., {et~al.} 2015, \aj,
  149, 153

\bibitem[{{Meyer} \& {Truran}(2000)}]{meyer2000}
{Meyer}, B.~S. \& {Truran}, J.~W. 2000, \physrep, 333, 1

\bibitem[{{Michaud} {et~al.}(2015){Michaud}, {Alecian}, \&
  {Richer}}]{michaud2015}
{Michaud}, G., {Alecian}, G., \& {Richer}, J. 2015, {Atomic Diffusion in Stars}
  (Springer International Publishing)

\bibitem[{{Miglio} {et~al.}(2021){Miglio}, {Girardi}, {Grundahl}, {Mosser},
  {Bastian}, {Bragaglia}, {Brogaard}, {Buldgen}, {Chantereau}, {Chaplin},
  {Chiappini}, {Dupret}, {Eggenberger}, {Gieles}, {Izzard}, {Kawata}, {Karoff},
  {Lagarde}, {Mackereth}, {Magrin}, {Meynet}, {Michel}, {Montalb{\'a}n},
  {Nascimbeni}, {Noels}, {Piotto}, {Ragazzoni}, {Soszy{\'n}ski}, {Tolstoy},
  {Toonen}, {Triaud}, \& {Vincenzo}}]{Miglio2021ExA....51..963M}
{Miglio}, A., {Girardi}, L., {Grundahl}, F., {et~al.} 2021, Experimental
  Astronomy, 51, 963

\bibitem[{{Miglio} {et~al.}(2024){Miglio}, {Mosser}, {Girardi}, {Kawata},
  {Chiappini}, {Samadi}, {Moya}, {Garc{\'\i}a}, {Brogaard}, \&
  {Consortium}}]{Miglio2024tkas.confE..51M}
{Miglio}, A., {Mosser}, B., {Girardi}, L., {et~al.} 2024, in 8th TESS/15th
  Kepler Asteroseismic Science Consortium Workshop, 51

\bibitem[{{Milakovi{\'c}} {et~al.}(2023){Milakovi{\'c}}, {Lee}, {Molaro}, \&
  {Webb}}]{Milakovic2023MmSAI..94b.270M}
{Milakovi{\'c}}, D., {Lee}, C.~C., {Molaro}, P., \& {Webb}, J.~K. 2023, in
  Memorie della Societa Astronomica Italiana, Vol.~94, 270

\bibitem[{{Miller} {et~al.}(2026){Miller}, {Caiazzo}, {Heyl}, {Richer},
  {Hollands}, {Tremblay}, {El-Badry}, {Rodriguez}, \&
  {Vanderbosch}}]{Miller2026}
{Miller}, D.~R., {Caiazzo}, I., {Heyl}, J., {et~al.} 2026, \apj, 996, 69

\bibitem[{{Milone} \& {Marino}(2022)}]{milone2022}
{Milone}, A.~P. \& {Marino}, A.~F. 2022, Universe, 8, 359

\bibitem[{{Milone} {et~al.}(2015){Milone}, {Marino}, {Piotto}, {Renzini},
  {Bedin}, {Anderson}, {Cassisi}, {D'Antona}, {Bellini}, {Jerjen},
  {Pietrinferni}, \& {Ventura}}]{Milone2015}
{Milone}, A.~P., {Marino}, A.~F., {Piotto}, G., {et~al.} 2015, \apj, 808, 51

\bibitem[{{Milone} {et~al.}(2012){Milone}, {Piotto}, {Bedin}, {Marino},
  {Momany}, \& {Villanova}}]{Milone2012}
{Milone}, A.~P., {Piotto}, G., {Bedin}, L.~R., {et~al.} 2012, Memorie della
  Societa Astronomica Italiana Supplementi, 19, 173

\bibitem[{{Milone} {et~al.}(2017){Milone}, {Piotto}, {Renzini}, {Marino},
  {Bedin}, {Vesperini}, {D'Antona}, {Nardiello}, {Anderson}, {King}, {Yong},
  {Bellini}, {Aparicio}, {Barbuy}, {Brown}, {Cassisi}, {Ortolani}, {Salaris},
  {Sarajedini}, \& {van der Marel}}]{milone2017a}
{Milone}, A.~P., {Piotto}, G., {Renzini}, A., {et~al.} 2017, \mnras, 464, 3636

\bibitem[{{Minor}(2013)}]{minor2013}
{Minor}, Q.~E. 2013, \apj, 779, 116

\bibitem[{{Mishenina} {et~al.}(2016){Mishenina}, {Kovtyukh}, {Soubiran}, \&
  {Adibekyan}}]{mishenina2016}
{Mishenina}, T., {Kovtyukh}, V., {Soubiran}, C., \& {Adibekyan}, V.~Z. 2016,
  \mnras, 462, 1563

\bibitem[{{Mishenina} {et~al.}(2022){Mishenina}, {Pignatari}, {Gorbaneva},
  {C{\^o}t{\'e}}, {Yag{\"u}e L{\'o}pez}, {Thielemann}, \&
  {Soubiran}}]{Mishenina2022a}
{Mishenina}, T., {Pignatari}, M., {Gorbaneva}, T., {et~al.} 2022, \mnras, 516,
  3786

\bibitem[{{Mochejska} {et~al.}(2006){Mochejska}, {Stanek}, {Sasselov},
  {Szentgyorgyi}, {Adams}, {Cooper}, {Foster}, {Hartman}, {Hickox}, {Lai},
  {Westover}, \& {Winn}}]{Mochejska2006AJ....131.1090M}
{Mochejska}, B.~J., {Stanek}, K.~Z., {Sasselov}, D.~D., {et~al.} 2006, \aj,
  131, 1090

\bibitem[{{Mohorian} {et~al.}(2025){Mohorian}, {Kamath}, {Menon}, {Van
  Winckel}, {Jian}, {Amarsi}, \& {Andrych}}]{mohorian25}
{Mohorian}, M., {Kamath}, D., {Menon}, M., {et~al.} 2025, \pasa, 42, e151

\bibitem[{{Molaro} {et~al.}(2023{\natexlab{a}}){Molaro}, {Aguado}, {Caffau},
  {Allende Prieto}, {Bonifacio}, {Gonz{\'a}lez Hern{\'a}ndez}, {Rebolo},
  {Zapatero Osorio}, {Cristiani}, {Pepe}, {Santos}, {Alibert}, {Cupani}, {Di
  Marcantonio}, {D'Odorico}, {Lovis}, {Martins}, {Milakovi{\'c}}, {Murphy},
  {Nunes}, {Schmidt}, {Sousa}, {Sozzetti}, \& {Su{\'a}rez
  Mascare{\~n}o}}]{Molaro2023}
{Molaro}, P., {Aguado}, D.~S., {Caffau}, E., {et~al.} 2023{\natexlab{a}}, \aap,
  679, A72

\bibitem[{{Molaro} {et~al.}(2023{\natexlab{b}}){Molaro}, {Izzo}, {Selvelli},
  {Bonifacio}, {Aydi}, {Cescutti}, {Guido}, {Harvey}, {Hernanz}, \& {Della
  Valle}}]{molaro23LSb}
{Molaro}, P., {Izzo}, L., {Selvelli}, P., {et~al.} 2023{\natexlab{b}}, \mnras,
  518, 2614

\bibitem[{{Molero} {et~al.}(2023){Molero}, {Magrini}, {Matteucci}, {Romano},
  {Palla}, {Cescutti}, {Viscasillas V{\'a}zquez}, \& {Spitoni}}]{Molero2023}
{Molero}, M., {Magrini}, L., {Matteucci}, F., {et~al.} 2023, \mnras, 523, 2974

\bibitem[{{Monaco} {et~al.}(2014){Monaco}, {Boffin}, {Bonifacio}, {Villanova},
  {Carraro}, {Caffau}, {Steffen}, {Ahumada}, {Beletsky}, \&
  {Beccari}}]{monaco14LSb}
{Monaco}, L., {Boffin}, H.~M.~J., {Bonifacio}, P., {et~al.} 2014, \aap, 564, L6

\bibitem[{{Montalto} {et~al.}(2007){Montalto}, {Piotto}, {Desidera}, {de
  Marchi}, {Bruntt}, {Stetson}, {Arellano Ferro}, {Momany}, {Gratton},
  {Poretti}, {Aparicio}, {Barbieri}, {Claudi}, {Grundahl}, \&
  {Rosenberg}}]{Montalto2007A&A...470.1137M}
{Montalto}, M., {Piotto}, G., {Desidera}, S., {et~al.} 2007, \aap, 470, 1137

\bibitem[{{Monty} {et~al.}(2024){Monty}, {Belokurov}, {Sanders}, {Hansen},
  {Sakari}, {McKenzie}, {Myeong}, {Davies}, {Ardern-Arentsen}, \&
  {Massari}}]{Monty2024}
{Monty}, S., {Belokurov}, V., {Sanders}, J.~L., {et~al.} 2024, \mnras, 533,
  2420

\bibitem[{{Morbidelli} {et~al.}(2012){Morbidelli}, {Lunine}, {O'Brien},
  {Raymond}, \& {Walsh}}]{morbidelli2012}
{Morbidelli}, A., {Lunine}, J.~I., {O'Brien}, D.~P., {Raymond}, S.~N., \&
  {Walsh}, K.~J. 2012, Annual Review of Earth and Planetary Sciences, 40, 251

\bibitem[{{Morel} \& {Miglio}(2012)}]{Morel2012}
{Morel}, T. \& {Miglio}, A. 2012, \mnras, 419, L34

\bibitem[{{Morel} {et~al.}(2014){Morel}, {Miglio}, {Lagarde}, {Montalb{\'a}n},
  {Rainer}, {Poretti}, {Eggenberger}, {Hekker}, {Kallinger}, {Mosser},
  {Valentini}, {Carrier}, {Hareter}, \& {Mantegazza}}]{Morel2014}
{Morel}, T., {Miglio}, A., {Lagarde}, N., {et~al.} 2014, \aap, 564, A119

\bibitem[{{Moresco} {et~al.}(2022){Moresco}, {Amati}, {Amendola}, {Birrer},
  {Blakeslee}, {Cantiello}, {Cimatti}, {Darling}, {Della Valle}, {Fishbach},
  {Grillo}, {Hamaus}, {Holz}, {Izzo}, {Jimenez}, {Lusso}, {Meneghetti},
  {Piedipalumbo}, {Pisani}, {Pourtsidou}, {Pozzetti}, {Quartin}, {Risaliti},
  {Rosati}, \& {Verde}}]{Moresco2022LRR....25....6M}
{Moresco}, M., {Amati}, L., {Amendola}, L., {et~al.} 2022, Living Reviews in
  Relativity, 25, 6

\bibitem[{{Mori} {et~al.}(2025){Mori}, {Di Matteo}, {Salvadori}, {Mondelin},
  {Khoperskov}, {Haywood}, \& {Mastrobuono-Battisti}}]{mori2026}
{Mori}, A., {Di Matteo}, P., {Salvadori}, S., {et~al.} 2025, arXiv e-prints,
  arXiv:2509.13408

\bibitem[{{M{\"o}sta} {et~al.}(2018){M{\"o}sta}, {Roberts}, {Halevi}, {Ott},
  {Lippuner}, {Haas}, \& {Schnetter}}]{Mosta2018}
{M{\"o}sta}, P., {Roberts}, L.~F., {Halevi}, G., {et~al.} 2018, \apj, 864, 171

\bibitem[{{Mucciarelli} {et~al.}(2012{\natexlab{a}}){Mucciarelli},
  {Bellazzini}, {Ibata}, {Merle}, {Chapman}, {Dalessandro}, \&
  {Sollima}}]{Mucciarelli2012}
{Mucciarelli}, A., {Bellazzini}, M., {Ibata}, R., {et~al.} 2012{\natexlab{a}},
  \mnras, 426, 2889

\bibitem[{{Mucciarelli} {et~al.}(2012{\natexlab{b}}){Mucciarelli}, {Salaris},
  \& {Bonifacio}}]{mucciarelli12LSb}
{Mucciarelli}, A., {Salaris}, M., \& {Bonifacio}, P. 2012{\natexlab{b}},
  \mnras, 419, 2195

\bibitem[{Mulholland {et~al.}(2024)Mulholland, McElroy, McNeill, Sim, Ballance,
  \& Ramsbottom}]{Mulholland2024}
Mulholland, L.~P., McElroy, N.~E., McNeill, F.~L., {et~al.} 2024, Monthly
  Notices of the Royal Astronomical Society, 532, 2289

\bibitem[{{Mustill} \& {Villaver}(2012)}]{MustillVillaver2012}
{Mustill}, A.~J. \& {Villaver}, E. 2012, \apj, 761, 121

\bibitem[{{Mustill} {et~al.}(2018){Mustill}, {Villaver}, {Veras},
  {G{\"a}nsicke}, \& {Bonsor}}]{Mustill2018}
{Mustill}, A.~J., {Villaver}, E., {Veras}, D., {G{\"a}nsicke}, B.~T., \&
  {Bonsor}, A. 2018, \mnras, 476, 3939

\bibitem[{{Myers} {et~al.}(2022){Myers}, {Donor}, {Spoo}, {Frinchaboy},
  {Cunha}, {Price-Whelan}, {Majewski}, {Beaton}, {Zasowski}, {O'Connell},
  {Ray}, {Bizyaev}, {Chiappini}, {Garc{\'\i}a-Hern{\'a}ndez}, {Geisler},
  {J{\"o}nsson}, {Lane}, {Longa-Pe{\~n}a}, {Minchev}, {Minniti}, {Nitschelm},
  \& {Roman-Lopes}}]{myers2022}
{Myers}, N., {Donor}, J., {Spoo}, T., {et~al.} 2022, \aj, 164, 85

\bibitem[{{Naponiello} {et~al.}(2025){Naponiello}, {Bonomo}, {Mancini},
  {Steinmeyer}, {Biazzo}, {Polychroni}, {Dorn}, {Turrini}, {Lanza}, {Sozzetti},
  {Desidera}, {Damasso}, {Collins}, {Carleo}, {Collins}, {Colombo}, {D'Arpa},
  {Dumusque}, {Gonz{\'a}lez}, {Guilluy}, {Lorenzi}, {Mantovan}, {Nardiello},
  {Pinamonti}, {Schwarz}, {Singh}, {Watkins}, \& {Zingales}}]{Naponiello2025}
{Naponiello}, L., {Bonomo}, A.~S., {Mancini}, L., {et~al.} 2025, \aap, 693, A7

\bibitem[{{Nardiello} {et~al.}(2015){Nardiello}, {Bedin}, {Nascimbeni},
  {Libralato}, {Cunial}, {Piotto}, {Bellini}, {Borsato}, {Brogaard}, {Granata},
  {Malavolta}, {Marino}, {Milone}, {Ochner}, {Ortolani}, {Tomasella},
  {Clemens}, \& {Salaris}}]{Nardiello2015MNRAS.447.3536N}
{Nardiello}, D., {Bedin}, L.~R., {Nascimbeni}, V., {et~al.} 2015, \mnras, 447,
  3536

\bibitem[{{Nardiello} {et~al.}(2019){Nardiello}, {Borsato}, {Piotto},
  {Colombo}, {Manthopoulou}, {Bedin}, {Granata}, {Lacedelli}, {Libralato},
  {Malavolta}, {Montalto}, \& {Nascimbeni}}]{nardiello2019mnras.490.3806n}
{Nardiello}, D., {Borsato}, L., {Piotto}, G., {et~al.} 2019, \mnras, 490, 3806

\bibitem[{{Nardiello} {et~al.}(2016){Nardiello}, {Libralato}, {Bedin},
  {Piotto}, {Borsato}, {Granata}, {Malavolta}, \&
  {Nascimbeni}}]{Nardiello2016MNRAS.463.1831N}
{Nardiello}, D., {Libralato}, M., {Bedin}, L.~R., {et~al.} 2016, \mnras, 463,
  1831

\bibitem[{{Nardiello} {et~al.}(2022){Nardiello}, {Malavolta}, {Desidera},
  {Baratella}, {D'Orazi}, {Messina}, {Biazzo}, {Benatti}, {Damasso}, {Rajpaul},
  {Bonomo}, {Capuzzo Dolcetta}, {Mallonn}, {Cale}, {Plavchan}, {El Mufti},
  {Bignamini}, {Borsa}, {Carleo}, {Claudi}, {Covino}, {Lanza}, {Maldonado},
  {Mancini}, {Micela}, {Molinari}, {Pinamonti}, {Piotto}, {Poretti},
  {Scandariato}, {Sozzetti}, {Andreuzzi}, {Boschin}, {Cosentino}, {Fiorenzano},
  {Harutyunyan}, {Knapic}, {Pedani}, {Affer}, {Maggio}, \&
  {Rainer}}]{Nardiello2022}
{Nardiello}, D., {Malavolta}, L., {Desidera}, S., {et~al.} 2022, \aap, 664,
  A163

\bibitem[{{Nardiello} {et~al.}(2020){Nardiello}, {Piotto}, {Deleuil},
  {Malavolta}, {Montalto}, {Bedin}, {Borsato}, {Granata}, {Libralato}, \& {Ma
  nthopoulou}}]{2020MNRAS.495.4924N}
{Nardiello}, D., {Piotto}, G., {Deleuil}, M., {et~al.} 2020, \mnras, 495, 4924

\bibitem[{{Nascimbeni} {et~al.}(2012){Nascimbeni}, {Bedin}, {Piotto}, {De
  Marchi}, \& {Rich}}]{Nascimbeni2012A&A...541A.144N}
{Nascimbeni}, V., {Bedin}, L.~R., {Piotto}, G., {De Marchi}, F., \& {Rich},
  R.~M. 2012, \aap, 541, A144

\bibitem[{{Nascimbeni} {et~al.}(2025){Nascimbeni}, {Piotto}, {Cabrera},
  {Montalto}, {Marinoni}, {Marrese}, {Aerts}, {Altavilla}, {Benatti},
  {B{\"o}rner}, {Deleuil}, {Desidera}, {Gizon}, {Goupil}, {Granata}, {Heras},
  {Magrin}, {Malavolta}, {Mas-Hesse}, {Osborn}, {Pagano}, {Paproth},
  {Pollacco}, {Prisinzano}, {Ragazzoni}, {Ramsay}, {Rauer}, {Tkachenko}, \&
  {Udry}}]{Nascimbeni2025A&A...694A.313N}
{Nascimbeni}, V., {Piotto}, G., {Cabrera}, J., {et~al.} 2025, \aap, 694, A313

\bibitem[{Nave \& Clear(2021)}]{Nave2021}
Nave, G. \& Clear, C. 2021, Monthly Notices of the Royal Astronomical Society,
  502, 5679

\bibitem[{{Newton} {et~al.}(2021){Newton}, {Mann}, {Kraus}, {Livingston},
  {Vanderburg}, {Curtis}, {Thao}, {Hawkins}, {Wood}, {Rizzuto}, {Soubkiou},
  {Tofflemire}, {Zhou}, {Crossfield}, {Pearce}, {Collins}, {Conti}, {Tan},
  {Villeneuva}, {Spencer}, {Dragomir}, {Quinn}, {Jensen}, {Collins},
  {Stockdale}, {Cloutier}, {Hellier}, {Benkhaldoun}, {Ziegler}, {Brice{\~n}o},
  {Law}, {Benneke}, {Christiansen}, {Gorjian}, {Kane}, {Kreidberg}, {Morales},
  {Werner}, {Twicken}, {Levine}, {Ciardi}, {Guerrero}, {Hesse}, {Quintana},
  {Shiao}, {Smith}, {Torres}, {Ricker}, {Vanderspek}, {Seager}, {Winn},
  {Jenkins}, \& {Latham}}]{Newton2021}
{Newton}, E.~R., {Mann}, A.~W., {Kraus}, A.~L., {et~al.} 2021, \aj, 161, 65

\bibitem[{{Newton} {et~al.}(2019){Newton}, {Mann}, {Tofflemire}, {Pearce},
  {Rizzuto}, {Vanderburg}, {Martinez}, {Wang}, {Ruffio}, {Kraus}, {Johnson},
  {Thao}, {Wood}, {Rampalli}, {Nielsen}, {Collins}, {Dragomir}, {Hellier},
  {Anderson}, {Barclay}, {Brown}, {Feiden}, {Hart}, {Isopi}, {Kielkopf},
  {Mallia}, {Nelson}, {Rodriguez}, {Stockdale}, {Waite}, {Wright}, {Lissauer},
  {Ricker}, {Vanderspek}, {Latham}, {Seager}, {Winn}, {Jenkins}, {Bouma},
  {Burke}, {Davies}, {Fausnaugh}, {Li}, {Morris}, {Mukai}, {Villase{\~n}or},
  {Villeneuva}, {De Rosa}, {Macintosh}, {Mengel}, {Okumura}, \&
  {Wittenmyer}}]{Newton2019ApJ...880L..17N}
{Newton}, E.~R., {Mann}, A.~W., {Tofflemire}, B.~M., {et~al.} 2019, \apjl, 880,
  L17

\bibitem[{{Nguyen} {et~al.}(2025){Nguyen}, {Cescutti}, {Matteucci}, {Rizzuti},
  {Mucciarelli}, {Romano}, {Magrini}, {Korn}, {Bressan}, \&
  {Girardi}}]{ngueyen25LSb}
{Nguyen}, C.~T., {Cescutti}, G., {Matteucci}, F., {et~al.} 2025, \aap, 703,
  A204

\bibitem[{{Nguyen} \& {Adibekyan}(2025)}]{Nguyen2025AJ....170..334N}
{Nguyen}, M. \& {Adibekyan}, V. 2025, \aj, 170, 334

\bibitem[{Nilsson {et~al.}(2019)Nilsson, Andersson, Engstrom, Lundberg, \&
  Hartman}]{Nilsson2019a}
Nilsson, H., Andersson, J., Engstrom, L., Lundberg, H., \& Hartman, H. 2019,
  Astronomy and Astrophysics, 622

\bibitem[{Nilsson {et~al.}(2002)Nilsson, Zhang, Lundberg, Johansson, \&
  Nordstrom}]{Nilsson2002}
Nilsson, H., Zhang, Z.~G., Lundberg, H., Johansson, S., \& Nordstrom, B. 2002,
  Astronomy and Astrophysics, 382, 368

\bibitem[{{Nissen}(2015)}]{nissen2015}
{Nissen}, P.~E. 2015, \aap, 579, A52

\bibitem[{{Nissen} {et~al.}(2007){Nissen}, {Akerman}, {Asplund}, {Fabbian},
  {Kerber}, {Kaufl}, \& {Pettini}}]{nissen2007}
{Nissen}, P.~E., {Akerman}, C., {Asplund}, M., {et~al.} 2007, \aap, 469, 319

\bibitem[{{Nissen} {et~al.}(2024){Nissen}, {Amarsi}, {Sk{\'u}lad{\'o}ttir}, \&
  {Schuster}}]{nissen2024}
{Nissen}, P.~E., {Amarsi}, A.~M., {Sk{\'u}lad{\'o}ttir}, {\'A}., \& {Schuster},
  W.~J. 2024, \aap, 682, A116

\bibitem[{{Nissen} {et~al.}(2004){Nissen}, {Chen}, {Asplund}, \&
  {Pettini}}]{nissen2004sulphur}
{Nissen}, P.~E., {Chen}, Y.~Q., {Asplund}, M., \& {Pettini}, M. 2004, \aap,
  415, 993

\bibitem[{{Nissen} {et~al.}(2014){Nissen}, {Chen}, {Carigi}, {Schuster}, \&
  {Zhao}}]{nissen2014}
{Nissen}, P.~E., {Chen}, Y.~Q., {Carigi}, L., {Schuster}, W.~J., \& {Zhao}, G.
  2014, \aap, 568, A25

\bibitem[{{Nissen} \& {Gustafsson}(2018)}]{nissen2018}
{Nissen}, P.~E. \& {Gustafsson}, B. 2018, \aapr, 26, 6

\bibitem[{{Nissen} \& {Schuster}(2010)}]{nissen2010}
{Nissen}, P.~E. \& {Schuster}, W.~J. 2010, \aap, 511, L10

\bibitem[{{Nissen} \& {Schuster}(2011)}]{nissen2011}
{Nissen}, P.~E. \& {Schuster}, W.~J. 2011, \aap, 530, A15

\bibitem[{{Nomoto} {et~al.}(2013){Nomoto}, {Kobayashi}, \&
  {Tominaga}}]{Nomoto2013}
{Nomoto}, K., {Kobayashi}, C., \& {Tominaga}, N. 2013, \araa, 51, 457

\bibitem[{{Nordlander} {et~al.}(2024){Nordlander}, {Baratella}, {Spina}, \&
  {D'Orazi}}]{2024nordlander}
{Nordlander}, T., {Baratella}, M., {Spina}, L., \& {D'Orazi}, V. 2024, \mnras,
  535, 2863

\bibitem[{{Nordlander} {et~al.}(2012){Nordlander}, {Korn}, {Richard}, \&
  {Lind}}]{nordlander2012}
{Nordlander}, T., {Korn}, A.~J., {Richard}, O., \& {Lind}, K. 2012, \apj, 753,
  48

\bibitem[{{Nordlund} {et~al.}(2018){Nordlund}, {Ramsey}, {Popovas}, \&
  {K{\"u}ffmeier}}]{nordlund2018}
{Nordlund}, {\r{A}}., {Ramsey}, J.~P., {Popovas}, A., \& {K{\"u}ffmeier}, M.
  2018, \mnras, 477, 624

\bibitem[{{Norris} {et~al.}(2013){Norris}, {Bessell}, {Yong}, {Christlieb},
  {Barklem}, {Asplund}, {Murphy}, {Beers}, {Frebel}, \& {Ryan}}]{Norris2013}
{Norris}, J.~E., {Bessell}, M.~S., {Yong}, D., {et~al.} 2013, \apj, 762, 25

\bibitem[{{Noyes} {et~al.}(1984){Noyes}, {Hartmann}, {Baliunas}, {Duncan}, \&
  {Vaughan}}]{noyes1984}
{Noyes}, R.~W., {Hartmann}, L.~W., {Baliunas}, S.~L., {Duncan}, D.~K., \&
  {Vaughan}, A.~H. 1984, \apj, 279, 763

\bibitem[{{Nunnari} {et~al.}(2025){Nunnari}, {D'Orazi}, {Fiorentino}, {Braga},
  {Bono}, {Fabrizio}, {J{\"o}nsson}, {Kudritzki}, {da Silva}, {Bergemann},
  {Poggio}, {Otto}, {Baeza-Villagra}, {Bragaglia}, {Ceci}, {Dall'Ora}, {Inno},
  {Lardo}, {Matsunaga}, {Monelli}, {S{\'a}nchez-Benavente}, {Sneden},
  {Tantalo}, \& {Th{\'e}v{\'e}nin}}]{nunnari25}
{Nunnari}, A., {D'Orazi}, V., {Fiorentino}, G., {et~al.} 2025, arXiv e-prints,
  arXiv:2511.22491

\bibitem[{{Odert} {et~al.}(2017){Odert}, {Leitzinger}, {Hanslmeier}, \&
  {Lammer}}]{odert2017}
{Odert}, P., {Leitzinger}, M., {Hanslmeier}, A., \& {Lammer}, H. 2017, \mnras,
  472, 876

\bibitem[{{Oliveira} {et~al.}(2020){Oliveira}, {Souza}, {Kerber}, {Barbuy},
  {Ortolani}, {Piotto}, {Nardiello}, {P{\'e}rez-Villegas}, {Maia}, {Bica},
  {Cassisi}, {D'Antona}, {Lagioia}, {Libralato}, {Milone}, {Anderson},
  {Aparicio}, {Bedin}, {Brown}, {King}, {Marino}, {Pietrinferni}, {Renzini},
  {Sarajedini}, {van der Marel}, \& {Vesperini}}]{Oliveira2020ApJ...891...37O}
{Oliveira}, R.~A.~P., {Souza}, S.~O., {Kerber}, L.~O., {et~al.} 2020, \apj,
  891, 37

\bibitem[{{{\"O}nehag} {et~al.}(2014){{\"O}nehag}, {Gustafsson}, \&
  {Korn}}]{onehag2014}
{{\"O}nehag}, A., {Gustafsson}, B., \& {Korn}, A. 2014, \aap, 562, A102

\bibitem[{{Orosz} {et~al.}(2012){Orosz}, {Welsh}, {Carter}, {Fabrycky},
  {Cochran}, {Endl}, {Ford}, {Haghighipour}, {MacQueen}, {Mazeh},
  {Sanchis-Ojeda}, {Short}, {Torres}, {Agol}, {Buchhave}, {Doyle}, {Isaacson},
  {Lissauer}, {Marcy}, {Shporer}, {Windmiller}, {Barclay}, {Boss}, {Clarke},
  {Fortney}, {Geary}, {Holman}, {Huber}, {Jenkins}, {Kinemuchi}, {Kruse},
  {Ragozzine}, {Sasselov}, {Still}, {Tenenbaum}, {Uddin}, {Winn}, {Koch}, \&
  {Borucki}}]{Orosz2012}
{Orosz}, J.~A., {Welsh}, W.~F., {Carter}, J.~A., {et~al.} 2012, Science, 337,
  1511

\bibitem[{{Osterbrock} \&
  {Ferland}(2006)}]{Osterbrock_and_Ferland2006agna.book.....O}
{Osterbrock}, D.~E. \& {Ferland}, G.~J. 2006, {Astrophysics of gaseous nebulae
  and active galactic nuclei} (University Science Books)

\bibitem[{{Owen} \& {Wu}(2013)}]{owen2013}
{Owen}, J.~E. \& {Wu}, Y. 2013, \apj, 775, 105

\bibitem[{{P{\~o}der} {et~al.}(2025){P{\~o}der}, {Pata}, {Benito}, {Alonso
  Asensio}, \& {Dalla Vecchia}}]{Poder2025}
{P{\~o}der}, S., {Pata}, J., {Benito}, M., {Alonso Asensio}, I., \& {Dalla
  Vecchia}, C. 2025, \aap, 693, A227

\bibitem[{{Pacetti} {et~al.}(2025){Pacetti}, {Schisano}, {Turrini},
  {Dullemond}, {Molinari}, {Walsh}, {Fonte}, {Lebreuilly}, {Klessen},
  {Hennebelle}, {Ivanovski}, {Politi}, {Polychroni}, {Simonetti}, \&
  {Testi}}]{Pacetti2025}
{Pacetti}, E., {Schisano}, E., {Turrini}, D., {et~al.} 2025, \aap, 701, A194

\bibitem[{{Pacetti} {et~al.}(2022){Pacetti}, {Turrini}, {Schisano}, {Molinari},
  {Fonte}, {Politi}, {Hennebelle}, {Klessen}, {Testi}, \&
  {Lebreuilly}}]{Pacetti2022}
{Pacetti}, E., {Turrini}, D., {Schisano}, E., {et~al.} 2022, \apj, 937, 36

\bibitem[{{Palanque-Delabrouille} {et~al.}(2016){Palanque-Delabrouille},
  {Magneville}, {Y{\`e}che}, {P{\^a}ris}, {Petitjean}, {Burtin}, {Dawson},
  {McGreer}, {Myers}, {Rossi}, {Schlegel}, {Schneider}, {Streblyanska}, \&
  {Tinker}}]{Palanque-Delabrouille2016A&A...587A..41P}
{Palanque-Delabrouille}, N., {Magneville}, C., {Y{\`e}che}, C., {et~al.} 2016,
  \aap, 587, A41

\bibitem[{{Palla} {et~al.}(2025){Palla}, {Molero}, {Romano}, \&
  {Mucciarelli}}]{Palla2025}
{Palla}, M., {Molero}, M., {Romano}, D., \& {Mucciarelli}, A. 2025, \aap, 699,
  A209

\bibitem[{{Pancino} {et~al.}(2017){Pancino}, {Romano}, {Tang},
  {Tautvai{\v{s}}ien{\.{e}}}, {Casey}, {Gruyters}, {Geisler}, {San Roman},
  {Randich}, {Alfaro}, {Bragaglia}, {Flaccomio}, {Korn}, {Recio-Blanco},
  {Smiljanic}, {Carraro}, {Bayo}, {Costado}, {Damiani}, {Jofr{\'e}}, {Lardo},
  {de Laverny}, {Monaco}, {Morbidelli}, {Sbordone}, {Sousa}, \&
  {Villanova}}]{pancino2017}
{Pancino}, E., {Romano}, D., {Tang}, B., {et~al.} 2017, \aap, 601, A112

\bibitem[{{Pascucci} {et~al.}(2022){Pascucci}, {Cabrit}, {Edwards}, {Gorti},
  {Gressel}, \& {Suzuki}}]{pascucci2022a}
{Pascucci}, I., {Cabrit}, S., {Edwards}, S., {et~al.} 2022, arXiv e-prints,
  arXiv:2203.10068

\bibitem[{{Pascucci} {et~al.}(2016){Pascucci}, {Testi}, {Herczeg}, {Long},
  {Manara}, {Hendler}, {Mulders}, {Krijt}, {Ciesla}, {Henning}, {Mohanty},
  {Drabek-Maunder}, {Apai}, {Sz{\H{u}}cs}, {Sacco}, \&
  {Olofsson}}]{Pascucci2016}
{Pascucci}, I., {Testi}, L., {Herczeg}, G.~J., {et~al.} 2016, \apj, 831, 125

\bibitem[{{Pasquini} {et~al.}(2012){Pasquini}, {Brucalassi}, {Ruiz},
  {Bonifacio}, {Lovis}, {Saglia}, {Melo}, {Biazzo}, {Randich}, \&
  {Bedin}}]{Pasquini2012A&A...545A.139P}
{Pasquini}, L., {Brucalassi}, A., {Ruiz}, M.~T., {et~al.} 2012, \aap, 545, A139

\bibitem[{{Patel} {et~al.}(2025){Patel}, {Metzger}, {Cehula}, {Burns},
  {Goldberg}, \& {Thompson}}]{Patel2025}
{Patel}, A., {Metzger}, B.~D., {Cehula}, J., {et~al.} 2025, \apjl, 984, L29

\bibitem[{{Paxton} {et~al.}(2011){Paxton}, {Bildsten}, {Dotter}, {Herwig},
  {Lesaffre}, \& {Timmes}}]{2011ApJS..192....3P}
{Paxton}, B., {Bildsten}, L., {Dotter}, A., {et~al.} 2011, \apjs, 192, 3

\bibitem[{{Paxton} {et~al.}(2013){Paxton}, {Cantiello}, {Arras}, {Bildsten},
  {Brown}, {Dotter}, {Mankovich}, {Montgomery}, {Stello}, {Timmes}, \&
  {Townsend}}]{2013ApJS..208....4P}
{Paxton}, B., {Cantiello}, M., {Arras}, P., {et~al.} 2013, \apjs, 208, 4

\bibitem[{{Paxton} {et~al.}(2015){Paxton}, {Marchant}, {Schwab}, {Bauer},
  {Bildsten}, {Cantiello}, {Dessart}, {Farmer}, {Hu}, {Langer}, {Townsend},
  {Townsley}, \& {Timmes}}]{2015ApJS..220...15P}
{Paxton}, B., {Marchant}, P., {Schwab}, J., {et~al.} 2015, \apjs, 220, 15

\bibitem[{Pehlivan {et~al.}(2015)Pehlivan, Nilsson, \& Hartman}]{Pehlivan2015}
Pehlivan, A., Nilsson, H., \& Hartman, H. 2015, Astronomy and Astrophysics, 582

\bibitem[{{Penny} {et~al.}(2019){Penny}, {Gaudi}, {Kerins}, {Rattenbury},
  {Mao}, {Robin}, \& {Calchi Novati}}]{penny2019}
{Penny}, M.~T., {Gaudi}, B.~S., {Kerins}, E., {et~al.} 2019, \apjs, 241, 3

\bibitem[{{Penoyre} {et~al.}(2022){Penoyre}, {Belokurov}, \&
  {Evans}}]{penoyre2022}
{Penoyre}, Z., {Belokurov}, V., \& {Evans}, N.~W. 2022, \mnras, 513, 5270

\bibitem[{{Pepe} {et~al.}(2021){Pepe}, {Cristiani}, {Rebolo}, {Santos},
  {Dekker}, {Cabral}, {Di Marcantonio}, {Figueira}, {Lo Curto}, {Lovis},
  {Mayor}, {M{\'e}gevand}, {Molaro}, {Riva}, {Zapatero Osorio}, {Amate},
  {Manescau}, {Pasquini}, {Zerbi}, {Adibekyan}, {Abreu}, {Affolter}, {Alibert},
  {Aliverti}, {Allart}, {Allende Prieto}, {{\'A}lvarez}, {Alves}, {Avila},
  {Baldini}, {Bandy}, {Barros}, {Benz}, {Bianco}, {Borsa}, {Bourrier},
  {Bouchy}, {Broeg}, {Calderone}, {Cirami}, {Coelho}, {Conconi}, {Coretti},
  {Cumani}, {Cupani}, {D'Odorico}, {Damasso}, {Deiries}, {Delabre},
  {Demangeon}, {Dumusque}, {Ehrenreich}, {Faria}, {Fragoso}, {Genolet},
  {Genoni}, {G{\'e}nova Santos}, {Gonz{\'a}lez Hern{\'a}ndez}, {Hughes},
  {Iwert}, {Kerber}, {Knudstrup}, {Landoni}, {Lavie}, {Lillo-Box}, {Lizon},
  {Maire}, {Martins}, {Mehner}, {Micela}, {Modigliani}, {Monteiro}, {Monteiro},
  {Moschetti}, {Murphy}, {Nunes}, {Oggioni}, {Oliveira}, {Oshagh}, {Pall{\'e}},
  {Pariani}, {Poretti}, {Rasilla}, {Rebord{\~a}o}, {Redaelli}, {Santana
  Tschudi}, {Santin}, {Santos}, {S{\'e}gransan}, {Schmidt}, {Segovia},
  {Sosnowska}, {Sozzetti}, {Sousa}, {Span{\`o}}, {Su{\'a}rez Mascare{\~n}o},
  {Tabernero}, {Tenegi}, {Udry}, \& {Zanutta}}]{Pepe2021A&A...645A..96P}
{Pepe}, F., {Cristiani}, S., {Rebolo}, R., {et~al.} 2021, \aap, 645, A96

\bibitem[{{Pepe} {et~al.}(2002){Pepe}, {Mayor}, {Galland}, {Naef}, {Queloz},
  {Santos}, {Udry}, \& {Burnet}}]{Pepe2002}
{Pepe}, F., {Mayor}, M., {Galland}, F., {et~al.} 2002, \aap, 388, 632

\bibitem[{{Perets}(2010)}]{Perets2010}
{Perets}, H.~B. 2010, arXiv e-prints, arXiv:1001.0581

\bibitem[{{Perryman} {et~al.}(2014){Perryman}, {Hartman}, {Bakos}, \&
  {Lindegren}}]{perryman2014}
{Perryman}, M., {Hartman}, J., {Bakos}, G.~{\'A}., \& {Lindegren}, L. 2014,
  \apj, 797, 14

\bibitem[{{Pian} {et~al.}(2017){Pian}, {D'Avanzo}, {Benetti}, {Branchesi},
  {Brocato}, {Campana}, {Cappellaro}, {Covino}, {D'Elia}, {Fynbo}, {Getman},
  {Ghirlanda}, {Ghisellini}, {Grado}, {Greco}, {Hjorth}, {Kouveliotou},
  {Levan}, {Limatola}, {Malesani}, {Mazzali}, {Melandri}, {M{\o}ller},
  {Nicastro}, {Palazzi}, {Piranomonte}, {Rossi}, {Salafia}, {Selsing},
  {Stratta}, {Tanaka}, {Tanvir}, {Tomasella}, {Watson}, {Yang}, {Amati},
  {Antonelli}, {Ascenzi}, {Bernardini}, {Bo{\"e}r}, {Bufano}, {Bulgarelli},
  {Capaccioli}, {Casella}, {Castro-Tirado}, {Chassande-Mottin}, {Ciolfi},
  {Copperwheat}, {Dadina}, {De Cesare}, {di Paola}, {Fan}, {Gendre},
  {Giuffrida}, {Giunta}, {Hunt}, {Israel}, {Jin}, {Kasliwal}, {Klose}, {Lisi},
  {Longo}, {Maiorano}, {Mapelli}, {Masetti}, {Nava}, {Patricelli}, {Perley},
  {Pescalli}, {Piran}, {Possenti}, {Pulone}, {Razzano}, {Salvaterra},
  {Schipani}, {Spera}, {Stamerra}, {Stella}, {Tagliaferri}, {Testa}, {Troja},
  {Turatto}, {Vergani}, \& {Vergani}}]{Pian2017}
{Pian}, E., {D'Avanzo}, P., {Benetti}, S., {et~al.} 2017, \nat, 551, 67

\bibitem[{{Pignatari} {et~al.}(2010){Pignatari}, {Gallino}, {Heil}, {Wiescher},
  {K{\"a}ppeler}, {Herwig}, \& {Bisterzo}}]{Pignatari2010}
{Pignatari}, M., {Gallino}, R., {Heil}, M., {et~al.} 2010, \apj, 710, 1557

\bibitem[{{Pignatari} {et~al.}(2016){Pignatari}, {Herwig}, {Hirschi},
  {Bennett}, {Rockefeller}, {Fryer}, {Timmes}, {Ritter}, {Heger}, {Jones},
  {Battino}, {Dotter}, {Trappitsch}, {Diehl}, {Frischknecht}, {Hungerford},
  {Magkotsios}, {Travaglio}, \& {Young}}]{Pignatari2016}
{Pignatari}, M., {Herwig}, F., {Hirschi}, R., {et~al.} 2016, \apjs, 225, 24

\bibitem[{{Pignatari} {et~al.}(2015){Pignatari}, {Zinner}, {Hoppe}, {Jordan},
  {Gibson}, {Trappitsch}, {Herwig}, {Fryer}, {Hirschi}, \&
  {Timmes}}]{Pignatari2015}
{Pignatari}, M., {Zinner}, E., {Hoppe}, P., {et~al.} 2015, \apjl, 808, L43

\bibitem[{{Pillitteri} {et~al.}(2014){Pillitteri}, {Wolk}, {Sciortino}, \&
  {Antoci}}]{pillitteri2014}
{Pillitteri}, I., {Wolk}, S.~J., {Sciortino}, S., \& {Antoci}, V. 2014, \aap,
  567, A128

\bibitem[{{Piotto} {et~al.}(2007){Piotto}, {Bedin}, {Anderson}, {King},
  {Cassisi}, {Milone}, {Villanova}, {Pietrinferni}, \& {Renzini}}]{Piotto2007}
{Piotto}, G., {Bedin}, L.~R., {Anderson}, J., {et~al.} 2007, \apjl, 661, L53

\bibitem[{{Pirani} {et~al.}(2019){Pirani}, {Johansen}, {Bitsch}, {Mustill}, \&
  {Turrini}}]{Pirani2019A&A...623A.169P}
{Pirani}, S., {Johansen}, A., {Bitsch}, B., {Mustill}, A.~J., \& {Turrini}, D.
  2019, \aap, 623, A169

\bibitem[{{Piskunov} \& {Valenti}(2017)}]{piskunov2017}
{Piskunov}, N. \& {Valenti}, J.~A. 2017, \aap, 597, A16

\bibitem[{{Placco} {et~al.}(2017){Placco}, {Holmbeck}, {Frebel}, {Beers},
  {Surman}, {Ji}, {Ezzeddine}, {Points}, {Kaleida}, {Hansen}, {Sakari}, \&
  {Casey}}]{Placco2017}
{Placco}, V.~M., {Holmbeck}, E.~M., {Frebel}, A., {et~al.} 2017, \apj, 844, 18

\bibitem[{{Plez}(2012)}]{turbospectrum}
{Plez}, B. 2012, {Turbospectrum: Code for spectral synthesis}

\bibitem[{{Pollack} {et~al.}(1996){Pollack}, {Hubickyj}, {Bodenheimer},
  {Lissauer}, {Podolak}, \& {Greenzweig}}]{pollack1996}
{Pollack}, J.~B., {Hubickyj}, O., {Bodenheimer}, P., {et~al.} 1996, \icarus,
  124, 62

\bibitem[{{Polychroni} {et~al.}(2023){Polychroni}, {Turrini}, \&
  {Pirani}}]{Polychroni2023}
{Polychroni}, D., {Turrini}, D., \& {Pirani}, S. 2023, {GroMiT: Planet Growth
  and Migration Track code}

\bibitem[{{Poppenhaeger} \& {Wolk}(2014)}]{poppenhaeger2014}
{Poppenhaeger}, K. \& {Wolk}, S.~J. 2014, \aap, 565, L1

\bibitem[{{Prantzos} {et~al.}(1996){Prantzos}, {Aubert}, \&
  {Audouze}}]{Prantzos1996}
{Prantzos}, N., {Aubert}, O., \& {Audouze}, J. 1996, \aap, 309, 760

\bibitem[{{Prantzos} {et~al.}(2007){Prantzos}, {Charbonnel}, \&
  {Iliadis}}]{Prantzos2007}
{Prantzos}, N., {Charbonnel}, C., \& {Iliadis}, C. 2007, \aap, 470, 179

\bibitem[{{Przy{\l}uski} {et~al.}(2025){Przy{\l}uski}, {Rickman}, {Wajer},
  {Wi{\'s}niowski}, {Turrini}, {Polychroni}, {Danielski}, {Kruijssen},
  {Longmore}, \& {Chevance}}]{Przyluski2025}
{Przy{\l}uski}, R., {Rickman}, H., {Wajer}, P., {et~al.} 2025, Universe, 11,
  240

\bibitem[{{Queloz} {et~al.}(1998){Queloz}, {Allain}, {Mermilliod}, {Bouvier},
  \& {Mayor}}]{queloz1998}
{Queloz}, D., {Allain}, S., {Mermilliod}, J.~C., {Bouvier}, J., \& {Mayor}, M.
  1998, \aap, 335, 183

\bibitem[{{Queloz} {et~al.}(2001){Queloz}, {Henry}, {Sivan}, {Baliunas},
  {Beuzit}, {Donahue}, {Mayor}, {Naef}, {Perrier}, \& {Udry}}]{Queloz2001}
{Queloz}, D., {Henry}, G.~W., {Sivan}, J.~P., {et~al.} 2001, \aap, 379, 279

\bibitem[{Quinet {et~al.}(2016)Quinet, Fivet, Palmeri, EngstrÃ¶m, Hartman,
  Lundberg, \& Nilsson}]{Quinet2016}
Quinet, P., Fivet, V., Palmeri, P., {et~al.} 2016, Monthly Notices of the Royal
  Astronomical Society, 462, 3912

\bibitem[{{Rajpaul} {et~al.}(2015){Rajpaul}, {Aigrain}, {Osborne}, {Reece}, \&
  {Roberts}}]{Rajpaul2015}
{Rajpaul}, V., {Aigrain}, S., {Osborne}, M.~A., {Reece}, S., \& {Roberts}, S.
  2015, \mnras, 452, 2269

\bibitem[{Ralchenko {et~al.}(2021)Ralchenko, Fontes,
  {et~al.}}]{NISTLANLLanthanideActinideOpacityDatabase}
Ralchenko, Y., Fontes, C.~J., {et~al.} 2021, NIST–LANL Lanthanide/Actinide
  Opacity Database, National Institute of Standards and Technology and Los
  Alamos National Laboratory

\bibitem[{{Ram{\'\i}rez} {et~al.}(2025){Ram{\'\i}rez}, {G{\"a}nsicke},
  {Koester}, {Lafarga}, \& {Gentile-Fusillo}}]{Ramirez2025}
{Ram{\'\i}rez}, S.~H., {G{\"a}nsicke}, B.~T., {Koester}, D., {Lafarga}, M., \&
  {Gentile-Fusillo}, N.~P. 2025, \mnras, 539, 2884

\bibitem[{{Randich}(2019)}]{Randich2019vltt.confE..11R}
{Randich}, S. 2019, in The Very Large Telescope in 2030, 11

\bibitem[{{Randich} {et~al.}(2026){Randich}, {Bianco}, {Caito}, {Fernandes
  Alvar}, {Gonzalez}, {Sousa}, \& {Hrmos Consortium}}]{randichVLT2030}
{Randich}, S., {Bianco}, A., {Caito}, L., {et~al.} 2026, in VLT Beyond 2030, 44

\bibitem[{{Randich} {et~al.}(2022){Randich}, {Gilmore}, {Magrini}, {Sacco},
  {Jackson}, {Jeffries}, {Worley}, {Hourihane}, {Gonneau}, {Viscasillas
  Vazquez}, {Franciosini}, {Lewis}, {Alfaro}, {Allende Prieto}, {Bensby},
  {Blomme}, {Bragaglia}, {Flaccomio}, {Fran{\c{c}}ois}, {Irwin}, {Koposov},
  {Korn}, {Lanzafame}, {Pancino}, {Recio-Blanco}, {Smiljanic}, {Van Eck},
  {Zwitter}, {Asplund}, {Bonifacio}, {Feltzing}, {Binney}, {Drew}, {Ferguson},
  {Micela}, {Negueruela}, {Prusti}, {Rix}, {Vallenari}, {Bayo}, {Bergemann},
  {Biazzo}, {Carraro}, {Casey}, {Damiani}, {Frasca}, {Heiter}, {Hill},
  {Jofr{\'e}}, {de Laverny}, {Lind}, {Marconi}, {Martayan}, {Masseron},
  {Monaco}, {Morbidelli}, {Prisinzano}, {Sbordone}, {Sousa}, {Zaggia},
  {Adibekyan}, {Bonito}, {Caffau}, {Daflon}, {Feuillet}, {Gebran}, {Gonzalez
  Hernandez}, {Guiglion}, {Herrero}, {Lobel}, {Maiz Apellaniz}, {Merle},
  {Mikolaitis}, {Montes}, {Morel}, {Soubiran}, {Spina}, {Tabernero},
  {Tautvai{\v{s}}iene}, {Traven}, {Valentini}, {Van der Swaelmen}, {Villanova},
  {Wright}, {Abbas}, {Aguirre B{\o}rsen-Koch}, {Alves}, {Balaguer-Nunez},
  {Barklem}, {Barrado}, {Berlanas}, {Binks}, {Bressan}, {Capuzzo-Dolcetta},
  {Casagrande}, {Casamiquela}, {Collins}, {D'Orazi}, {Dantas}, {Debattista},
  {Delgado-Mena}, {Di Marcantonio}, {Drazdauskas}, {Evans}, {Famaey},
  {Franchini}, {Fr{\'e}mat}, {Friel}, {Fu}, {Geisler}, {Gerhard}, {Gonzalez
  Solares}, {Grebel}, {Gutierrez Albarran}, {Hatzidimitriou}, {Held},
  {Jim{\'e}nez-Esteban}, {J{\"o}nsson}, {Jordi}, {Khachaturyants},
  {Kordopatis}, {Kos}, {Lagarde}, {Mahy}, {Mapelli}, {Marfil}, {Martell},
  {Messina}, {Miglio}, {Minchev}, {Moitinho}, {Montalban}, {Monteiro},
  {Morossi}, {Mowlavi}, {Mucciarelli}, {Murphy}, {Nardetto}, {Ortolani},
  {Paletou}, {Palou{\v{s}}}, {Paunzen}, {Pickering}, {Quirrenbach}, {Re
  Fiorentin}, {Read}, {Romano}, {Ryde}, {Sanna}, {Santos}, {Seabroke},
  {Spagna}, {Steinmetz}, {Stonkut{\'e}}, {Sutorius}, {Th{\'e}venin}, {Tosi},
  {Tsantaki}, {Vink}, {Wright}, {Wyse}, {Zoccali}, {Zorec}, {Zucker}, \&
  {Walton}}]{randich2022}
{Randich}, S., {Gilmore}, G., {Magrini}, L., {et~al.} 2022, \aap, 666, A121

\bibitem[{{Rastello} {et~al.}(2020){Rastello}, {Carraro}, \&
  {Capuzzo-Dolcetta}}]{Rastello2020}
{Rastello}, S., {Carraro}, G., \& {Capuzzo-Dolcetta}, R. 2020, \apj, 896, 152

\bibitem[{{Rauer} {et~al.}(2014){Rauer}, {Catala}, {Aerts}, {Appourchaux},
  {Benz}, {Brandeker}, {Christensen-Dalsgaard}, {Deleuil}, {Gizon}, {Goupil},
  {G{\"u}del}, {Janot-Pacheco}, {Mas-Hesse}, {Pagano}, {Piotto}, {Pollacco},
  {Santos}, {Smith}, {Su{\'a}rez}, {Szab{\'o}}, {Udry}, {Adibekyan}, {Alibert},
  {Almenara}, {Amaro-Seoane}, {Eiff}, {Asplund}, {Antonello}, {Barnes},
  {Baudin}, {Belkacem}, {Bergemann}, {Bihain}, {Birch}, {Bonfils}, {Boisse},
  {Bonomo}, {Borsa}, {Brand{\~a}o}, {Brocato}, {Brun}, {Burleigh}, {Burston},
  {Cabrera}, {Cassisi}, {Chaplin}, {Charpinet}, {Chiappini}, {Church},
  {Csizmadia}, {Cunha}, {Damasso}, {Davies}, {Deeg}, {D{\'\i}az}, {Dreizler},
  {Dreyer}, {Eggenberger}, {Ehrenreich}, {Eigm{\"u}ller}, {Erikson}, {Farmer},
  {Feltzing}, {de Oliveira Fialho}, {Figueira}, {Forveille}, {Fridlund},
  {Garc{\'\i}a}, {Giommi}, {Giuffrida}, {Godolt}, {Gomes da Silva}, {Granzer},
  {Grenfell}, {Grotsch-Noels}, {G{\"u}nther}, {Haswell}, {Hatzes},
  {H{\'e}brard}, {Hekker}, {Helled}, {Heng}, {Jenkins}, {Johansen},
  {Khodachenko}, {Kislyakova}, {Kley}, {Kolb}, {Krivova}, {Kupka}, {Lammer},
  {Lanza}, {Lebreton}, {Magrin}, {Marcos-Arenal}, {Marrese}, {Marques},
  {Martins}, {Mathis}, {Mathur}, {Messina}, {Miglio}, {Montalban}, {Montalto},
  {Monteiro}, {Moradi}, {Moravveji}, {Mordasini}, {Morel}, {Mortier},
  {Nascimbeni}, {Nelson}, {Nielsen}, {Noack}, {Norton}, {Ofir}, {Oshagh},
  {Ouazzani}, {P{\'a}pics}, {Parro}, {Petit}, {Plez}, {Poretti}, {Quirrenbach},
  {Ragazzoni}, {Raimondo}, {Rainer}, {Reese}, {Redmer}, {Reffert},
  {Rojas-Ayala}, {Roxburgh}, {Salmon}, {Santerne}, {Schneider}, {Schou},
  {Schuh}, {Schunker}, {Silva-Valio}, {Silvotti}, {Skillen}, {Snellen}, {Sohl},
  {Sousa}, {Sozzetti}, {Stello}, {Strassmeier}, {{\v{S}}vanda}, {Szab{\'o}},
  {Tkachenko}, {Valencia}, {Van Grootel}, {Vauclair}, {Ventura}, {Wagner},
  {Walton}, {Weingrill}, {Werner}, {Wheatley}, \& {Zwintz}}]{Rauer2014}
{Rauer}, H., {Catala}, C., {Aerts}, C., {et~al.} 2014, Experimental Astronomy,
  38, 249

\bibitem[{{Reggiani} {et~al.}(2019){Reggiani}, {Amarsi}, {Lind}, {Barklem},
  {Zatsarinny}, {Bartschat}, {Fursa}, {Bray}, {Spina}, \&
  {Mel{\'e}ndez}}]{reggiani19}
{Reggiani}, H., {Amarsi}, A.~M., {Lind}, K., {et~al.} 2019, \aap, 627, A177

\bibitem[{Rehse \& Ryder(2009)}]{Rehse2009}
Rehse, S.~J. \& Ryder, C.~A. 2009, Spectrochimica Acta - Part B Atomic
  Spectroscopy, 64, 974

\bibitem[{{Reichert} {et~al.}(2020){Reichert}, {Hansen}, {Hanke},
  {Sk{\'u}lad{\'o}ttir}, {Arcones}, \& {Grebel}}]{reichert2020}
{Reichert}, M., {Hansen}, C.~J., {Hanke}, M., {et~al.} 2020, \aap, 641, A127

\bibitem[{{Reichert} {et~al.}(2023){Reichert}, {Obergaulinger}, {Aloy},
  {Gabler}, {Arcones}, \& {Thielemann}}]{Reichert2023}
{Reichert}, M., {Obergaulinger}, M., {Aloy}, M.~{\'A}., {et~al.} 2023, \mnras,
  518, 1557

\bibitem[{{Rein} \& {Liu}(2012)}]{Rein2012}
{Rein}, H. \& {Liu}, S.-F. 2012, \aap, 537, A128

\bibitem[{{Reiners} \& {Mohanty}(2012)}]{reiners2012}
{Reiners}, A. \& {Mohanty}, S. 2012, \apj, 746, 43

\bibitem[{{Reiners} {et~al.}(2014){Reiners}, {Sch{\"u}ssler}, \&
  {Passegger}}]{Reiners2014}
{Reiners}, A., {Sch{\"u}ssler}, M., \& {Passegger}, V.~M. 2014, \apj, 794, 144

\bibitem[{{R{\'e}ville} {et~al.}(2015){R{\'e}ville}, {Brun}, {Matt},
  {Strugarek}, \& {Pinto}}]{reville2015}
{R{\'e}ville}, V., {Brun}, A.~S., {Matt}, S.~P., {Strugarek}, A., \& {Pinto},
  R.~F. 2015, \apj, 798, 116

\bibitem[{Rhodin {et~al.}(2017)Rhodin, Belmonte, Engstrom, Lundberg, Nilsson,
  Hartman, Pickering, Clear, Quinet, Fivet, \& Palmeri}]{Rhodin2017b}
Rhodin, A.~P., Belmonte, M.~T., Engstrom, L., {et~al.} 2017, Monthly Notices of
  the Royal Astronomical Society, 472, 3337

\bibitem[{{Richard} {et~al.}(2005){Richard}, {Michaud}, \&
  {Richer}}]{richard2005}
{Richard}, O., {Michaud}, G., \& {Richer}, J. 2005, \apj, 619, 538

\bibitem[{{Ricker} {et~al.}(2015){Ricker}, {Winn}, {Vanderspek}, {Latham},
  {Bakos}, {Bean}, {Berta-Thompson}, {Brown}, {Buchhave}, {Butler}, {Butler},
  {Chaplin}, {Charbonneau}, {Christensen-Dalsgaard}, {Clampin}, {Deming},
  {Doty}, {De Lee}, {Dressing}, {Dunham}, {Endl}, {Fressin}, {Ge}, {Henning},
  {Holman}, {Howard}, {Ida}, {Jenkins}, {Jernigan}, {Johnson}, {Kaltenegger},
  {Kawai}, {Kjeldsen}, {Laughlin}, {Levine}, {Lin}, {Lissauer}, {MacQueen},
  {Marcy}, {McCullough}, {Morton}, {Narita}, {Paegert}, {Palle}, {Pepe},
  {Pepper}, {Quirrenbach}, {Rinehart}, {Sasselov}, {Sato}, {Seager},
  {Sozzetti}, {Stassun}, {Sullivan}, {Szentgyorgyi}, {Torres}, {Udry}, \&
  {Villasenor}}]{Ricker2015JATIS...1a4003R}
{Ricker}, G.~R., {Winn}, J.~N., {Vanderspek}, R., {et~al.} 2015, Journal of
  Astronomical Telescopes, Instruments, and Systems, 1, 014003

\bibitem[{{Rickman} {et~al.}(2023){Rickman}, {Wajer}, {Przy{\l}uski},
  {Wi{\'s}niowski}, {Nesvorn{\'y}}, \& {Morbidelli}}]{Rickman2023}
{Rickman}, H., {Wajer}, P., {Przy{\l}uski}, R., {et~al.} 2023, \mnras, 520, 637

\bibitem[{{Rigault} {et~al.}(2020){Rigault}, {Brinnel}, {Aldering},
  {Antilogus}, {Aragon}, {Bailey}, {Baltay}, {Barbary}, {Bongard}, {Boone},
  {Buton}, {Childress}, {Chotard}, {Copin}, {Dixon}, {Fagrelius}, {Feindt},
  {Fouchez}, {Gangler}, {Hayden}, {Hillebrandt}, {Howell}, {Kim}, {Kowalski},
  {Kuesters}, {Leget}, {Lombardo}, {Lin}, {Nordin}, {Pain}, {Pecontal},
  {Pereira}, {Perlmutter}, {Rabinowitz}, {Runge}, {Rubin}, {Saunders},
  {Smadja}, {Sofiatti}, {Suzuki}, {Taubenberger}, {Tao}, \&
  {Thomas}}]{rigault2020}
{Rigault}, M., {Brinnel}, V., {Aldering}, G., {et~al.} 2020, \aap, 644, A176

\bibitem[{{Riva} {et~al.}(2026){Riva}, {Bianco}, {Randich}, {Tozzi},
  {Brucalassi}, {Munari}, {Gonzales}, {Harris}, {De Caprio}, {D'Auria},
  {Joven}, {Giro}, \& {Caito}}]{riva26}
{Riva}, M., {Bianco}, A., {Randich}, S., {et~al.} 2026, in VLT Beyond 2030, 54

\bibitem[{{Rizzuti} {et~al.}(2025){Rizzuti}, {Cescutti}, {Molaro}, {Roberti},
  {Chieffi}, {Limongi}, {Magrini}, \& {Matteucci}}]{Rizzuti2025}
{Rizzuti}, F., {Cescutti}, G., {Molaro}, P., {et~al.} 2025, \aap, 698, A118

\bibitem[{{Rizzuto} {et~al.}(2020){Rizzuto}, {Newton}, {Mann}, {Tofflemire},
  {Vanderburg}, {Kraus}, {Wood}, {Quinn}, {Zhou}, {Thao}, {Law}, {Ziegler}, \&
  {Brice{\~n}o}}]{Rizzuto2020AJ....160...33R}
{Rizzuto}, A.~C., {Newton}, E.~R., {Mann}, A.~W., {et~al.} 2020, \aj, 160, 33

\bibitem[{{Roberts} {et~al.}(2010){Roberts}, {Woosley}, \&
  {Hoffman}}]{Roberts2010}
{Roberts}, L.~F., {Woosley}, S.~E., \& {Hoffman}, R.~D. 2010, \apj, 722, 954

\bibitem[{{Rodr{\'\i}guez D{\'\i}az} {et~al.}(2022){Rodr{\'\i}guez D{\'\i}az},
  {Bigot}, {Aguirre B{\o}rsen-Koch}, {Lund}, {R{\o}rsted}, {Kallinger},
  {Sulis}, \& {Mary}}]{rodriguezdiaz2022}
{Rodr{\'\i}guez D{\'\i}az}, L.~F., {Bigot}, L., {Aguirre B{\o}rsen-Koch}, V.,
  {et~al.} 2022, \mnras, 514, 1741

\bibitem[{{Rodr{\'\i}guez D{\'\i}az} {et~al.}(2024){Rodr{\'\i}guez D{\'\i}az},
  {Lagae}, {Amarsi}, {Bigot}, {Zhou}, {Aguirre B{\o}rsen-Koch}, {Lind},
  {Trampedach}, \& {Collet}}]{rodriguezdiaz2024}
{Rodr{\'\i}guez D{\'\i}az}, L.~F., {Lagae}, C., {Amarsi}, A.~M., {et~al.} 2024,
  \aap, 688, A212

\bibitem[{{Roederer}(2017)}]{roederer2017}
{Roederer}, I.~U. 2017, \apj, 835, 23

\bibitem[{{Roederer} {et~al.}(2024){Roederer}, {Beers}, {Hattori}, {Placco},
  {Hansen}, {Ezzeddine}, {Frebel}, {Holmbeck}, \& {Sakari}}]{Roederer2024}
{Roederer}, I.~U., {Beers}, T.~C., {Hattori}, K., {et~al.} 2024, \apj, 971, 158

\bibitem[{{Roederer} {et~al.}(2016){Roederer}, {Karakas}, {Pignatari}, \&
  {Herwig}}]{Roederer2016}
{Roederer}, I.~U., {Karakas}, A.~I., {Pignatari}, M., \& {Herwig}, F. 2016,
  \apj, 821, 37

\bibitem[{{Rogers} {et~al.}(2024){Rogers}, {Debes}, {Anslow}, {Bonsor},
  {Casewell}, {Dos Santos}, {Dufour}, {G{\"a}nsicke}, {Gentile Fusillo},
  {Koester}, {Nielsen}, {Penoyre}, {Rickman}, {Sahlmann}, {Tremblay},
  {Vanderburg}, {Xu}, {Dennihy}, {Farihi}, {Hermes}, {Hodgkin}, {Kilic},
  {Kowalski}, {Sanderson}, \& {Toonen}}]{Rogers2024}
{Rogers}, L.~K., {Debes}, J., {Anslow}, R.~J., {et~al.} 2024, \mnras, 527, 977

\bibitem[{{Romano}(2022)}]{Romano2022}
{Romano}, D. 2022, \aapr, 30, 7

\bibitem[{{Romano} \& {Matteucci}(2003)}]{RomanoMatteucci2003}
{Romano}, D. \& {Matteucci}, F. 2003, \mnras, 342, 185

\bibitem[{{Rossi} {et~al.}(2021){Rossi}, {Salvadori}, \&
  {Sk{\'u}lad{\'o}ttir}}]{rossi2021}
{Rossi}, M., {Salvadori}, S., \& {Sk{\'u}lad{\'o}ttir}, {\'A}. 2021, \mnras,
  503, 6026

\bibitem[{{Rossi} {et~al.}(2025){Rossi}, {Salvadori}, {Sk{\'u}lad{\'o}ttir},
  {Vanni}, \& {Koutsouridou}}]{rossi2025}
{Rossi}, M., {Salvadori}, S., {Sk{\'u}lad{\'o}ttir}, {\'A}., {Vanni}, I., \&
  {Koutsouridou}, I. 2025, \apj, 987, 121

\bibitem[{{Royer} {et~al.}(2024){Royer}, {Merle}, {Dsilva}, {Sekaran}, {Van
  Winckel}, {Fr{\'e}mat}, {Van der Swaelmen}, {Gebruers}, {Tkachenko},
  {Laverick}, {Dirickx}, {Raskin}, {Hensberge}, {Abdul-Masih}, {Acke},
  {Alonso}, {Bandhu Mahato}, {Beck}, {Behara}, {Bloemen}, {Buysschaert}, {Cox},
  {Debosscher}, {De Cat}, {Degroote}, {De Nutte}, {De Smedt}, {de Vries},
  {Dumortier}, {Escorza}, {Exter}, {Goriely}, {Gorlova}, {Hillen}, {Homan},
  {Jorissen}, {Kamath}, {Karjalainen}, {Karjalainen}, {Lampens}, {Lobel},
  {Lombaert}, {Marcos-Arenal}, {Menu}, {Merges}, {Moravveji}, {Nemeth},
  {Neyskens}, {Ostensen}, {P{\'a}pics}, {Perez}, {Prins}, {Royer},
  {Samadi-Ghadim}, {Sana}, {Sans Fuentes}, {Scaringi}, {Schmid}, {Siess},
  {Siopis}, {Smolders}, {S{\'o}dor}, {Thoul}, {Triana}, {Vandenbussche}, {Van
  de Sande}, {Van De Steene}, {Van Eck}, {van Hoof}, {Van Marle}, {Van Reeth},
  {Vermeylen}, {Volpi}, {Vos}, \& {Waelkens}}]{Royer2024A&A...681A.107R}
{Royer}, P., {Merle}, T., {Dsilva}, K., {et~al.} 2024, \aap, 681, A107

\bibitem[{{Ruchti} {et~al.}(2013){Ruchti}, {Bergemann}, {Serenelli},
  {Casagrande}, \& {Lind}}]{rutchi2013MNRAS.429..126R}
{Ruchti}, G.~R., {Bergemann}, M., {Serenelli}, A., {Casagrande}, L., \& {Lind},
  K. 2013, \mnras, 429, 126

\bibitem[{Ruffoni {et~al.}(2014)Ruffoni, Hartog, Lawler, Brewer, Lind, Nave, \&
  Pickering}]{Ruffoni2014}
Ruffoni, M.~P., Hartog, E. A.~D., Lawler, J.~E., {et~al.} 2014, Monthly Notices
  of the Royal Astronomical Society, 3127

\bibitem[{{Rybicki} \& {Hummer}(1992)}]{rybicki1992}
{Rybicki}, G.~B. \& {Hummer}, D.~G. 1992, \aap, 262, 209

\bibitem[{{Sackmann} \& {Boothroyd}(1992)}]{sackmann92LSb}
{Sackmann}, I.-J. \& {Boothroyd}, A.~I. 1992, \apjl, 392, L71

\bibitem[{{Saffe} {et~al.}(2017){Saffe}, {Jofr{\'e}}, {Martioli}, {Flores},
  {Petrucci}, \& {Jaque Arancibia}}]{saffe2017}
{Saffe}, C., {Jofr{\'e}}, E., {Martioli}, E., {et~al.} 2017, \aap, 604, L4

\bibitem[{{Sairam} {et~al.}(2024){Sairam}, {Triaud}, {Baycroft}, {Orosz},
  {Boisse}, {Heidari}, {Sebastian}, {Dransfield}, {Martin}, {Santerne}, \&
  {Standing}}]{Sairam2024}
{Sairam}, L., {Triaud}, A. H.~M.~J., {Baycroft}, T.~A., {et~al.} 2024, \mnras,
  527, 2261

\bibitem[{{Salpeter}(1952)}]{Salpeter1952}
{Salpeter}, E.~E. 1952, \apj, 115, 326

\bibitem[{{Salvadori} {et~al.}(2019){Salvadori}, {Bonifacio}, {Caffau},
  {Korotin}, {Andreevsky}, {Spite}, \& {Sk{\'u}lad{\'o}ttir}}]{salvadori2019}
{Salvadori}, S., {Bonifacio}, P., {Caffau}, E., {et~al.} 2019, \mnras, 487,
  4261

\bibitem[{{Salvadori} {et~al.}(2015){Salvadori}, {Sk{\'u}lad{\'o}ttir}, \&
  {Tolstoy}}]{salvadori2015}
{Salvadori}, S., {Sk{\'u}lad{\'o}ttir}, {\'A}., \& {Tolstoy}, E. 2015, \mnras,
  454, 1320

\bibitem[{{Saracino} {et~al.}(2020){Saracino}, {Kamann}, {Usher}, {Bastian},
  {Martocchia}, {Lardo}, {Latour}, {Cabrera-Ziri}, {Dreizler}, {Giesers},
  {Husser}, {Kacharov}, \& {Salaris}}]{Saracino2020}
{Saracino}, S., {Kamann}, S., {Usher}, C., {et~al.} 2020, \mnras, 498, 4472

\bibitem[{Sarma {et~al.}(2025)Sarma, Belmonte, Llorente, \& Mar}]{Sarma2025}
Sarma, P. R.~S., Belmonte, M.~T., Llorente, S., \& Mar, S. 2025, Spectrochimica
  Acta Part B: Atomic Spectroscopy, 107342

\bibitem[{Sarma {et~al.}(2024)Sarma, Belmonte, \& Mar}]{SenSarma2024}
Sarma, P. R.~S., Belmonte, M.~T., \& Mar, S. 2024, European Physical Journal D,
  78

\bibitem[{{Sato} {et~al.}(2007){Sato}, {Izumiura}, {Toyota}, {Kambe}, {Takeda},
  {Masuda}, {Omiya}, {Murata}, {Itoh}, {Ando}, {Yoshida}, {Ikoma}, {Kokubo}, \&
  {Ida}}]{sato2007}
{Sato}, B., {Izumiura}, H., {Toyota}, E., {et~al.} 2007, \apj, 661, 527

\bibitem[{{Saunders} {et~al.}(2025){Saunders}, {Grunblatt}, {Huber}, {Ong},
  {Schlaufman}, {Hey}, {Li}, {Butler}, {Crane}, {Shectman}, {Teske}, {Quinn},
  {Yee}, {Brahm}, {Trifonov}, {Jord{\'a}n}, {Henning}, {Sing}, {MacGregor},
  {Clark}, {Littlefield}, {Deveny}, {Howell}, {Page}, {Rapetti}, {Falk},
  {Levine}, {Huang}, {Lund}, {Ricker}, {Seager}, {Winn}, \&
  {Jenkins}}]{Saunders2025}
{Saunders}, N., {Grunblatt}, S.~K., {Huber}, D., {et~al.} 2025, \aj, 169, 75

\bibitem[{{Savvidou} \& {Bitsch}(2023)}]{Savvidou2023}
{Savvidou}, S. \& {Bitsch}, B. 2023, \aap, 679, A42

\bibitem[{{Sayeed} {et~al.}(2024){Sayeed}, {Ness}, {Montet}, {Cantiello},
  {Casey}, {Buder}, {Bedell}, {Breivik}, {Metzger}, {Martell}, \&
  {McGee-Gold}}]{sayeed24lsb}
{Sayeed}, M., {Ness}, M.~K., {Montet}, B.~T., {et~al.} 2024, \apj, 964, 42

\bibitem[{{Sbordone}(2005)}]{sbordone05LSb}
{Sbordone}, L. 2005, Memorie della Societa Astronomica Italiana Supplementi, 8,
  61

\bibitem[{{Sbordone} {et~al.}(2004){Sbordone}, {Bonifacio}, {Castelli}, \&
  {Kurucz}}]{sbordone04LSb}
{Sbordone}, L., {Bonifacio}, P., {Castelli}, F., \& {Kurucz}, R.~L. 2004,
  Memorie della Societa Astronomica Italiana Supplementi, 5, 93

\bibitem[{{Sbordone} {et~al.}(2009){Sbordone}, {Limongi}, {Chieffi}, {Caffau},
  {Ludwig}, \& {Bonifacio}}]{sbordone2009}
{Sbordone}, L., {Limongi}, M., {Chieffi}, A., {et~al.} 2009, \aap, 503, 121

\bibitem[{{Schatz} {et~al.}(2022){Schatz}, {Becerril Reyes}, {Best}, {Brown},
  {Chatziioannou}, {Chipps}, {Deibel}, {Ezzeddine}, {Galloway}, {Hansen},
  {Herwig}, {Ji}, {Lugaro}, {Meisel}, {Norman}, {Read}, {Roberts}, {Spyrou},
  {Tews}, {Timmes}, {Travaglio}, {Vassh}, {Abia}, {Adsley}, {Agarwal},
  {Aliotta}, {Aoki}, {Arcones}, {Aryan}, {Bandyopadhyay}, {Banu}, {Bardayan},
  {Barnes}, {Bauswein}, {Beers}, {Bishop}, {Boztepe}, {C{\^o}t{\'e}}, {Caplan},
  {Champagne}, {Clark}, {Couder}, {Couture}, {de Mink}, {Debnath}, {deBoer},
  {den Hartogh}, {Denissenkov}, {Dexheimer}, {Dillmann}, {Escher}, {Famiano},
  {Farmer}, {Fisher}, {Fr{\"o}hlich}, {Frebel}, {Fryer}, {Fuller}, {Ganguly},
  {Ghosh}, {Gibson}, {Gorda}, {Gourgouliatos}, {Graber}, {Gupta}, {Haxton},
  {Heger}, {Hix}, {Ho}, {Holmbeck}, {Hood}, {Huth}, {Imbriani}, {Izzard},
  {Jain}, {Jayatissa}, {Johnston}, {Kajino}, {Kankainen}, {Kiss},
  {Kwiatkowski}, {La Cognata}, {Laird}, {Lamia}, {Landry}, {Laplace}, {Launey},
  {Leahy}, {Leckenby}, {Lennarz}, {Longfellow}, {Lovell}, {Lynch}, {Lyons},
  {Maeda}, {Masha}, {Matei}, {Merc}, {Messer}, {Montes}, {Mukherjee},
  {Mumpower}, {Neto}, {Nevins}, {Newton}, {Nguyen}, {Nishikawa}, {Nishimura},
  {Nunes}, {O'Connor}, {O'Shea}, {Ong}, {Pain}, {Pajkos}, {Pignatari},
  {Pizzone}, {Placco}, {Plewa}, {Pritychenko}, {Psaltis}, {Puentes}, {Qian},
  {Radice}, {Rapagnani}, {Rebeiro}, {Reifarth}, {Richard}, {Rijal}, {Roederer},
  {Rojo}, {J S K}, {Saito}, {Schwenk}, {Sergi}, {Sidhu}, {Simon}, {Sivarani},
  {Sk{\'u}lad{\'o}ttir}, {Smith}, {Spiridon}, {Sprouse}, {Starrfield},
  {Steiner}, {Strieder}, {Sultana}, {Surman}, {Sz{\"u}cs}, {Tawfik},
  {Thielemann}, {Trache}, {Trappitsch}, {Tsang}, {Tumino}, {Upadhyayula},
  {Valle Mart{\'\i}nez}, {Van der Swaelmen}, {Viscasillas V{\'a}zquez},
  {Watts}, {Wehmeyer}, {Wiescher}, {Wrede}, {Yoon}, {Zegers}, {Zermane}, \&
  {Zingale}}]{schatz2022}
{Schatz}, H., {Becerril Reyes}, A.~D., {Best}, A., {et~al.} 2022, Journal of
  Physics G Nuclear Physics, 49, 110502

\bibitem[{{Schiappacasse-Ulloa} {et~al.}(2022){Schiappacasse-Ulloa},
  {Lucatello}, {Rain}, \& {Pietrinferni}}]{Schiappacasse-Ulloa2022}
{Schiappacasse-Ulloa}, J., {Lucatello}, S., {Rain}, M.~J., \& {Pietrinferni},
  A. 2022, \mnras, 511, 231

\bibitem[{{Schlegel} {et~al.}(1998){Schlegel}, {Finkbeiner}, \&
  {Davis}}]{Schlegel1998}
{Schlegel}, D.~J., {Finkbeiner}, D.~P., \& {Davis}, M. 1998, \apj, 500, 525

\bibitem[{{Schmidt} \& {Schlaufman}(2026)}]{Schmidt2026AJ....171..157S}
{Schmidt}, S.~P. \& {Schlaufman}, K.~C. 2026, \aj, 171, 157

\bibitem[{{Schmidt-May} {et~al.}(2024){Schmidt-May}, {Barklem}, {Grumer},
  {Amarsi}, {Bj{\"o}rkhage}, {Blom}, {Dochain}, {Ji}, {Martini}, {Reinhed},
  {Ros{\'e}n}, {Simonsson}, {Zettergren}, {Cederquist}, \&
  {Schmidt}}]{schmidt-may24}
{Schmidt-May}, A.~F., {Barklem}, P.~S., {Grumer}, J., {et~al.} 2024, \pra, 109,
  052820

\bibitem[{{Schneider}(2015)}]{schneider2015}
{Schneider}, A. 2015, \mnras, 451, 3117

\bibitem[{{Schofield} {et~al.}(2022){Schofield}, {Pignatari}, {Stancliffe}, \&
  {Hoppe}}]{Schofield2022}
{Schofield}, J., {Pignatari}, M., {Stancliffe}, R.~J., \& {Hoppe}, P. 2022,
  \mnras, 517, 1803

\bibitem[{{Schuessler} \& {Solanki}(1992)}]{schuessler1992}
{Schuessler}, M. \& {Solanki}, S.~K. 1992, \aap, 264, L13

\bibitem[{{See} {et~al.}(2018){See}, {Jardine}, {Vidotto}, {Donati}, {Boro
  Saikia}, {Fares}, {Folsom}, {Jeffers}, {Marsden}, {Morin}, {Petit}, \& {BCool
  Collaboration}}]{see2018}
{See}, V., {Jardine}, M., {Vidotto}, A.~A., {et~al.} 2018, \mnras, 474, 536

\bibitem[{{Shah} {et~al.}(2026{\natexlab{a}}){Shah}, {Ezzeddine}, {Holmbeck},
  {Ji}, {Placco}, {Roederer}, {Mardini}, {Usman}, {Bandyopadhyay}, {Beers},
  {Frebel}, {Hansen}, {Sakari}, \& {Sneden}}]{shivani2026}
{Shah}, S.~P., {Ezzeddine}, R., {Holmbeck}, E.~M., {et~al.} 2026{\natexlab{a}},
  arXiv e-prints, arXiv:2604.12892

\bibitem[{{Shah} {et~al.}(2026{\natexlab{b}}){Shah}, {Ezzeddine}, {Holmbeck},
  {Ji}, {Placco}, {Roederer}, {Mardini}, {Usman}, {Bandyopadhyay}, {Beers},
  {Frebel}, {Hansen}, {Sakari}, \& {Sneden}}]{Shah2026arXiv260412892S}
{Shah}, S.~P., {Ezzeddine}, R., {Holmbeck}, E.~M., {et~al.} 2026{\natexlab{b}},
  arXiv e-prints, arXiv:2604.12892

\bibitem[{{Siegel} {et~al.}(2019){Siegel}, {Barnes}, \& {Metzger}}]{Siegel2019}
{Siegel}, D.~M., {Barnes}, J., \& {Metzger}, B.~D. 2019, \nat, 569, 241

\bibitem[{Silva {et~al.}(2022)Silva, Sampaio, Amaro, FlÃ¶rs,
  MartÃ­nez-Pinedo, \& Marques}]{Silva2022NdIII}
Silva, R.~F., Sampaio, J.~M., Amaro, P., {et~al.} 2022, Atoms, 10, 18

\bibitem[{{Simon}(2019)}]{simon2019}
{Simon}, J.~D. 2019, \araa, 57, 375

\bibitem[{{Singh} {et~al.}(2021){Singh}, {Reddy}, {Campbell}, {Kumar}, \&
  {Vrard}}]{Singh2021}
{Singh}, R., {Reddy}, B.~E., {Campbell}, S.~W., {Kumar}, Y.~B., \& {Vrard}, M.
  2021, \apjl, 913, L4

\bibitem[{{Sirono} \& {Turrini}(2025)}]{Sirono2025}
{Sirono}, S.-i. \& {Turrini}, D. 2025, Scientific Reports, 15, 30919

\bibitem[{{Sitnova} {et~al.}(2025){Sitnova}, {Lombardo}, {Mashonkina},
  {Rizzuti}, {Cescutti}, {Hansen}, {Bonifacio}, {Caffau}, {Koch-Hansen},
  {Meynet}, \& {Fernandes de Melo}}]{Sitnova2025a}
{Sitnova}, T.~M., {Lombardo}, L., {Mashonkina}, L.~I., {et~al.} 2025, \aap,
  699, A262

\bibitem[{{Sitnova} {et~al.}(2022){Sitnova}, {Yakovleva}, {Belyaev}, \&
  {Mashonkina}}]{sitnova2022}
{Sitnova}, T.~M., {Yakovleva}, S.~A., {Belyaev}, A.~K., \& {Mashonkina}, L.~I.
  2022, \mnras, 515, 1510

\bibitem[{{Sk{\'u}lad{\'o}ttir} {et~al.}(2015){Sk{\'u}lad{\'o}ttir},
  {Andrievsky}, {Tolstoy}, {Hill}, {Salvadori}, {Korotin}, \&
  {Pettini}}]{skuladottir2015sulphur}
{Sk{\'u}lad{\'o}ttir}, {\'A}., {Andrievsky}, S.~M., {Tolstoy}, E., {et~al.}
  2015, \aap, 580, A129

\bibitem[{{Sk{\'u}lad{\'o}ttir} {et~al.}(2025){Sk{\'u}lad{\'o}ttir},
  {Ernandes}, {Feuillet}, {Mori}, {Feltzing}, {Lucchesi}, \& {Di
  Matteo}}]{skuladottir2025}
{Sk{\'u}lad{\'o}ttir}, {\'A}., {Ernandes}, H., {Feuillet}, D.~K., {et~al.}
  2025, \apjl, 986, L21

\bibitem[{{Sk{\'u}lad{\'o}ttir} {et~al.}(2019){Sk{\'u}lad{\'o}ttir}, {Hansen},
  {Salvadori}, \& {Choplin}}]{skuladottir2019}
{Sk{\'u}lad{\'o}ttir}, {\'A}., {Hansen}, C.~J., {Salvadori}, S., \& {Choplin},
  A. 2019, \aap, 631, A171

\bibitem[{{Sk{\'u}lad{\'o}ttir}
  {et~al.}(2024{\natexlab{a}}){Sk{\'u}lad{\'o}ttir}, {Koutsouridou}, {Vanni},
  {Amarsi}, {Lucchesi}, {Salvadori}, \& {Aguado}}]{skuladottir2024b}
{Sk{\'u}lad{\'o}ttir}, {\'A}., {Koutsouridou}, I., {Vanni}, I., {et~al.}
  2024{\natexlab{a}}, \apjl, 968, L23

\bibitem[{{Sk{\'u}lad{\'o}ttir} {et~al.}(2023){Sk{\'u}lad{\'o}ttir}, {Puls},
  {Amarsi}, {Battaglia}, {Buder}, {Campbell}, {Cardona-Barrero}, {Christlieb},
  {Feuillet}, {Gelli}, {Hansen}, {Hill}, {Ibata}, {Jablonka}, {Kacharov},
  {Karakas}, {Koch-Hansen}, {Lind}, {Lombardo}, {Lucchesi}, {Lugaro}, {Martin},
  {Massari}, {Nordlander}, {Reichert}, {Rossi}, {Ruiter}, {Salvadori},
  {Seitenzahl}, {Tolstoy}, {Xylakis-Dornbusch}, \& {Youakim}}]{skuladottir2023}
{Sk{\'u}lad{\'o}ttir}, {\'A}., {Puls}, A.~A., {Amarsi}, A.~M., {et~al.} 2023,
  The Messenger, 190, 19

\bibitem[{{Sk{\'u}lad{\'o}ttir} \&
  {Salvadori}(2020)}]{skuladottirsalvadori2020}
{Sk{\'u}lad{\'o}ttir}, {\'A}. \& {Salvadori}, S. 2020, \aap, 634, L2

\bibitem[{{Sk{\'u}lad{\'o}ttir}
  {et~al.}(2024{\natexlab{b}}){Sk{\'u}lad{\'o}ttir}, {Vanni}, {Salvadori}, \&
  {Lucchesi}}]{skuladottir2024}
{Sk{\'u}lad{\'o}ttir}, {\'A}., {Vanni}, I., {Salvadori}, S., \& {Lucchesi}, R.
  2024{\natexlab{b}}, \aap, 681, A44

\bibitem[{{Skumanich}(1972)}]{skumanich1972}
{Skumanich}, A. 1972, \apj, 171, 565

\bibitem[{{Smiljanic} {et~al.}(2009){Smiljanic}, {Gauderon}, {North}, {Barbuy},
  {Charbonnel}, \& {Mowlavi}}]{Smiljanic2009}
{Smiljanic}, R., {Gauderon}, R., {North}, P., {et~al.} 2009, \aap, 502, 267

\bibitem[{{Sneden} {et~al.}(2008){Sneden}, {Cowan}, \& {Gallino}}]{Sneden-2008}
{Sneden}, C., {Cowan}, J.~J., \& {Gallino}, R. 2008, \araa, 46, 241

\bibitem[{{Sneden} {et~al.}(1996){Sneden}, {McWilliam}, {Preston}, {Cowan},
  {Burris}, \& {Armosky}}]{Sneden1996}
{Sneden}, C., {McWilliam}, A., {Preston}, G.~W., {et~al.} 1996, \apj, 467, 819

\bibitem[{{Solanki} \& {Unruh}(2004)}]{solanki2004}
{Solanki}, S.~K. \& {Unruh}, Y.~C. 2004, \mnras, 348, 307

\bibitem[{{Sollima}(2020)}]{Sollima2020}
{Sollima}, A. 2020, \mnras, 495, 2222

\bibitem[{{Sollima} {et~al.}(2019){Sollima}, {Baumgardt}, \&
  {Hilker}}]{Sollima2019}
{Sollima}, A., {Baumgardt}, H., \& {Hilker}, M. 2019, \mnras, 485, 1460

\bibitem[{{Souto} {et~al.}(2019){Souto}, {Allende Prieto}, {Cunha},
  {Pinsonneault}, {Smith}, {Garcia-Dias}, {Bovy}, {Garc{\'\i}a-Hern{\'a}ndez},
  {Holtzman}, {Johnson}, {J{\"o}nsson}, {Majewski}, {Shetrone}, {Sobeck},
  {Zamora}, {Pan}, \& {Nitschelm}}]{Souto2019}
{Souto}, D., {Allende Prieto}, C., {Cunha}, K., {et~al.} 2019, \apj, 874, 97

\bibitem[{{Sozzetti} {et~al.}(2006){Sozzetti}, {Torres}, {Latham}, {Carney},
  {Stefanik}, {Boss}, {Laird}, \& {Korzennik}}]{sozzetti2006}
{Sozzetti}, A., {Torres}, G., {Latham}, D.~W., {et~al.} 2006, \apj, 649, 428

\bibitem[{{Spencer} {et~al.}(2018){Spencer}, {Mateo}, {Olszewski}, {Walker},
  {McConnachie}, \& {Kirby}}]{spencer2018}
{Spencer}, M.~E., {Mateo}, M., {Olszewski}, E.~W., {et~al.} 2018, \aj, 156, 257

\bibitem[{{Spina} {et~al.}(2020){Spina}, {Nordlander}, {Casey}, {Bedell},
  {D'Orazi}, {Mel{\'e}ndez}, {Karakas}, {Desidera}, {Baratella}, {Yana
  Galarza}, \& {Casali}}]{spina2020}
{Spina}, L., {Nordlander}, T., {Casey}, A.~R., {et~al.} 2020, \apj, 895, 52

\bibitem[{{Spina} {et~al.}(2021){Spina}, {Sharma}, {Mel{\'e}ndez}, {Bedell},
  {Casey}, {Carlos}, {Franciosini}, \& {Vallenari}}]{Spina2021NatAs...5.1163S}
{Spina}, L., {Sharma}, P., {Mel{\'e}ndez}, J., {et~al.} 2021, Nature Astronomy,
  5, 1163

\bibitem[{Standing {et~al.}(2026)Standing, Barnes, Haswell, Stevenson, Faria,
  Quintin, Ross, Fossati, Jenkins, Alves, \& Staab}]{Standing2026}
Standing, M.~R., Barnes, J.~R., Haswell, C.~A., {et~al.} 2026, Monthly Notices
  of the Royal Astronomical Society, 547

\bibitem[{{Standing} {et~al.}(2023){Standing}, {Sairam}, {Martin}, {Triaud},
  {Correia}, {Coleman}, {Baycroft}, {Kunovac}, {Boisse}, {Cameron},
  {Dransfield}, {Faria}, {Gillon}, {Hara}, {Hellier}, {Howard}, {Lane},
  {Mardling}, {Maxted}, {Miller}, {Nelson}, {Orosz}, {Pepe}, {Santerne},
  {Sebastian}, {Udry}, \& {Welsh}}]{Standing2023}
{Standing}, M.~R., {Sairam}, L., {Martin}, D.~V., {et~al.} 2023, Nature
  Astronomy, 7, 702

\bibitem[{{Standing} {et~al.}(2022){Standing}, {Triaud}, {Faria}, {Martin},
  {Boisse}, {Correia}, {Deleuil}, {Dransfield}, {Gillon}, {H{\'e}brard},
  {Hellier}, {Kunovac}, {Maxted}, {Mardling}, {Santerne}, {Sairam}, \&
  {Udry}}]{Standing2022}
{Standing}, M.~R., {Triaud}, A. H.~M.~J., {Faria}, J.~P., {et~al.} 2022,
  \mnras, 511, 3571

\bibitem[{{Stauffer} {et~al.}(1994){Stauffer}, {Caillault}, {Gagne}, {Prosser},
  \& {Hartmann}}]{stauffer1994}
{Stauffer}, J.~R., {Caillault}, J.-P., {Gagne}, M., {Prosser}, C.~F., \&
  {Hartmann}, L.~W. 1994, \apjs, 91, 625

\bibitem[{{Stein} {et~al.}(2024){Stein}, {Nordlund}, {Collet}, \&
  {Trampedach}}]{stein2024}
{Stein}, R.~F., {Nordlund}, {\r{A}}., {Collet}, R., \& {Trampedach}, R. 2024,
  \apj, 970, 24

\bibitem[{{Steinmetz} {et~al.}(2020){Steinmetz}, {Matijevi{\v{c}}}, {Enke},
  {Zwitter}, {Guiglion}, {McMillan}, {Kordopatis}, {Valentini}, {Chiappini},
  {Casagrande}, {Wojno}, {Anguiano}, {Bienaym{\'e}}, {Bijaoui}, {Binney},
  {Burton}, {Cass}, {de Laverny}, {Fiegert}, {Freeman}, {Fulbright}, {Gibson},
  {Gilmore}, {Grebel}, {Helmi}, {Kunder}, {Munari}, {Navarro}, {Parker},
  {Ruchti}, {Recio-Blanco}, {Reid}, {Seabroke}, {Siviero}, {Siebert}, {Stupar},
  {Watson}, {Williams}, {Wyse}, {Anders}, {Antoja}, {Birko}, {Bland-Hawthorn},
  {Bossini}, {Garc{\'\i}a}, {Carrillo}, {Chaplin}, {Elsworth}, {Famaey},
  {Gerhard}, {Jofre}, {Just}, {Mathur}, {Miglio}, {Minchev}, {Monari},
  {Mosser}, {Ritter}, {Rodrigues}, {Scholz}, {Sharma}, {Sysoliatina}, \& {RAVE
  Collaboration}}]{Steinmetz2020}
{Steinmetz}, M., {Matijevi{\v{c}}}, G., {Enke}, H., {et~al.} 2020, \aj, 160, 82

\bibitem[{{Storm} {et~al.}(2024){Storm}, {Barklem}, {Yakovleva}, {Belyaev},
  {Palmeri}, {Quinet}, {Lodders}, {Bergemann}, \& {Hoppe}}]{storm2024}
{Storm}, N., {Barklem}, P.~S., {Yakovleva}, S.~A., {et~al.} 2024, \aap, 683,
  A200

\bibitem[{{Storm} {et~al.}(2025){Storm}, {Bergemann}, {Eitner}, {Hoppe},
  {Kemp}, {Ruiter}, {Janka}, {Sieverding}, {de Mink}, {Seitenzahl}, \&
  {Owusu}}]{Storm2025}
{Storm}, N., {Bergemann}, M., {Eitner}, P., {et~al.} 2025, \mnras, 538, 3284

\bibitem[{{Strassmeier}(2002)}]{strassmeier2002}
{Strassmeier}, K.~G. 2002, Astronomische Nachrichten, 323, 309

\bibitem[{{Strassmeier}(2009)}]{strassmeier2009}
{Strassmeier}, K.~G. 2009, \aapr, 17, 251

\bibitem[{{Su{\'a}rez-Andr{\'e}s} {et~al.}(2016){Su{\'a}rez-Andr{\'e}s},
  {Israelian}, {Gonz{\'a}lez Hern{\'a}ndez}, {Adibekyan}, {Delgado Mena},
  {Santos}, \& {Sousa}}]{suarezandres2016}
{Su{\'a}rez-Andr{\'e}s}, L., {Israelian}, G., {Gonz{\'a}lez Hern{\'a}ndez},
  J.~I., {et~al.} 2016, \aap, 591, A69

\bibitem[{{Su{\'a}rez-Andr{\'e}s} {et~al.}(2017){Su{\'a}rez-Andr{\'e}s},
  {Israelian}, {Gonz{\'a}lez Hern{\'a}ndez}, {Adibekyan}, {Delgado Mena},
  {Santos}, \& {Sousa}}]{suarezandres2017}
{Su{\'a}rez-Andr{\'e}s}, L., {Israelian}, G., {Gonz{\'a}lez Hern{\'a}ndez},
  J.~I., {et~al.} 2017, \aap, 599, A96

\bibitem[{{Tabone} {et~al.}(2022){Tabone}, {Rosotti}, {Cridland}, {Armitage},
  \& {Lodato}}]{tabone2022a}
{Tabone}, B., {Rosotti}, G.~P., {Cridland}, A.~J., {Armitage}, P.~J., \&
  {Lodato}, G. 2022, \mnras, 512, 2290

\bibitem[{{Takahashi} {et~al.}(2018){Takahashi}, {Yoshida}, \&
  {Umeda}}]{takahashi2018}
{Takahashi}, K., {Yoshida}, T., \& {Umeda}, H. 2018, \apj, 857, 111

\bibitem[{{Takeda}(1991)}]{takeda1991}
{Takeda}, Y. 1991, \aap, 242, 455

\bibitem[{{Takeda} \& {Takada-Hidai}(2011)}]{takeda2011}
{Takeda}, Y. \& {Takada-Hidai}, M. 2011, \pasj, 63, 537

\bibitem[{Tanaka {et~al.}(2020)Tanaka, Kato, Gaigalas, \&
  Kawaguchi}]{Tanaka2020}
Tanaka, M., Kato, D., Gaigalas, G., \& Kawaguchi, K. 2020, Monthly Notices of
  the Royal Astronomical Society, 496, 1369

\bibitem[{Tanaka {et~al.}(2018)Tanaka, Kato, Gaigalas, Rynkun,
  RadÅ¾iÅ«tÄ—, Wanajo, Sekiguchi, Nakamura, Tanuma, Murakami, \&
  Sakaue}]{Tanaka2018}
Tanaka, M., Kato, D., Gaigalas, G., {et~al.} 2018, The Astrophysical Journal,
  852, 109

\bibitem[{{Tanvir} {et~al.}(2017){Tanvir}, {Levan},
  {Gonz{\'a}lez-Fern{\'a}ndez}, {Korobkin}, {Mandel}, {Rosswog}, {Hjorth},
  {D'Avanzo}, {Fruchter}, {Fryer}, {Kangas}, {Milvang-Jensen}, {Rosetti},
  {Steeghs}, {Wollaeger}, {Cano}, {Copperwheat}, {Covino}, {D'Elia}, {de Ugarte
  Postigo}, {Evans}, {Even}, {Fairhurst}, {Figuera Jaimes}, {Fontes}, {Fujii},
  {Fynbo}, {Gompertz}, {Greiner}, {Hodosan}, {Irwin}, {Jakobsson},
  {J{\o}rgensen}, {Kann}, {Lyman}, {Malesani}, {McMahon}, {Melandri},
  {O'Brien}, {Osborne}, {Palazzi}, {Perley}, {Pian}, {Piranomonte}, {Rabus},
  {Rol}, {Rowlinson}, {Schulze}, {Sutton}, {Th{\"o}ne}, {Ulaczyk}, {Watson},
  {Wiersema}, \& {Wijers}}]{Tanvir2017}
{Tanvir}, N.~R., {Levan}, A.~J., {Gonz{\'a}lez-Fern{\'a}ndez}, C., {et~al.}
  2017, \apjl, 848, L27

\bibitem[{{Tassis} {et~al.}(2012){Tassis}, {Gnedin}, \&
  {Kravtsov}}]{tassis2012}
{Tassis}, K., {Gnedin}, N.~Y., \& {Kravtsov}, A.~V. 2012, \apj, 745, 68

\bibitem[{{Tatischeff} \& {Gabici}(2018)}]{TatischeffGabici2018}
{Tatischeff}, V. \& {Gabici}, S. 2018, Annual Review of Nuclear and Particle
  Science, 68, 377

\bibitem[{{Temple} {et~al.}(2019){Temple}, {Hellier}, {Anderson}, {Barkaoui},
  {Bouchy}, {Brown}, {Burdanov}, {Collier Cameron}, {Delrez}, {Ducrot},
  {Evans}, {Gillon}, {Jehin}, {Lendl}, {Maxted}, {McCormac}, {Murray},
  {Nielsen}, {Pepe}, {Pollacco}, {Queloz}, {S{\'e}gransan}, {Smalley},
  {Thompson}, {Triaud}, {Turner}, {Udry}, {West}, \& {Zouhair}}]{temple2019}
{Temple}, L.~Y., {Hellier}, C., {Anderson}, D.~R., {et~al.} 2019, \mnras, 490,
  2467

\bibitem[{{Temple} {et~al.}(2024){Temple}, {Rankine}, {Banerji}, {Hennawi},
  {Hewett}, {Matthews}, {Nanni}, {Ricci}, \&
  {Richards}}]{Temple2024MNRAS.532..424T}
{Temple}, M.~J., {Rankine}, A.~L., {Banerji}, M., {et~al.} 2024, \mnras, 532,
  424

\bibitem[{{Testi} {et~al.}(2022){Testi}, {Natta}, {Manara}, {de Gregorio
  Monsalvo}, {Lodato}, {Lopez}, {Muzic}, {Pascucci}, {Sanchis}, {Miranda},
  {Scholz}, {De Simone}, \& {Williams}}]{Testi2022}
{Testi}, L., {Natta}, A., {Manara}, C.~F., {et~al.} 2022, \aap, 663, A98

\bibitem[{{Thibodeaux} {et~al.}(2024){Thibodeaux}, {Ji}, {Cerny}, {Kirby}, \&
  {Simon}}]{thibodeux2024}
{Thibodeaux}, P., {Ji}, A.~P., {Cerny}, W., {Kirby}, E.~N., \& {Simon}, J.~D.
  2024, The Open Journal of Astrophysics, 7, 66

\bibitem[{{Thompson}(1975)}]{Thompson1975ApL....16....3T}
{Thompson}, R.~I. 1975, \aplett, 16, 3

\bibitem[{{Thygesen} {et~al.}(2017){Thygesen}, {Kirby}, {Gallagher}, {Ludwig},
  {Caffau}, {Bonifacio}, \& {Sbordone}}]{Thygesen2017}
{Thygesen}, A.~O., {Kirby}, E.~N., {Gallagher}, A.~J., {et~al.} 2017, \apj,
  843, 144

\bibitem[{{Thygesen} {et~al.}(2016){Thygesen}, {Sbordone}, {Ludwig}, {Ventura},
  {Yong}, {Collet}, {Christlieb}, {Melendez}, \& {Zaggia}}]{Thygesen2016}
{Thygesen}, A.~O., {Sbordone}, L., {Ludwig}, H.-G., {et~al.} 2016, \aap, 588,
  A66

\bibitem[{{Tiongco} {et~al.}(2019){Tiongco}, {Vesperini}, \&
  {Varri}}]{Tiongco2019}
{Tiongco}, M.~A., {Vesperini}, E., \& {Varri}, A.~L. 2019, \mnras, 487, 5535

\bibitem[{{Tolstoy}(2019)}]{Tolstoy2019vltt.confE...8T}
{Tolstoy}, E. 2019, in The Very Large Telescope in 2030, 8

\bibitem[{{Tomasetti} {et~al.}(2026){Tomasetti}, {Chiappini}, {Nepal},
  {Moresco}, {Lardo}, {Cimatti}, {Anders}, {Queiroz}, \&
  {Limberg}}]{Tomasetti2026}
{Tomasetti}, E., {Chiappini}, C., {Nepal}, S., {et~al.} 2026, \aap, 707, A111

\bibitem[{Townley-Smith {et~al.}(2016)Townley-Smith, Nave, Pickering, \&
  Blackwell-Whitehead}]{Townley-Smith2016}
Townley-Smith, K., Nave, G., Pickering, J.~C., \& Blackwell-Whitehead, R.~J.
  2016, Monthly Notices of the Royal Astronomical Society, 461, 73

\bibitem[{{Travaglio} {et~al.}(2001){Travaglio}, {Gallino}, {Busso}, \&
  {Gratton}}]{Travaglio2001lead}
{Travaglio}, C., {Gallino}, R., {Busso}, M., \& {Gratton}, R. 2001, \apj, 549,
  346

\bibitem[{{Tremblay} {et~al.}(2024){Tremblay}, {B{\'e}dard}, {O'Brien},
  {Munday}, {Elms}, {Gentillo Fusillo}, \& {Sahu}}]{Tremblay2024}
{Tremblay}, P.-E., {B{\'e}dard}, A., {O'Brien}, M.~W., {et~al.} 2024, \nar, 99,
  101705

\bibitem[{{Trippe} {et~al.}(2010){Trippe}, {Davies}, {Eisenhauer}, {F{\"o}rster
  Schreiber}, {Fritz}, \& {Genzel}}]{Trippe2010MNRAS.402.1126T}
{Trippe}, S., {Davies}, R., {Eisenhauer}, F., {et~al.} 2010, \mnras, 402, 1126

\bibitem[{{Tsantaki} {et~al.}(2025){Tsantaki}, {Magrini}, {Danielski},
  {Bossini}, {Turrini}, {Moedas}, {Folsom}, {Ramler}, {Biazzo}, {Campante},
  {Delgado-Mena}, {da Silva}, {Sousa}, {Benatti}, {Casali}, {He{\l}miniak},
  {Rainer}, \& {Sanna}}]{Tsantaki2025}
{Tsantaki}, M., {Magrini}, L., {Danielski}, C., {et~al.} 2025, \aap, 697, A102

\bibitem[{{Tsuji}(2016)}]{Tsuji2016}
{Tsuji}, T. 2016, \pasj, 68, 84

\bibitem[{Turrini(2023)}]{Turrini2023}
Turrini, D. 2023, The Compositional Dimension of Planet Formation (World
  Scientific Publishing), 1--47

\bibitem[{{Turrini} {et~al.}(2022){Turrini}, {Codella}, {Danielski}, {Fedele},
  {Fonte}, {Garufi}, {Guarcello}, {Helled}, {Ikoma}, {Kama}, {Kimura},
  {Kruijssen}, {Maldonado}, {Miguel}, {Molinari}, {Nikolaou}, {Oliva},
  {Pani{\'c}}, {Pignatari}, {Podio}, {Rickman}, {Schisano}, {Shibata}, {Vazan},
  \& {Wolkenberg}}]{Turrini2022}
{Turrini}, D., {Codella}, C., {Danielski}, C., {et~al.} 2022, Experimental
  Astronomy, 53, 225

\bibitem[{{Turrini} {et~al.}(2018){Turrini}, {Miguel}, {Zingales}, {Piccialli},
  {Helled}, {Vazan}, {Oliva}, {Sindoni}, {Pani{\'c}}, {Leconte}, {Min},
  {Pirani}, {Selsis}, {Coud{\'e} du Foresto}, {Mura}, \&
  {Wolkenberg}}]{Turrini2018}
{Turrini}, D., {Miguel}, Y., {Zingales}, T., {et~al.} 2018, Experimental
  Astronomy, 46, 45

\bibitem[{{Ubachs}(2018)}]{Ubachs2018SSRv..214....3U}
{Ubachs}, W. 2018, \ssr, 214, 3

\bibitem[{{Usman} {et~al.}(2025){Usman}, {Ji}, {Rodriguez}, {Simpson},
  {Martell}, {Li}, {Bonaca}, {Shah}, \& {McKenzie}}]{usman2025}
{Usman}, S.~A., {Ji}, A.~P., {Rodriguez}, J., {et~al.} 2025, The Open Journal
  of Astrophysics, 8, 86

\bibitem[{Uylings \& Raassen(2019)}]{atoms7040102}
Uylings, P. \& Raassen, T. 2019, Atoms, 7

\bibitem[{{Uzan}(2025)}]{Uzan2025LRR....28....6U}
{Uzan}, J.-P. 2025, Living Reviews in Relativity, 28, 6

\bibitem[{{Valenti} {et~al.}(2017){Valenti}, {Sand}, {Yang}, {Cappellaro},
  {Tartaglia}, {Corsi}, {Jha}, {Reichart}, {Haislip}, \&
  {Kouprianov}}]{Valenti2017}
{Valenti}, S., {Sand}, D.~J., {Yang}, S., {et~al.} 2017, \apjl, 848, L24

\bibitem[{{Van der Swaelmen} {et~al.}(2023){Van der Swaelmen}, {Viscasillas
  V{\'a}zquez}, {Cescutti}, {Magrini}, {Cristallo}, {Vescovi}, {Randich},
  {Tautvai{\v{s}}ien{\.{e}}}, {Bagdonas}, {Bensby}, {Bergemann}, {Bragaglia},
  {Drazdauskas}, {Jim{\'e}nez-Esteban}, {Guiglion}, {Korn}, {Masseron},
  {Minkevii{\={u}}t{\.{e}}}, {Smiljanic}, {Spina}, {Stonkut{\.{e}}}, \&
  {Zaggia}}]{VanderSwaelmen2023}
{Van der Swaelmen}, M., {Viscasillas V{\'a}zquez}, C., {Cescutti}, G., {et~al.}
  2023, \aap, 670, A129

\bibitem[{{Van Eck} {et~al.}(2022){Van Eck}, {Shetye}, \& {Siess}}]{vanEck2022}
{Van Eck}, S., {Shetye}, S., \& {Siess}, L. 2022, Universe, 8, 220

\bibitem[{{van Regemorter}(1962)}]{vanregemorter1962}
{van Regemorter}, H. 1962, \apj, 136, 906

\bibitem[{{Vanderburg} {et~al.}(2020){Vanderburg}, {Rappaport}, {Xu},
  {Crossfield}, {Becker}, {Gary}, {Murgas}, {Blouin}, {Kaye}, {Palle}, {Melis},
  {Morris}, {Kreidberg}, {Gorjian}, {Morley}, {Mann}, {Parviainen}, {Pearce},
  {Newton}, {Carrillo}, {Zuckerman}, {Nelson}, {Zeimann}, {Brown},
  {Tronsgaard}, {Klein}, {Ricker}, {Vanderspek}, {Latham}, {Seager}, {Winn},
  {Jenkins}, {Adams}, {Benneke}, {Berardo}, {Buchhave}, {Caldwell},
  {Christiansen}, {Collins}, {Col{\'o}n}, {Daylan}, {Doty}, {Doyle},
  {Dragomir}, {Dressing}, {Dufour}, {Fukui}, {Glidden}, {Guerrero}, {Guo},
  {Heng}, {Henriksen}, {Huang}, {Kaltenegger}, {Kane}, {Lewis}, {Lissauer},
  {Morales}, {Narita}, {Pepper}, {Rose}, {Smith}, {Stassun}, \&
  {Yu}}]{Vanderburg2020}
{Vanderburg}, A., {Rappaport}, S.~A., {Xu}, S., {et~al.} 2020, \nat, 585, 363

\bibitem[{{Vangioni} \& {Olive}(2019)}]{VangioniOlive2019}
{Vangioni}, E. \& {Olive}, K.~A. 2019, \mnras, 484, 3561

\bibitem[{{Vanni} {et~al.}(2024){Vanni}, {Salvadori}, {D'Odorico}, {Becker}, \&
  {Cupani}}]{vanni2024}
{Vanni}, I., {Salvadori}, S., {D'Odorico}, V., {Becker}, G.~D., \& {Cupani}, G.
  2024, \apjl, 967, L22

\bibitem[{{Vanzella} {et~al.}(2023){Vanzella}, {Claeyssens}, {Welch}, {Adamo},
  {Coe}, {Diego}, {Mahler}, {Khullar}, {Kokorev}, {Oguri}, {Ravindranath},
  {Furtak}, {Hsiao}, {Abdurro'uf}, {Mandelker}, {Brammer}, {Bradley},
  {Brada{\v{c}}}, {Conselice}, {Dayal}, {Nonino}, {Andrade-Santos},
  {Windhorst}, {Pirzkal}, {Sharon}, {de Mink}, {Fujimoto}, {Zitrin},
  {Eldridge}, \& {Norman}}]{Vanzella2023}
{Vanzella}, E., {Claeyssens}, A., {Welch}, B., {et~al.} 2023, \apj, 945, 53

\bibitem[{{Vasiliev} \& {Baumgardt}(2021)}]{Vasiliev2021}
{Vasiliev}, E. \& {Baumgardt}, H. 2021, \mnras, 505, 5978

\bibitem[{{Velichko} {et~al.}(2011){Velichko}, {Mashonkina}, \&
  {Nilsson}}]{velichko2011}
{Velichko}, A.~B., {Mashonkina}, L.~I., \& {Nilsson}, H. 2011, Astronomy
  Letters, 37, 440

\bibitem[{{Ventura} {et~al.}(2018){Ventura}, {D'Antona}, {Imbriani}, {Di
  Criscienzo}, {Dell'Agli}, \& {Tailo}}]{Ventura2018}
{Ventura}, P., {D'Antona}, F., {Imbriani}, G., {et~al.} 2018, \mnras, 477, 438

\bibitem[{{Ventura} {et~al.}(2001){Ventura}, {D'Antona}, {Mazzitelli}, \&
  {Gratton}}]{Ventura2001}
{Ventura}, P., {D'Antona}, F., {Mazzitelli}, I., \& {Gratton}, R. 2001, \apjl,
  550, L65

\bibitem[{{Veras}(2016)}]{Veras2016}
{Veras}, D. 2016, Royal Society Open Science, 3, 150571

\bibitem[{{Veras}(2021)}]{Veras2021}
{Veras}, D. 2021, in Oxford Research Encyclopedia of Planetary Science (Oxford
  University Press), 1

\bibitem[{{Veras} {et~al.}(2020){Veras}, {Tremblay}, {Hermes}, {McDonald},
  {Kennedy}, {Meru}, \& {G{\"a}nsicke}}]{Veras2020}
{Veras}, D., {Tremblay}, P.-E., {Hermes}, J.~J., {et~al.} 2020, \mnras, 493,
  765

\bibitem[{{Veras} {et~al.}(2011){Veras}, {Wyatt}, {Mustill}, {Bonsor}, \&
  {Eldridge}}]{Veras2011}
{Veras}, D., {Wyatt}, M.~C., {Mustill}, A.~J., {Bonsor}, A., \& {Eldridge},
  J.~J. 2011, \mnras, 417, 2104

\bibitem[{{Vidotto} {et~al.}(2014){Vidotto}, {Gregory}, {Jardine}, {Donati},
  {Petit}, {Morin}, {Folsom}, {Bouvier}, {Cameron}, {Hussain}, {Marsden},
  {Waite}, {Fares}, {Jeffers}, \& {do Nascimento}}]{vidotto2014}
{Vidotto}, A.~A., {Gregory}, S.~G., {Jardine}, M., {et~al.} 2014, \mnras, 441,
  2361

\bibitem[{{Vietri} {et~al.}(2025){Vietri}, {Rodr{\'\i}guez Hidalgo}, {Rankine},
  {Zappacosta}, {Piconcelli}, {Flores}, {Saccheo}, {Melandri}, {Testa}, {Hall},
  {Sarnari}, {D'Odorico}, {Lanzuisi}, {Misawa}, {Onken}, {Vignali}, \&
  {Wolf}}]{Vietri2025A&A...704A.166V}
{Vietri}, G., {Rodr{\'\i}guez Hidalgo}, P., {Rankine}, A., {et~al.} 2025, \aap,
  704, A166

\bibitem[{{Villaver} \& {Livio}(2007)}]{VillaverLivio2007}
{Villaver}, E. \& {Livio}, M. 2007, \apj, 661, 1192

\bibitem[{{Viscasillas V{\'a}zquez} {et~al.}(2022){Viscasillas V{\'a}zquez},
  {Magrini}, {Casali}, {Tautvai{\v{s}}ien{\.{e}}}, {Spina}, {Van der Swaelmen},
  {Randich}, {Bensby}, {Bragaglia}, {Friel}, {Feltzing}, {Sacco}, {Turchi},
  {Jim{\'e}nez-Esteban}, {D'Orazi}, {Delgado-Mena}, {Mikolaitis},
  {Drazdauskas}, {Minkevi{\v{c}}i{\={u}}t{\.{e}}}, {Stonkut{\.{e}}},
  {Bagdonas}, {Montes}, {Guiglion}, {Baratella}, {Tabernero}, {Gilmore},
  {Alfaro}, {Francois}, {Korn}, {Smiljanic}, {Bergemann}, {Franciosini},
  {Gonneau}, {Hourihane}, {Worley}, \& {Zaggia}}]{ViscasillasVazquez2022}
{Viscasillas V{\'a}zquez}, C., {Magrini}, L., {Casali}, G., {et~al.} 2022,
  \aap, 660, A135

\bibitem[{{Vogt} {et~al.}(1994){Vogt}, {Allen}, {Bigelow}, {Bresee}, {Brown},
  {Cantrall}, {Conrad}, {Couture}, {Delaney}, {Epps}, {Hilyard}, {Hilyard},
  {Horn}, {Jern}, {Kanto}, {Keane}, {Kibrick}, {Lewis}, {Osborne},
  {Pardeilhan}, {Pfister}, {Ricketts}, {Robinson}, {Stover}, {Tucker}, {Ward},
  \& {Wei}}]{Vogt1994}
{Vogt}, S.~S., {Allen}, S.~L., {Bigelow}, B.~C., {et~al.} 1994, in Society of
  Photo-Optical Instrumentation Engineers (SPIE) Conference Series, Vol. 2198,
  Instrumentation in Astronomy VIII, ed. D.~L. {Crawford} \& E.~R. {Craine},
  362

\bibitem[{{Vogt} {et~al.}(1987){Vogt}, {Penrod}, \& {Hatzes}}]{vogt1987}
{Vogt}, S.~S., {Penrod}, G.~D., \& {Hatzes}, A.~P. 1987, \apj, 321, 496

\bibitem[{{Wagner-Carena} {et~al.}(2024){Wagner-Carena}, {Lee}, {Pennington},
  {Aalbers}, {Birrer}, \& {Wechsler}}]{WagnerCarena2024}
{Wagner-Carena}, S., {Lee}, J., {Pennington}, J., {et~al.} 2024, \apj, 975, 297

\bibitem[{{Wan} {et~al.}(2021){Wan}, {Oliver}, {Baumgardt}, {Lewis}, {Gieles},
  {H{\'e}nault-Brunet}, {de Boer}, {Balbinot}, {Da Costa}, \&
  {Mackey}}]{Wan2021}
{Wan}, Z., {Oliver}, W.~H., {Baumgardt}, H., {et~al.} 2021, \mnras, 502, 4513

\bibitem[{{Wang} {et~al.}(2021){Wang}, {Nordlander}, {Asplund}, {Amarsi},
  {Lind}, \& {Zhou}}]{exwang2021}
{Wang}, E.~X., {Nordlander}, T., {Asplund}, M., {et~al.} 2021, \mnras, 500,
  2159

\bibitem[{{Wang} {et~al.}(2020){Wang}, {Iwasawa}, {Nitadori}, \&
  {Makino}}]{Wang2020}
{Wang}, L., {Iwasawa}, M., {Nitadori}, K., \& {Makino}, J. 2020, \mnras, 497,
  536

\bibitem[{Ward {et~al.}(2023)Ward, Li, Schwartz, Nave, Raassen, \&
  Uylings}]{Ward2023}
Ward, J.~W., Li, J.~J., Schwartz, J., {et~al.} 2023, The Astrophysical Journal,
  959, 8

\bibitem[{{Watson} {et~al.}(2019){Watson}, {Hansen}, {Selsing}, {Koch},
  {Malesani}, {Andersen}, {Fynbo}, {Arcones}, {Bauswein}, {Covino}, {Grado},
  {Heintz}, {Hunt}, {Kouveliotou}, {Leloudas}, {Levan}, {Mazzali}, \&
  {Pian}}]{watson2019}
{Watson}, D., {Hansen}, C.~J., {Selsing}, J., {et~al.} 2019, \nat, 574, 497

\bibitem[{{Waugh} \& {Jardine}(2022)}]{waugh2022}
{Waugh}, R. F.~P. \& {Jardine}, M.~M. 2022, \mnras, 514, 5465

\bibitem[{{Webb} {et~al.}(1999){Webb}, {Flambaum}, {Churchill}, {Drinkwater},
  \& {Barrow}}]{Webb1999PhRvL..82..884W}
{Webb}, J.~K., {Flambaum}, V.~V., {Churchill}, C.~W., {Drinkwater}, M.~J., \&
  {Barrow}, J.~D. 1999, \prl, 82, 884

\bibitem[{{Webb} {et~al.}(2021){Webb}, {Lee}, {Carswell}, \&
  {Milakovi{\'c}}}]{Webb2021MNRAS.501.2268W}
{Webb}, J.~K., {Lee}, C.-C., {Carswell}, R.~F., \& {Milakovi{\'c}}, D. 2021,
  \mnras, 501, 2268

\bibitem[{{Webb} {et~al.}(2022){Webb}, {Lee}, \&
  {Milakovi{\'c}}}]{Webb2022Univ....8..266W}
{Webb}, J.~K., {Lee}, C.-C., \& {Milakovi{\'c}}, D. 2022, Universe, 8, 266

\bibitem[{{Weber} \& {Davis}(1967)}]{weber1967}
{Weber}, E.~J. \& {Davis}, Leverett, J. 1967, \apj, 148, 217

\bibitem[{{Wehrhahn} {et~al.}(2023){Wehrhahn}, {Piskunov}, \&
  {Ryabchikova}}]{wehrhahn2023}
{Wehrhahn}, A., {Piskunov}, N., \& {Ryabchikova}, T. 2023, \aap, 671, A171

\bibitem[{{Welsh} {et~al.}(2012){Welsh}, {Orosz}, {Carter}, {Fabrycky}, {Ford},
  {Lissauer}, {Pr{\v{s}}a}, {Quinn}, {Ragozzine}, {Short}, {Torres}, {Winn},
  {Doyle}, {Barclay}, {Batalha}, {Bloemen}, {Brugamyer}, {Buchhave},
  {Caldwell}, {Caldwell}, {Christiansen}, {Ciardi}, {Cochran}, {Endl},
  {Fortney}, {Gautier}, {Gilliland}, {Haas}, {Hall}, {Holman}, {Howard},
  {Howell}, {Isaacson}, {Jenkins}, {Klaus}, {Latham}, {Li}, {Marcy}, {Mazeh},
  {Quintana}, {Robertson}, {Shporer}, {Steffen}, {Windmiller}, {Koch}, \&
  {Borucki}}]{Welsh2012}
{Welsh}, W.~F., {Orosz}, J.~A., {Carter}, J.~A., {et~al.} 2012, \nat, 481, 475

\bibitem[{{Werk} {et~al.}(2016){Werk}, {Prochaska}, {Cantalupo}, {Fox},
  {Oppenheimer}, {Tumlinson}, {Tripp}, {Lehner}, \&
  {McQuinn}}]{Werk2016ApJ...833...54W}
{Werk}, J.~K., {Prochaska}, J.~X., {Cantalupo}, S., {et~al.} 2016, \apj, 833,
  54

\bibitem[{{Wheeler} {et~al.}(2023){Wheeler}, {Abruzzo}, {Casey}, \&
  {Ness}}]{Wheeler2023}
{Wheeler}, A.~J., {Abruzzo}, M.~W., {Casey}, A.~R., \& {Ness}, M.~K. 2023, \aj,
  165, 68

\bibitem[{{Wiedeking} {et~al.}(2025){Wiedeking}, {Goriely}, {Guttormsen},
  {Herwig}, {Larsen}, {Liddick}, {M{\"u}cher}, {Richard}, {Siem}, \&
  {Spyrou}}]{Wiedeking2025}
{Wiedeking}, M., {Goriely}, S., {Guttormsen}, M., {et~al.} 2025, Nature Reviews
  Physics, 7, 696

\bibitem[{{Wiescher} {et~al.}(2010){Wiescher}, {G{\"o}rres}, {Uberseder},
  {Imbriani}, \& {Pignatari}}]{Wiescher2010}
{Wiescher}, M., {G{\"o}rres}, J., {Uberseder}, E., {Imbriani}, G., \&
  {Pignatari}, M. 2010, Annual Review of Nuclear and Particle Science, 60, 381

\bibitem[{Wiese \& Martin(1980)}]{Wiese1980}
Wiese, W.~L. \& Martin, G.~A. 1980, Wavelengths and transition probabilities
  for atoms and atomic ions :, Tech. rep., National Bureau of Standards

\bibitem[{{Winn} \& {Fabrycky}(2015)}]{Winn2015}
{Winn}, J.~N. \& {Fabrycky}, D.~C. 2015, \araa, 53, 409

\bibitem[{{Wolthoff} {et~al.}(2022){Wolthoff}, {Reffert}, {Quirrenbach},
  {Jones}, {Wittenmyer}, \& {Jenkins}}]{Wolthoff2022}
{Wolthoff}, V., {Reffert}, S., {Quirrenbach}, A., {et~al.} 2022, \aap, 661, A63

\bibitem[{{Wood} {et~al.}(2005){Wood}, {M{\"u}ller}, {Zank}, {Linsky}, \&
  {Redfield}}]{wood2005}
{Wood}, B.~E., {M{\"u}ller}, H.~R., {Zank}, G.~P., {Linsky}, J.~L., \&
  {Redfield}, S. 2005, \apjl, 628, L143

\bibitem[{{Woosley} \& {Hoffman}(1992)}]{Woosley1992}
{Woosley}, S.~E. \& {Hoffman}, R.~D. 1992, \apj, 395, 202

\bibitem[{{Wright} {et~al.}(2011){Wright}, {Drake}, {Mamajek}, \&
  {Henry}}]{wright2011}
{Wright}, N.~J., {Drake}, J.~J., {Mamajek}, E.~E., \& {Henry}, G.~W. 2011,
  \apj, 743, 48

\bibitem[{{Wright} {et~al.}(2018){Wright}, {Newton}, {Williams}, {Drake}, \&
  {Yadav}}]{wright2018}
{Wright}, N.~J., {Newton}, E.~R., {Williams}, P. K.~G., {Drake}, J.~J., \&
  {Yadav}, R.~K. 2018, \mnras, 479, 2351

\bibitem[{{Xiao} {et~al.}(2024){Xiao}, {Teng}, {Zhou}, {Sato}, {Liu}, {Bi},
  {Takarada}, {Kuzuhara}, {Hon}, {Wang}, {Omiya}, {Harakawa}, {Zhao}, {Zhao},
  {Kambe}, {Izumiura}, {Ando}, {Noguchi}, {Wang}, {Zhai}, {Song}, {Yang}, {Li},
  {Brandt}, {Yoshida}, {Itoh}, \& {Kokubo}}]{Xiao2024}
{Xiao}, G.-Y., {Teng}, H.-Y., {Zhou}, J., {et~al.} 2024, \aj, 167, 59

\bibitem[{{Xuan} {et~al.}(2024){Xuan}, {Wang}, {Finnerty}, {Horstman}, {Grimm},
  {Peck}, {Nielsen}, {Knutson}, {Mawet}, {Isaacson}, {Howard}, {Liu}, {Walker},
  {Phillips}, {Blake}, {Ruffio}, {Zhang}, {Inglis}, {Wallack}, {Sanghi},
  {Gonzales}, {Dai}, {Baker}, {Bartos}, {Bond}, {Bryan}, {Calvin}, {Cetre},
  {Delorme}, {Doppmann}, {Echeverri}, {Fitzgerald}, {Jovanovic}, {Liberman},
  {L{\'o}pez}, {Martin}, {Morris}, {Pezzato}, {Ruane}, {Sappey}, {Schofield},
  {Skemer}, {Venenciano}, {Wallace}, {Wang}, {Wizinowich}, {Xin}, {Agrawal},
  {Do {\'O}}, {Hsu}, \& {Phillips}}]{Xuan2024}
{Xuan}, J.~W., {Wang}, J., {Finnerty}, L., {et~al.} 2024, \apj, 962, 10

\bibitem[{{Yakovleva} {et~al.}(2020){Yakovleva}, {Belyaev}, \&
  {Bergemann}}]{yakovleva2020}
{Yakovleva}, S.~A., {Belyaev}, A.~K., \& {Bergemann}, M. 2020, Atoms, 8, 34

\bibitem[{{Yakovleva} {et~al.}(2022){Yakovleva}, {Belyaev}, \&
  {Mashonkina}}]{yakovleva2022}
{Yakovleva}, S.~A., {Belyaev}, A.~K., \& {Mashonkina}, L.~I. 2022, Atoms, 10,
  33

\bibitem[{{Yan} {et~al.}(2022){Yan}, {Shi}, {Wang}, {Yan}, {Zhou}, {Zhou},
  {Fang}, {Li}, {Chen}, \& {Xie}}]{Yan2022Li}
{Yan}, T.-S., {Shi}, J.-R., {Wang}, L., {et~al.} 2022, \apjl, 929, L14

\bibitem[{{Yong} {et~al.}(2006){Yong}, {Aoki}, \& {Lambert}}]{Yong2006b}
{Yong}, D., {Aoki}, W., \& {Lambert}, D.~L. 2006, \apj, 638, 1018

\bibitem[{{Yong} {et~al.}(2003){Yong}, {Grundahl}, {Lambert}, {Nissen}, \&
  {Shetrone}}]{Yong2003}
{Yong}, D., {Grundahl}, F., {Lambert}, D.~L., {Nissen}, P.~E., \& {Shetrone},
  M.~D. 2003, \aap, 402, 985

\bibitem[{{Yong} {et~al.}(2021){Yong}, {Kobayashi}, {Da Costa}, {Bessell},
  {Chiti}, {Frebel}, {Lind}, {Mackey}, {Nordlander}, {Asplund}, {Casey},
  {Marino}, {Murphy}, \& {Schmidt}}]{Yong2021}
{Yong}, D., {Kobayashi}, C., {Da Costa}, G.~S., {et~al.} 2021, \nat, 595, 223

\bibitem[{{Yong} {et~al.}(2013){Yong}, {Mel{\'e}ndez}, {Grundahl}, {Roederer},
  {Norris}, {Milone}, {Marino}, {Coelho}, {McArthur}, {Lind}, {Collet}, \&
  {Asplund}}]{Yong2013}
{Yong}, D., {Mel{\'e}ndez}, J., {Grundahl}, F., {et~al.} 2013, \mnras, 434,
  3542

\bibitem[{Yuce {et~al.}(2011)Yuce, Castelli, \& Hubrig}]{2011yuce}
Yuce, K., Castelli, F., \& Hubrig, S. 2011, Astronomy and Astrophysics, 528, 1

\bibitem[{{Zatsarinny} {et~al.}(2019){Zatsarinny}, {Bartschat},
  {Fernandez-Menchero}, \& {Tayal}}]{zatsarinny2019}
{Zatsarinny}, O., {Bartschat}, K., {Fernandez-Menchero}, L., \& {Tayal}, S.~S.
  2019, \pra, 99, 023430

\bibitem[{{Zhao} {et~al.}(2012){Zhao}, {Zhao}, {Chu}, {Jing}, \&
  {Deng}}]{lamost2012}
{Zhao}, G., {Zhao}, Y.-H., {Chu}, Y.-Q., {Jing}, Y.-P., \& {Deng}, L.-C. 2012,
  Research in Astronomy and Astrophysics, 12, 723

\bibitem[{{Zhou} {et~al.}(2018){Zhou}, {Shi}, {Yan}, {Gao}, {Zhang}, {Zhao},
  {Pan}, \& {Kumar}}]{zhou18LSb}
{Zhou}, Y.~T., {Shi}, J.~R., {Yan}, H.~L., {et~al.} 2018, \aap, 615, A74

\bibitem[{{Zuckerman} {et~al.}(2010){Zuckerman}, {Melis}, {Klein}, {Koester},
  \& {Jura}}]{Zuckerman2010}
{Zuckerman}, B., {Melis}, C., {Klein}, B., {Koester}, D., \& {Jura}, M. 2010,
  \apj, 722, 725

\end{thebibliography}

\clearpage
\clearpage
\fancyhead[R]{\textsf{Affiliations}}
\addcontentsline{toc}{chapter}{Affiliations}
\section*{Affilitations}

\begingroup
\small
\setlength{\parindent}{0pt}
\setlength{\parskip}{0.4em}

\begingroup
\raggedright
[1] INAF - Osservatorio Astrofisico di Arcetri, Largo E. Fermi 5, 50125 Firenze, Italy\label{oaa}\par
[2] Lund Observatory, Division of Astrophysics, Department of Physics, Lund University, Sweden\par
[3] INAF - Osservatorio Astronomico di Brera, Via Bianchi 46, 23807, Merate, Italy\par
[4] UK Astronomy Technology Centre, Edinburgh, United Kingdom\par
[5] Facultad de F\'isica, Universidad de La Laguna, Avda. Astrof\'isico Fco. S\'anchez s/n, 38206, La Laguna, Santa Cruz de Tenerife, Spain\par
[6] Instituto de Astrof\'isica de Canarias, C. Vía L\`actea, s/n, 38205 La Laguna, Santa Cruz de Tenerife, Spain\par
[7] Instituto de Astrof\'isica e Ci\^encias do Espa\c{c}o, Universidade do Porto, CAUP, Rua das Estrelas, 4150-762 Porto, Portugal\par
[8] INAF- Direzione Scientifica- Viale del Parco Mellini, 84, Roma, Italy\par
[9] Departamento de Física e Astronomia, FCUP, Rua do Campo Alegre 687, Porto, Portugal\par
[10] Theoretical Astrophysics, Department of Physics and Astronomy, Uppsala University, Uppsala, Sweden\par
[11] Departamento de F\'isica Te\'orica, At\'omica y \'Optica, Universidad de Valladolid, Valladolid, Spain\par
[12] Department of Physics, Imperial College London, London, UK\par
[13] Departament d'Astronomia i Astrof\'isica, Universitat de Val\`encia, Burjassot, Spain\par
[14] Dept. of Physics, University of Rome Tor Vergata, Rome, Italy\par
[15] Goethe University Frankfurt, Institute for Applied Physics, Max-von-Laue-Str. 12, 60438, Frankfurt am Main, Germany\par
[16] Observatoire de la C\^ote d'Azur, Universit\'e C\^ote d'Azur, CNRS, Laboratoire Lagrange, Nice, France\par
[17] Keele University, Keele, ST5 5BG, UK\par
[18] Dipartimento di Fisica e Astronomia, Universita di Bologna, Via Gobetti 93/2, Bologna, Italy\par
[19] INAF - Osservatorio di Astrofisica e Scienza dello Spazio di Bologna, Via Gobetti 93/3, Bologna, Italy\par
[20] INAF - Osservatorio Astronomico di Trieste, via Tiepolo 11, 34143, Trieste, Italy\par
[21] IFPU -- Institute for Fundamental Physics of the Universe, via Beirut 2, I-34151 Trieste, Italy\par
[22] Dipartimento di Fisica e Astronomia, Universita degli Studi di Firenze, Via G. Sansone 1, Sesto Fiorentino, Italy\par
[23] Nicolaus Copernicus Astronomical Center, Polish Academy of Sciences, ul. Bartycka 18, Warsaw, Poland\par
[24] Technische Universit\"at Darmstadt, Department of Physics, 64289 Darmstadt, Germany \par
[25] GSI Helmholtzzentrum f\"ur Schwerionenforschung GmbH, 64291 Darmstadt, Germany\par
[26] Max-Planck-Institut f\"ur Kernphysik, Saupfercheckweg 1, 69117 Heidelberg, Germany\par
[27] ESO European Southern Observatory, Alonso de C\'ordova 3107, Vitacura, Santiago, Chile\par
[28] Universidade de S\~ao Paulo, IAG, Rua do Mat\~ao 1226, S\~ao Paulo, Brazil\par
[29] School of Physical Sciences, The Open University, Walton Hall, Milton Keynes, MK7 6AA, UK\par
[30] School of Mathematics and Physics, The University of Queensland, St Lucia, QLD 4072, Australia\par
[31] INAF - Osservatorio Astronomico di Roma, Via Frascati 88, I-00040 Monte Porzio Catone (RM), Italy\par
[32] Dipartimento di F\'isica ``Enrico Fermi'', Universit\`a di Pisa, Largo Bruno Pontecorvo 3, Pisa, I-56127, Italy\par
[33] Interdisciplinary Center for Scientific Computing (IWR), University of Heidelberg, Im Neuenheimer Feld 205, 69120, Heidelberg, Germany\par
[34] Universit\"at Heidelberg, Zentrum f\"ur Astronomie, Institut f\"ur Theoretische Astrophysik, Albert-Ueberle-Strasse 2, 69120, Heidelberg, Germany\par
[35] Australian National University; ARC Centre of Excellence for All Sky Astrophysics in 3 Dimensions (ASTRO 3D)\par
[36] Space sciences, Technologies and Astrophysics Research (STAR) Institute, Universit\'e de Liège, Li\`ege, Belgium\par
[37] Dipartimento di Fisica, Universit\`a degli Studi di Trieste, via Tiepolo 11, 34143, Trieste, Italy\par
[38] SUPA School of Physics and Astronomy, University of St Andrews, North Haugh, St Andrews KY16 9SS, UK\par
[39] Centre for Extragalactic Astronomy, Durham University, South Road, Durham DH1 3LE, UK \par
[40] Department of Physics, Durham University, South Road, Durham DH1 3LE, UK\par
[41] INAF, Osservatorio Astronomico d'Abruzzo, Via Mentore Maggini Snc, 64100 Teramo, Italy\par
[42] Sezione di Perugia, Istituto Nazionale di Fisica Nucleare, Perugia, Italy\par
[43] INAF - Osservatorio Astronomico di Palermo G.S.Vaiana, Piazza del Parlamento 1, 90134, Palermo, Italy\par
[44] Vilnius University, Faculty of Physics, Institute of Theoretical Physics and Astronomy, Sauletekio av. 3, 10257 Vilnius, Lithuania \par
[45] INAF - Osservatorio Astrofisico di Catania, via S. Sofia, 78, 95128, Catania, Italy\par
[46] ICREA, Pg. Lluis Companys 28, E-08010 Barcelona, Spain; Institut de Ciencies del Cosmos (ICCUB), Universitat de Barcelona (UB), c. Marti i Franques, 1, E-08028 Barcelona, Spain\par
[47] Institut d'Estudis Espacials de Catalunya (IEEC), Edifici RDIT, Campus UPC, E-08860 Castelldefels (Barcelona), Spain\par
[48] Department of Physics and Astronomy, University of Notre Dame, 225 Nieuwland Science Hall, Notre Dame, IN 46556, USA\par
[49] Department of Physics, University of Warwick, Gibbet Hill Road, Coventry CV4 7AL, UK\par
[50] Centre for Exoplanets and Habitability, University of Warwick, Gibbet Hill Road, Coventry CV4 7AL, UK\par
[51] Laboratoire d'Astrophysique de Bordeaux, Universit\'e Bordeaux, CNRS, B18N, All\'ee Geoffroy Saint-Hilaire, 33615, Pessac, France\par
[52] Dipartimento di Fisica e Astronomia "Galileo Galilei", Universit\`a degli Studi di Padova, Vicolo dell’Osservatorio 3, 35122 Padova, Italy\par
[53] Astronomisches Rechen-Institut, Zentrum f\"ur Astronomie der Universit\"at Heidelberg, M\"onchhofstra{\ss}e 12-14, 69120, Heidelberg, Germany\par
[54] Institut d'Astronomie et d'Astrophysique, Universit\'e Libre de Bruxelles (ULB), CP 226, Boulevard du Triomphe, 1050, Bruxelles, Belgium\par
[55] Royal Observatory of Belgium, Avenue Circulaire 3, 1180 Brussels, Belgium\par
[56] Space Telescope Science Institute\par
[57] INAF - Osservatorio Astronomico di Padova, Vicolo dell'Osservatorio 5, 85122, Padova, Italy\par
[58] INAF - Osservatorio Astrofisico di Torino, Via Osservatorio 20, 10025, Pino Torinese (TO), Italy\par
[59] HUN REN Konkoly Observatory, CSFK, H-1121, Budapest, Konkoly Thege M. ut 15-17, Hungary\par
[60] CSFK, MTA Centre of Excellence, Budapest, Konkoly Thege Miklos ut 15-17, H-1121, Hungary\par
[61] University of Bayreuth, BGI, Universitatsstrasse 30, 95447 Bayreuth, Germany\par
[62] Heidelberger Institut f\"ur Theoretische Studien, Schloss-Wolfsbrunnenweg 35, D-69118 Heidelberg, Germany\par
[63] European Space Agency (ESA), European Space Astronomy Centre (ESAC), Camino Bajo del Castillo s/n, E-28692 Villanueva de la Ca\~{n}ada, Madrid, Spain\par
[64] Department of Astronomy, The Ohio State University, Columbus, OH 43210, USA\par
[65] Center for Cosmology and AstroParticle Physics (CCAPP), The Ohio State University, Columbus, OH 43210, USA\par
[66] Max-Planck-Institut f\"ur Astronomie, K\"onigstuhl 17, D-69117 Heidelberg, Germany\par
[67] University of Victoria (Canada)\par
[68] School of Physical and Chemical Sciences - Te Kura Matu, University of Canterbury, Private Bag 4800, Christchurch, 8140, New Zealand\par
[69] Centre for Astrophysics Research, University of Hertfordshire, Hatfield, AL10 9AB, UK \par
[70] Department of Physics, University of Surrey, Stag Hill, Guildford, GU2 7XH, UK\par
[71] Department of Astrophysics, University of Vienna, Turkenschanzstrasse 17, 1180, Vienna\par
[72] University of Southern Queensland\par
[73] Research School of Astronomy and Astrophysics, The Australian National University, Canberra, ACT 2611, Australia\par
[74] Centro de Astrobiolog\'ia (CAB), CSIC-INTA, Camino Bajo del Castillo s/n, 28692, Villanueva de la Ca\~nada (Madrid), Spain\par
[75] Instituto de Astrof\'isica, Facultad de F\'isica, Pontificia Universidad Cat\'olica de Chile, Campus San Joaqu\'in, Av. Vicu\~na Mackenna 4860, Macul, Santiago, 7820436, Chile \par
[76] INAF-IAPS, Via del Fosso del Cavaliere 100, Roma, Italy\par
[77] Purple Mountain Observatory, Chinese Academy of Sciences, Nanjing, 210023, PR China\par
[78] Department of Astronomy, Stockholm University, AlbaNova University Centre, Stockholm, Sweden\par
[79] Materials Science and Applied Mathematics, M\"alm\"o University, 20506, M\"alm\"o, Sweden\par
[80] Department of Astronomy, The University of Texas at Austin, 2515 Speedway Boulevard, Austin, TX 78712, USA\par
[81] Institut d'astrophysique de Paris, CNRS \par
[82] Sorbonne Universit\'e, UMR 7095, 98bis Boulevard Arago, 75014, Paris, France\par
[83] Laboratoire d'astrophysique, \'Ecole Polytechnique F\'ed\'erale de Lausanne (EPFL), Observatoire, 1290, Versoix, Switzerland\par 
[84] GEPI, Observatoire de Paris, CNRS, Universit\'e de Paris Diderot, 92195, Meudon Cedex, France\par
[85] Centre for Astrophysics Research, Department of Physics, Astronomy and Mathematics University of Hertfordshire, College Lane, AL10 9AB, Hatfield, UK\par
[86] Sapienza - University of Rome, Physics department, Piazzale Aldo Moro 5, 00185, Rome, Italy\par
[87] Tartu Observatory, University of Tartu, T\~oravere, Estonia\par
[88] INAF - Osservatorio Astronomico di Capodimonte, Via Salita Moiariello 16, 80131, Napoli, Italy\par
[89] University of New South Wales - School of Physics, Sydney, Australia\par
[90] Institute of Astronomy, University of Cambridge, Madingley Road, Cambridge CB3 0HA, UK\par
[91] Centre for Astrophysics and Supercomputing, Swinburne University of Technology, Hawthorn, Victoria 3122, Australia\par
[92] Centre for Astrophysics, University of Southern Queensland, Toowoomba, QLD 4350, Australia\par
[93] Sub-department of Astrophysics, University of Oxford, Keble Rd, Oxford OX13RH, UK\par
[94] Physique Atomique et Astrophysique, Universit\'e de Mons - UMONS, 7000, Mons, Belgium\par
[95] European Southern Observatory, Garching, Germany\par
[96] Physique Atomique et Astrophysique, Universit\'e de Mons, B-7000 Mons, Belgium\par
[97] IPNAS, Universit\'e de Li\`ege, Sart Tilman, B-4000 Li\`ege, Belgium\par
[98] Universit\'e de Strasbourg, CNRS UMR 7550, Observatoire astronomique de Strasbourg, 11 rue de l'Universit\'e, 67000 Strasbourg, France\par
[99] University of Insubria\par
[100] Institute of Astrophysics and Space Sciences (IA)\par
[101] Centre for Advanced Instrumentation, Department of Physics, Durham University, South Road, Durham\par



\endgroup

\clearpage
\fancyhead[R]{\textsf{Acknowledgements}}
\addcontentsline{toc}{chapter}{Acknowledgements}
\section*{Acknowledgements}

\begingroup
\small
\setlength{\parindent}{0pt}
\setlength{\parskip}{0.4em}

L.M., S.R., R.E.G., J.S.H., M.T. R.E.G., F.R., G.C. acknowledge support from INAF through the Large Grants EPOCH and WST, funding for the WEAVE project, the Mini-Grants Checs (1.05.23.04.02), and financial support under the National Recovery and Resilience Plan (PNRR), Mission 4, Component 2, Investment 1.1, Call for tender No. 104 published on 2 February 2022 by the Italian Ministry of University and Research (MUR), funded by the European Union – NextGenerationEU, through the Project ‘Cosmic POT’ (Grant Assignment Decree No. 2022X4TM3H, MUR).
M.R.S. acknowledges support from the European Space Agency as an ESA Research Fellow; 
D.N. acknowledges support from PLATO ASI-INAF agreements n. 2022-28-HH.0; 
S.G.S. acknowledge support from FCT through FCT contract nr. CEECIND/00826/2018 and POPH/FSE (EC) and through the Scientific Employment Stimulus programme (reference 2024.08008.CEECIND).  This work was supported by FCT - Funda\c{c}\~{a}o para a Ci\^{e}ncia e a Tecnologia through national funds and by FEDER through COMPETE2020 through through national funds under the research grant UID/04434/2025 (DOI 10.54499/UID/04434/2025);   
V.A. acknowledges support from FCT through a work contract funded by the FCT Scientific Employment Stimulus program (reference 2023.06055.CEECIND/CP2839/CT0005, DOI:10.54499/2023.06055.CEECIND/CP2839/CT0005);
C.D. acknowledges financial support from the grant RYC2023-044903-I funded by MCIU/AEI/10.13039/501100011033 and by the ESF+;  
D.T. acknowledge support from the ASI-INAF, grant no. 2021-5-HH.0, plus addenda no. 2021-5-HH.1-2022 and 2021-5-HH.2-2024, and from the European Research Council via the Horizon 2020 Framework Programme ERC Synergy “ECOGAL” Project GA-855130;  
D.P. is supported by the Fondazione ICSC, Spoke 3 ``Astrophysics and Cosmos Observations'', National Recovery and Resilience Plan (Piano Nazionale di Ripresa e Resilienza, PNRR) Project ID CN\_00000013 ``Italian Research Center on High-Performance Computing, Big Data and Quantum Computing'' funded by MUR Missione 4 Componente 2 Investimento 1.4: Potenziamento strutture di ricerca e creazione di “campioni nazionali di R\&S (M4C2-19)'' - Next Generation EU (NGEU);  
J.S.U., thanks to INAF for its support through the Mini-Grant (1.05.24.07.02);  
M.M. and A.A. thank the financial support from the Deutsche Forschungsgemeinschaft (DFG, German Research Foundation) – Project-ID 279384907 – SFB 1245;  
A.C.C. acknowledges support from UKRI/ERC Synergy Grant EP/Z000181/1 (REVEAL);
V.G. acknowledges financial support from INAF under the program “Giovani Astrofisiche ed Astrofisici di Eccellenza – IAF: INAF Astrophysics Fellowships in Italy” (Project: GalacticA, “Galactic Archaeology: reconstructing the history of the Galaxy”)
F.Z.M. acknowledges support from the ASI-INAF Agreement no. 2021-12-HH.0 and Addendum 2021-12-HH.1-2024 "Missione Solar-C EUVST-Supporto scientifico di Fase B/C/D";
M.M.J. acknowledges support from STFC consolidated grant
number ST/R000824/1;
S.B. acknowledges support from the Australian Research Council under grant number DE240100150; 
T.B.’s contribution to this project was made possible by funding from the Carl-Zeiss-Stiftung. T.B. is funded by the Deutsche Forschungsgemeinschaft (DFG, German Research Foundation) – 573580088.
AE received the support of a fellowship from “La Caixa” Foundation (ID 100010434) with fellowship code LCF/BQ/PI23/11970031;
T.C.~is supported by FCT in the form of a work contract (2023.08117.CEECIND/CP2839/CT0004). This work was supported by national funds through FCT under project 2024.15303.PEX;
IMC acknowledges funding from Agencia Nacional de Investigaci\'on y Desarrollo / Fondo ALMA 2025 / 31250023;
This study has received funding from the Research Council of Lithuania, agreement No P-MIP-24-652;  
H.R. acknowledges ESA Estonia research infrastructure funded by the Estonian Research Council grant TARISTU24-TK3;
T.B. acknowledges financial support by grants No. 2018-04857 and 2024-04990 from the Swedish Research Council;  
E.F.A and G.B.  acknowledge support from the Agencia Estatal de Investigaci\'on del Ministerio de Ciencia, Innovaci\'on y Universidades (MCIU/AEI) under grant "En la Frontera De La Arqueolog\'ia Gal\'actica: Evolución De La Materia Luminosa y Oscura De La Vía Láctea Y Las Galaxias Enanas Del Grupo Local En La Era De Gaia: los aspectos qu\'imicos" (FOGALERA-Chem) and the European Regional Development Fund (ERDF) with reference PID2023-150319NB-C22/10.13039/501100011033. This work is supported by the ERC Grant (ERC Advanced Grant ChronoGal, GA 101201412. Funded by the European Union;  
A.M.A. acknowledges support from the Swedish Research Council (VR 2020-03940, VR 2025-05167) and the Crafoord Foundation via the Royal Swedish Academy of Sciences (CR 2024-0015); 
M.T.B. thanks the Ministerio de Ciencia, Innovaci\'on y Universidades of the Spanish Government for her Beatriz Galindo Fellowship and acknowledges funding from MICIU/AEI /10.13039/501100011033 and by FEDER (UE) under project PID2021-127786NA-100;
A.M. acknowledges support from the ERC Consolidator Grant funding scheme (project ASTEROCHRONOMETRY, G.A. n. 772293); 
M.M. acknowledges the financial contribution from the grant PRIN-MUR 2022 2022NY2ZRS 001 ``Optimizing the extraction of cosmological information from Large Scale Structure analysis in view of the next large spectroscopic surveys'' supported by NextGenerationEU and the financial contribution from the grant ASI n. 2024-10-HH.0 ``Attività scientifiche per la missione Euclid – fase E''
A.F. acknowledges funding from the Large Grant INAF-2024 ``Spectral Key features of Young stellar objects: Wind-Accretion LinKs Explored in the infraRed (SKYWALKER)''.%
M.B. is funded by the European Union (ERA Fellowship, DiaLoGues, 101180670).
M.P. acknowledges support from HORIZON-INFRA-2024-DEV-01-01 – Research Infrastructure Concept Development, through the project WST: The Wide-Field Spectroscopic Telescope (Grant No. 101183153). 
T.M.\ is granted by the BELSPO Belgian federal research program FED-tWIN under the research profile Prf-2020-033\_BISTRO. 
S.C. and D.V. acknowledge funding by the European Union - NextGenerationEU RFF M4C2 1.1 PRIN 2022 project ``2022RJLWHN URKA'' and  by INAF 2023 Theory Grant ObFu 1.05.23.06.06 ``Understanding R-process \& Kilonovae Aspects''.
M.P. thanks the European Union’s Horizon 2020 research and innovation programme (ChETEC-INFRA -- Project no. 101008324), the Lend\"ulet Program LP2023-10 of the Hungarian Academy of Sciences, the Hungarian NKFIH via K-project 138031 and NKKP Advanced grant 153697, the IReNA network by NSF AccelNet (Grant No. OISE-1927130) and the ERC Synergy Grant Programme (Geoastronomy, grant agreement number 101166936, Germany)%
I.C. acknowledges funding from the Italian National Recovery and Resilience Plan (PNRR), funded by the European Union -- NextGenerationEU (Mission~4, Component~2, Investment~1.2; CUP: C23C24001360006).
I.K. acknowledges ERC support (grant agreement No. 101117455).
F.R. is a fellow of the Alexander von Humboldt Foundation, and acknowledges support by the Klaus Tschira Foundation. F.R. also acknowledges I.N.A.F. for the 1.05.24.07.02 Mini grant 2024 ``GALoMS - Galactic Archaeology for Low Mass Stars'' (PI: C.T. Nguyen). 
%
C.L. acknowledges funding from a UKRI Future Leader Fellowship (grant numbers MR/S035214/1 and MR/Y011759/1).
D.R. research was supported by the Italian National Institute for Astrophysics (INAF) through the program “Finanziamento della Ricerca Fondamentale”, projects \emph{``An in-depth theoretical study of CNO element evolution in galaxies''} Fu.~Ob.~1.05.12.06.08 (PI: D.~Romano).
R.S. acknowledges support from the National Science Centre, Poland, research grant 2022/47/I/ST9/02358.
K.L. acknowledges funds from the Knut and Alice Wallenberg Foundation.
J.R.B. was funded by STFC under consolidated grants ST/T000295/1 and ST/X001164/1.

%
%

\endgroup

\end{document}